\documentclass[11pt,showpacs,preprintnumbers]{revtex4}

\usepackage{graphicx}% Include figure files
\usepackage{dcolumn}% Align table columns on decimal point
\usepackage{bm}% bold math
\usepackage{amssymb}
\usepackage{amsmath}
\usepackage{epsfig}    
\def\beq{\begin{equation}}
\def\eeq{\end{equation}}
\def\ba{\begin{array}}
\def\ea{\end{array}}
\def\bea{\begin{eqnarray}}
\def\eea{\end{eqnarray}}

\def\sq2{\sqrt{2}}
\def\End{\end{document}}
\newcommand{\gm}{\gamma^\mu}
\newcommand{\smn}{\sigma^{\mu \nu}}
\newcommand{\Wmn}{W_{\mu \nu}}

\newcommand{\sqd}{\sqrt{2}}
\newcommand{\es}{\ensuremath{\emptyset}}
\renewcommand{\P}{\ensuremath{\mathcal{P}_e}}
\renewcommand{\S}{\ensuremath{\mathcal{S}}}
\newcommand{\bbar}{\ensuremath{\overline{b}}}
\newcommand{\tbar}{\ensuremath{\overline{t}}}

\newcommand{\met}{\ensuremath{\not\!\!E_T}}

\begin{document}

\title{Analysis of the top-quark charged-current coupling at the LHeC}

\author{% 
{I.~A.~Sarmiento-Alvarado},
{Antonio~O.~Bouzas}~~and~~{F.~Larios}\footnote{
larios@mda.cinvestav.mx, corresponding author.}}

\affiliation{%
%\address{\vspace*{5mm}
\vspace*{2mm} 
Departamento de F\'{\i}sica Aplicada,
CINVESTAV-M\'erida, A.P. 73, 97310 M\'erida,
Yucat\'an, M\'exico}

\begin{abstract}
  In the context of $SU(2)_L\times U(1)$ dimension six operators we
  study the potential of the LHeC to provide information on top
  quark effective interactions.  We focus on single antitop
  production and how it is affected not only by the effective $tbW$
  coupling but also by four-fermion operators. Compared to the LHC,
  the LHeC provides a cleaner environment to make a precise
  measurement of the top quark production cross section.  Therefore,
  this machine would give a much better assesment of $V_{tb}$ in the
  context of the SM or $V_L$ in the context of higher dimension
  operators.  The LHeC could also give a slightly better measurement
  for $V_R$.  For $g_R$ the HL-LHC precise measurements
  of $F_L$ and $F_R$ (the $W$-boson helicity decay ratios of top)
  would yield better constraints than those obtained by the LHeC.
  Lepton-quark contact interactions would also be
  significantly better probed by the LHeC, since the only way of
  measuring them at the LHC would be through leptonic top decay which
  is hardly sensitive to these interactions.
\end{abstract}

\maketitle

\section{Introduction}

The Large Hadron Electron Collider (LHeC) is the proposal of a new
electron beam with an energy $E_e=60$ GeV or higher, to collide with
one of the $7$ TeV LHC proton beams.  The expected luminosity of the
LHeC could reach $100\;{\rm fb}^{-1}$ as the machine would run
simultaneously with the high luminosity phase of the LHC (HL-LHC) that
is expected to achieve a luminosity of $3000\;{\rm fb}^{-1}$
\cite{lhecreport,bruening}.  Such a facility would be very useful in
understanding, among other issues, parton and gluon interactions at
very low $x$ and very high $Q^2$, thus providing much needed
complementary information for the physics program of the LHC.  It
could also be used to discover new resonances such as leptoquarks
\cite{lptqrks} and heavy Majorana neutrinos \cite{dua15}.  Moreover,
compared to the LHC the LHeC gives us a much cleaner environment that
could furnish very accurate information on Higgs physics
\cite{higgsmellado,higgslhec} and trilinear gauge boson couplings
\cite{bosonslhec}.
%%%%%%%%%%%%%%%%%%%% cambio 1 arbitro 1
The proposed detector of the LHeC will have to meet special
requirements such as \cite{lhecreport}: 1) being able to detect a
backward scattered electron at angles up to $179\deg$ and a forward
scattered final state at $1\deg$; 2) optimum scale calibration for the
measurement of $\alpha_s$; 3) high resolution for the reconstruction
of multi-jet final states; 4) good electron-hadron separation as
required for electron identification at high rapidity; 5) hermetic
calorimetry as required for a precise determination of $\not\!\!E_T$
in charged current processes (single top production).  We believe that
these requirements will enable the LHeC to measure single top
production, on the one hand with a precision in the few percent range
for the leptonic mode, and on the other hand with a somewhat lower
precision in the hadronic mode.  This latter mode has not been
measured in either the Tevatron or the LHC due to the enormous
multi-jet activity.  Therefore, the LHeC could provide a unique window
in this case.
%%%%%%%%%%%%%%%%%%%%%%%%%%%%%%%%%%%%%%%%%%%%%%%%%%%%%%%%%%%

In this work we focus on single top quark production at the LHeC and
on finding out how much information we can obtain on the dimension six
$SU(2)_L \times U(1)$ gauge-invariant operators that involve the top
quark. Excluding flavor changing and CP violating effects, there are
31 independent operators.  The LHC by itself will not be able to probe
the effects of all these interactions.  The input from future
colliders like the LHeC and the ILC would be essential in making a
complete analysis of the top quark.

Whenever a study of single top production as a probe of effective 
top-quark interactions is made, the framework of choice is the set of four
independent on-shell $tbW$ couplings with the structures $\gamma_\mu
P_{L,R}$ and $\sigma_{\mu \nu} q^\nu P_{L,R}$.  However, it has been
pointed out the importance of including the effects of the off-shell
$W$ to work out a truly complete description \cite{bach}.  Those
effects may be taken into account by means of an appropriate set of
four-fermion operators, which should therefore be included in a
complete study of single top production.  In this paper we adopt this
more general framework and base our analysis on the full basis of
$SU(2)_L \times U(1)$ operators including the relevant four-fermion
ones.

Before discussing the LHeC sensitivity to the effective couplings we
address the question of how well will this machine perform in
comparison with the HL-LHC, with which it will run concurrently.  We
do this by means of a simplified evaluation of the limits on the
couplings obtained from $W$-boson helicity fractions in top decays at
the LHC, and from single top production at both the LHC and the LHeC.

In our study of the LHeC sensitivity to the effective couplings in
single top production we obtain bounds on those couplings from global
observables (as opposed to differential ones) such as the cross
section and several kinematical asymmetries, computed at leading order
(LO) at the partonic level.  The experimental errors we assume are
based on a detailed study of the Standard Model (SM) backgrounds.  All
of our computations and estimates are carried out at three
electron-beam energies $E_e=60$ and 140 GeV, which are typical
proposed operational energies \cite{lhecreport}, and the higher
$E_e=300$ GeV.  This allows us to ascertain the energy dependence and
the stability of our results.  Another parameter we take into account
is beam polarization, on which the sensitivity of observables to
effective interactions involving right-handed electrons depends
strongly.

In a previous report of single antitop production at the LHeC
constraints were obtained for the four on-shell $tbW$ couplings at a
fixed $E_e=60$ GeV, and based on a set of four kinematical distributions
for the leptonic channel and six for the hadronic channel \cite{mellado}.
Our work is an extension of that report in several ways:
(1) we adopt the general framework of gauge invariant operators,
that include the four $tbW$ couplings plus four lepton-quark
contact interactions, (2) we consider three values of $E_e=60$, $140$
and $300$ GeV, and (3) we obtain bounds based on the cross section as
well as six kinematic observables for the leptonic channel and
twenty-two for the hadronic channel (not all improve the bounds from
the cross section, we show results only for those that do).

The paper is organized as follows.  In section~\ref{secoperators} we
list the dimension six operators that are relevant for single antitop
production at the LHeC.  Since the LHeC will run along with the
HL-LHC, in Section~\ref{tdecandsingle} we present an estimate of the
sensitivity to the top-quark effective couplings of the LHeC as
compared to the HL-LHC.  In section~\ref{sec:sm} we make a systematic
study of the SM signal and background processes in antitop production
and decay at the LHeC, discuss the appropriate cuts and their
efficiency for background suppression, and estimate the experimental
errors expected in the leptonic and hadronic channels.  In
section~\ref{sec:eff.oper} we present the bounds we obtain on the
anomalous couplings.  We discuss bounds obtained from the cross
section, and make an extensive survey of asymmetries of different
kinematical observables to find the most sensitive among them.  In
this section we also discuss in detail the impact of electron-beam
energy and polarization.  Finally, in
section~\ref{conclusions} we present our conclusions.

%%%%%%%%%%%%%%%%%%%%%%%%%%%%%%%%%%%%%%%%%%%%%%%%%%
\section{ Top quark dimension six operators}
\label{secoperators}

The discovery of the Higgs boson at the LHC, although still to be
thoroughly tested in the future, suggests that the SM is indeed the
correct model that explains all collider experiments up to the
electroweak scale.  However, the SM may be seen as an effective theory
that is valid below a certain scale $\Lambda$ \cite{wells}.  At and
above this scale the heavy degrees of freedom of a larger theory
become apparent.  Therefore, new physics effects may be properly
described by an effective Lagrangian of the form: 
\bea 
{\cal L} = {\cal
L}_{\rm SM} + \frac{1}{\Lambda^2} \sum_k \left( C_k O^{(6)}_k + {\rm
  h.c.} \right) \; + \cdots\;,  \nonumber 
\eea 
where the ellipsis stands for operators of dimension higher than
six.  Many years ago a long list of gauge invariant dimension-six
operators was introduced in Ref.~\cite{buchmuller86}.  Some time
afterwards it became apparent that some of the operators involving the
top quark were in fact redundant \cite{wudka04}.  In particular, from
the original list of 14 top-gauge boson operators it has been shown
that only 8 are truly independent \cite{aguilaroperators}.  A revised
general list of all the operators including those without the top
quark, appears in Ref.~\cite{rosiek}.  It has been pointed out that
this list in particular satisfies a so-called criterion of
Potential-Tree-Generated operators, so that they may have the largest
possible coefficients \cite{einhorn}.  The complete list of dimension
six operators for the top quark can be divided into three classes: (1)
Operators that involve gauge bosons \cite{aguilaroperators}, (2)
Operators that involve the Higgs but no gauge bosons
\cite{aguilartophiggs}, and (3) Four-fermion contact-interaction
operators \cite{aguilar4}.  There are eight of the first type, one of
the second type and 22 quartic operators.  However, when the different
flavor combinations as well as the CP even and odd parts are
considered we end up with many independent variations of these
operators.  In the context of the single top production process $ep\to
\nu \bar t$ at the LHeC, however, we have only four gauge boson-top
quark and four lepton-quark operators that are relevant.  The
description of new physics effects in single antitop quark
production at the LHeC will thus be described by an effective
Lagrangian of the form ${\cal L}_\mathrm{eff}={\cal L}_{\rm SM}+{\cal
  L}_{4f}+{\cal L}_{tbW}$.  We discuss in detail the Lagrangians
${\cal L}_{4f}$ and ${\cal L}_{tbW}$ below.

%%%%%%%%%%%%%%%%%%%%%%%%%%%%%%%%%%%%%%%%%%%%%%%%%%%%%%%%%%%%%%%%%%
\subsection{Effective four fermion couplings of the top quark}
\label{quarticops}
%%%%%%%%%%%%%%%%%%%%%%%%%%%%%%%%%%%%%%%%%%%%%%%%%%%%%%%%%%%%%%%%%%%

Operators for four-fermion vertices have been analyzed in full
generality in \cite{aguilar4} in the context of the effective $SU(2)_L
\times U(1)$ gauge-invariant Lagrangian.  Furthermore, a recent study
on quark-quark operators in single top production at the LHC has been
carried out in \cite{cao,bach4}. Apart from flavor assignments, a
complete list of 25 independent operators is given in Eqs. (2)-(5) of
\cite{aguilar4}.  In this list, 3 operators are for lepton-lepton, 10
are for lepton-quark and 12 are for quark-quark interactions.  As
expected, if we take into account the different flavor combinations,
we can find hundreds of variations.  For the LHeC single-top process
we are interested in flavor-diagonal lepton-quark operators involving
the top quark and the first family of leptons, which selects only
eight operators out of the 25 listed in \cite{aguilar4}.  They are
shown in Table~\ref{leptonquark}, where $q_{Li}$ and $\ell_{Li}$ are
the left-handed quark and lepton doublets, and $e_{R1}$, $u_{Rj}$,
$d_{Rj}$ are the right-handed electron and up and down quark singlets.
%%%%%%%%%%%%%%%%%%%%%%%%%%%%%%%%%%%%%%%%%%%%%%%%%%%%%%%%%%
\begin{table}
\begin{center}
  \begin{tabular}{|c|c|}\hline
    Charged Current & Neutral Current \\\hline\hline 
${\cal O}^{1331}_{\ell q'} = {\bar \ell}_{L1} \gamma^\mu q_{L3} 
\; \bar q_{L3} \gamma_\mu  \ell_{L1}$
&
 ${\cal O}^{1133}_{\ell q} = {\bar \ell}_{L1} \gamma^\mu \ell_{L1}
\; \bar q_{L3} \gamma_\mu q_{L3}$  
\\\hline
${\cal O}^{1133}_{qde} = {\bar \ell}_{L1} e_{R1}
\; \bar d_{R3} q_{L3}$
&
${\cal O}^{1331}_{\ell u} = {\bar \ell}_{L1} u_{R3}
\; \bar u_{R3} \ell_{L1}$
\\\hline
${\cal O}^{3113}_{q\ell \epsilon} = \bar q_{L3} e_{R1} \;
{( {\bar \ell}_{L1} \epsilon )}^T u_{R3}$
&
${\cal O}^{1331}_{qe} = \bar q_{L3} e_{R1}
\; \bar e_{R1} q_{L3}$
\\\hline
${\cal O}^{1133}_{\ell q\epsilon} = {\bar \ell}_{L1} e_{R1}
\; {( \bar q_{L3} \epsilon )}^T u_{R3}$
&
${\cal O}^{1133}_{eu} = \bar e_{R1} \gamma^\mu e_{R1}
\; \bar u_{R3} \gamma_\mu u_{R3}$
\\\hline
  \end{tabular}
\end{center}
\caption{Flavor diagonal quartic Lepton-Quark operators
that involve top quarks and electrons. The ones on the
left column contribute to single top production at
the LHeC.}
\label{leptonquark}
\end{table}
%%%%%%%%%%%%%%%%%%%%%%%%%%%%%%%%%%%%%%%%%%%%%%%%%%%%%%%%%%

The operators on the left column of Table~\ref{leptonquark} are of
CC type, therefore relevant to the LHeC, whereas those on the
right column are of NC type, relevant for $t\bar t$ production at a
$e^+e^-$ collider like the ILC \cite{aguilar4}.  The CC operator
${\cal O}^{1331}_{\ell q'}$, which contains terms of the form $\bar
\nu_L \gamma^\mu e_L \bar b_L \gamma_\mu t_L$, is of special interest
as it is the only one that can lead to an interference term with the
SM amplitude.  It can be generated by a heavy $W'$ or a vector
leptoquark \cite{wprime,hewett}.  The other three operators, that
involve a right-handed electron can be generated by a heavy charged
Higgs or a scalar leptoquark \cite{delepine}.
The four-fermion Lagrangian defined by the CC type
operators in Table~\ref{leptonquark} is then: 
\bea 
\Lambda^{2} {\cal  L}_{4f} &=&
C_{1} O^{1331}_{\ell q'} + \left[ C_2 O^{1133}_{qde} +
C_3 O^{3113}_{q \ell \epsilon} + C_4 O^{1133}_{\ell q\epsilon} \; +
  h.c. \right]
\label{fourfermion} \\
&=&
C_{1} (\bar \nu_L \gamma^\mu t_L \bar b_L \gamma_\mu e_L + h.c.) +
[C_2 \bar \nu_L e_R \bar b_R t_L + C_3 \bar b_L e_R \bar \nu_L t_R 
+ C_4 \bar \nu_L e_R \bar b_L t_R \; + h.c.], 
\nonumber
\eea
where ${\cal O}^{1331}_{\ell q'}$ is already Hermitian, but the other
three operators are not.  The coefficient $C_1$ must therefore be real, and
$C_{2,3,4}$ are complex, their imaginary parts giving rise to CP-odd
interactions.

%%%%%%%%%%%%%%%%%%%%%%%%%%%%%%%%%%%%%%%%%%%%%%%%%%%%%%%%%%%%%%%%%%
\subsection{Effective $tbW$ couplings of the top quark}
\label{tbwops}
%%%%%%%%%%%%%%%%%%%%%%%%%%%%%%%%%%%%%%%%%%%%%%%%%%%%%%%%%%%%%%%%%%%

The complete list of top-gauge boson operators is given in Eqs.\ (3)
and (4) of \cite{aguilaroperators}.  Notice that not all of the
operators in that list are independent.  The operators that modify the
CC effective $tbW$ coupling are:
\begin{equation}
  \label{gaugeoperators}
  \begin{aligned}
O_{\phi q}^{(3,ij)} &= \frac{i}{2}
\phi^\dagger \tau^I D_\mu \phi \; \bar q_{Li} \gm \tau^I q_{Lj} \,,
&
O_{\phi \phi}^{ij} &= i {\tilde \phi}^\dagger D_\mu \phi \; \bar
u_{Ri} \gm d_{Rj} \,,\\
O_{uW}^{ij} &= \bar q_{Li} \smn \tau^I u_{Rj} \;
\tilde \phi \, \Wmn^I \,,
&
O_{dW}^{ij} &= \bar q_{Li} \smn \tau^I d_{Rj} \; \phi \, \Wmn^I \,.   
  \end{aligned}
\end{equation}
We use standard notation with $I$, $J$, $K$ $SU(2)$ gauge indices,
$\tau^I$ Pauli matrices, and $\phi$ the SM Higgs doublet with $\tilde
\phi = i\tau^2 \phi^*$ \cite{aguilaroperators}.  For every operator in
(\ref{gaugeoperators}) there are 3 or more variations depending on the
flavor content.  Throughout this paper we consider only
flavor-diagonal interactions, corresponding to flavor indices $ij=33$
in (\ref{gaugeoperators}).  The flavor-changing (FC) combinations $13$
and $23$ do contribute to single-top production but we will not
consider them.  The operator $O_{\phi q}^{(3,33)}$ is Hermitian, and
the other three flavor-diagonal operators can be decomposed into
Hermitian and anti-Hermitian parts.  In Table I of \cite{bouzaslhec}
the CP-even and CP-odd parts of these operators are displayed in
detail. In Table II of \cite{bouzaslhec} the constraints from
electroweak data and $b\to s\gamma$ observables can be found.  Those
constraints were found by taking into account only one operator at a
time, but in general there are correlations among coefficients
\cite{willenbrock,bouzas13,drobnak}.

We write the effective $tbW$ Lagrangian as:
\bea
\Lambda^{2} {\cal L}_{tbW} &=& 
C_{\phi q} O^{(3,33)}_{\phi q} + [C_{\phi \phi} O^{33}_{\phi \phi}
+ C_{tw} O^{33}_{uW} + C_{bW} O^{33}_{dW} + \; h.c.], 
\label{tbwlagrangian}
\eea
where $C_{\phi q}$ is real and $C_{\phi \phi}$, $C_{tw}$, $C_{bW}$ are
complex, their imaginary parts multiplying the anti-Hermitian parts of
the corresponding operators in (\ref{tbwlagrangian}) and giving rise
to CP-odd interactions.  A recent phenomenological study on the
imaginary parts of the effective $tbW$ couplings at the LHC is given
in \cite{onofre}.

We have not included in (\ref{tbwlagrangian}) the operator
$O_{qW}^{ij} = \bar q_{Li} \gm \tau^I D^\nu q_{Lj} W_\mu^I$, that is
independent of the set of operators (\ref{gaugeoperators}) and gives
an important contribution to the single-top production process.  In
fact, as pointed out in \cite{bach}, a complete parametrization of new
physics effects from the trilinear $tbW$ coupling should involve the
contribution from this operator.  However, $O_{qW}^{ij}$ can be
written as a linear combination of the operators
(\ref{gaugeoperators}) and the four-fermion operators of CC type in
Table \ref{leptonquark}.  We can use the equation of motion for the $W$
field \cite{bach}:
\bea
(D^\nu W_{\nu \mu})^I = \frac{g}{2} \sum^3_{k=1}
\left( \bar q_{Lk} \gamma_\mu \tau^I q_{Lk}
+ \bar {\ell}_{Lk} \gamma_\mu \tau^I {\ell}_{Lk} \right)
+ i\frac{g}{2} \left[
\phi^\dagger \tau^I D_\mu \phi - (D_\mu \phi)^\dagger
\tau^I \phi  \right] \, ,\nonumber
\eea
to apply it in the expression $O_{qW}^{ij} +
( O_{qW}^{ji} )^\dagger = \left( 
\bar q_{Li} \gamma^\mu \tau^I q_{Lj} \right) (D^\nu W_{\nu \mu})^I$.
Then, by means of a Fierz rearrangement
of the field operators, we can write
$
\bar q_{Li} \gamma^\mu \tau^I q_{Lj} \;
\bar {\ell}_{Lk} \gamma_\mu \tau^I {\ell}_{Lk}
= 2{O}^{kjik}_{\ell q'}-{O}^{kkij}_{\ell q} \;,
$
where the operators $O^{ijkl}_{\ell q(')}$ and $O^{ijkl}_{qq(')}$
are four-fermion lepton-quark and quark-quark contact
interactions \cite{aguilar4}.  Therefore:
\bea
O_{qW}^{ij} + ( O_{qW}^{ji} )^\dagger &=&  \frac{g}{2}
\left(   O_{\phi q}^{(3,ij)} + ( O_{\phi q}^{(3,ji)} )^\dagger
\right)  \label{oqwrelation} \\
&+& g \sum^3_{k=1} \left( {O}^{kjik}_{\ell q'} - 
\frac{1}{2} O^{kkij}_{\ell q} \right)  + 2g
\sum^3_{k=1} \left( {O}^{kjik}_{qq'} - 
\frac{1}{2} O^{kkij}_{qq} \right) \, ,\nonumber
\eea
which in the case of single-top production at the LHeC of interest to
us reduces to $O_{qW}^{33} + ( O_{qW}^{33} )^\dagger = g O_{\phi
  q}^{(3,33)} + g O^{1331}_{\ell q'} + \cdots$.  
We take advantage of this relation and choose to perform our
study with the quartic terms instead of $O_{qW}^{ij}$.
Notice that these terms can enter in both the top decay and
production process.

As is common practice in the literature, we can write down
the effective $tbW$ couplings in terms of form factors.
Let us separate those terms in ${\cal L}_{tbW}$ that generate
the effective $tbW$ vertex \cite{bach,aguilaroperators}:
\bea
{\cal L}_{tbW} &=& -\frac{g}{\sqd}
\bar b  \left(  \gamma^\mu (V_L P_L + V_R P_R)
- \frac{i\sigma^{\mu \nu}q_\nu}{m_W} 
(g_L P_L + g_R P_R) \right) t  W^{-}_\mu \; + h.c. 
\label{bachfactors}
\eea
Notice that $V_{L(R)} \equiv F_1^{L(R)}$ and
$g_{L(R)} \equiv -F_2^{R(L)}$ as in Ref.~\cite{bouzaslhec}.
The relation between the form factors and the operator
coefficients in (\ref{tbwlagrangian}) is given by: 
\bea 
V_L = V_{tb}+ \frac{1}{2} \frac{v^2}{\Lambda^2} C_{\phi q}
\, , \;\;
V_R = \frac{1}{2} \frac{v^2}{\Lambda^2} C_{\phi \phi}
\, , \;\;
g_R = \sqd \frac{v^2}{\Lambda^2} C_{tW} \, , \;\;
g_L = \sqd \frac{v^2}{\Lambda^2} C_{bW} \; .
\label{factors}
\eea
For concreteness we set $\Lambda \equiv 1$ TeV, and write the
dimensionful parameters in the operators in units of TeV, namely,
$v=0.246$, $m_t=0.173$ and $m_W=0.08$.  We can go back to a general
$\Lambda$ by just replacing the anomalous coupling constants $C$ by
$C/\Lambda^2$.

%%%%%%%%%%%%%%%%%%%%%%%%%%%%%%%%%%%%%%%%%%%%%%%%%%%%%%%%%%%
\section{Top quark decay and single top production}
\label{tdecandsingle}

The LHeC will run along with the high luminosity phase of the LHC, so
that by the time the LHeC experiment delivers useful data so will the
other experiments at the LHC.  Precision measurements of $t\bar t$
production, top decay, LHC and LHeC single-top production, and more,
will be analysed simultaneously.  In this section we present a
broad-brush picture of what to expect from the LHeC as compared to the
performance of the LHC by the time the data from the very high
luminosity phase becomes available.  For that purpose, we use three
independent observables relevant to the study of the $tbW$ and top
quark quartic couplings, namely, the $W$-boson helicity fractions, the
LHC and the LHeC single-top production cross sections.
%%%%%%%%%%%%%%%%%%%%%%%%%%%  cambio 3 arbitro 1
Below, we proceed in two steps.  First, we compute the observables in
the approximation of two-body final states, in order to estimate the
bounds on the anomalous couplings from recent measurements by CMS of
$W$ helicity fractions in top decays, and of single top production
\cite{cms2013,cms2012,cms2014}.
Measurements on single top production by the ATLAS collaboration are
also available but with a larger error \cite{cmsatlas2014}; for instance,
at $\sqrt{s}=8$ TeV
ATLAS measures $\sigma^{t-chan}(t+\bar t)=82.6$ with $14.2\%$ error,
whereas CMS measures $\sigma^{t-chan}(t+\bar t)=83.6$ with $9.3\%$.
We will only use the CMS results, which provide with the most
constraining bounds.  We also compare our results with more
%%%%%%%%%%%%%%%%%%%%%%%%%%%%%%%%%%%%%%%%%%%%%%%%
precise analyses done in the literature to test whether our formulas
yield good approximations. Finally, we apply this same approach to
make a conservative estimate of what the bounds will be like once the
data from the HL-LHC and the LHeC are available.  In the remainder of
this section we take the anomalous couplings to be real for
simplicity.  We should bear in mind that the estimated bounds for
the LHeC and the HL-LHC in this section are only conservative
approximations.  We expect both machines to actually yield better
constraints based on all the by then available data and on more
powerful techniques of data analysis.

\subsection{The polarized $t\to b W$ decay ratios $F_0$, $F_L$ and $F_R$}

It is well known that the helicity of the $W$ boson in the $t\to bW$
decay can be used to study the effective $tbW$ coupling
\cite{whelicity,aguilarwhel}.  The CMS collaboration has published
precise measurements of the decay ratios and have used their results
to set bounds on the effective $g_L$ and $g_R$ parameters in
(\ref{bachfactors}) \cite{cms2013}.  Another recent study based on CMS
measurements can be found in \cite{fabbrichesi}.  In our study, we
would like to use a simplified set of formulas to help us make a
conservative estimate of the possible bounds that the future HL-LHC
could achieve and compare them with the estimated bounds from the LHeC
experiment.

The tree level decay $t\to b W$  with the
general $tbW$ vertex has been analyzed in \cite{whelicity,aguilarwhel}.
At second order in the anomalous couplings,  the longitudinal, left-handed
and right-handed  $W$ polarization fractions are:
\begin{equation}
  \label{ratios}
  \begin{aligned}
    F_0 &= F_0^\mathrm{SM}-0.926
  g_L^2+0.709g_R+0.457g_R^2+0.709g_LV_R-0.709g_R \, \delta V_L,\\
    F_L &= F_L^\mathrm{SM}-0.468g_L^2-0.709g_R-0.457g_R^2 + 
       0.591g_LV_R-0.303V_R^2+0.709g_R \, \delta V_L,\\
    F_R &= F_R^\mathrm{SM}+1.394g_L^2-1.300g_LV_R+0.303 V_R^2,
  \end{aligned}
\end{equation}
where we have set $m_t=172.5$ GeV and $m_t/m_W \; =2.145$, and we have
neglected terms proportional to $m_b$.  In this approximation the SM
helicity fractions are $F_0^\mathrm{SM}=0.697$, $F_L^\mathrm{SM}=0.303$
and $F_R^\mathrm{SM}=0$. At higher order and with $m_b$ terms there is a
small but non-zero fraction $F_R=0.0017$, and a slight $2.6\%$ increase
in the $F_L$ value as shown in Table~\ref{expresults}.
In the limit where $V_R=g_R=g_L=0$,
the expressions in (\ref{ratios}) are independent of $V_L$.  Hence,
these quantities can not be used to probe that coupling (or the CKM
$V_{tb}$ coefficient).  It is also apparent from (\ref{ratios}) that
$F_L$ is mostly sensitive to $g_R$, and $F_R$ is mostly sensitive to
$g_L$ and $V_R$.

In principle, there are also contributions to the top quark width
coming from the quartic operators that we are considering here.
However, they are about two orders of magnitude lower than the
contribution from the $tbW$ coefficients \cite{celine}.  In this work
we do not include the negligible effects of four-fermion operators
on top decay.  

Below, we will show the bounds obtained using the latest (published)
measurement by CMS on $F_L$ and $F_R$ \cite{cms2013}.  In doing so, we
will use for $F_{0,L,R}^\mathrm{SM}$ in (\ref{ratios}) the SM values with
full QCD corrections and $m_b$ terms as shown in Table~\ref{expresults}.
The experimental data on the branching ratios $F_{0,L,R}$ are from last
year CMS publication \cite{cms2013}.  The errors presented there are
evenly distributed in statistical and systematic origins.  For the
purpose of making an estimate of a future measurement in the HL-LHC era
we will assume that the errors then will be of mainly systematic origin
and about half the size of current results (see Table~\ref{smvalues}).

Notice that there is a recent CMS internal report with new results
(and with lower errors) \cite{cmsinter}. These values have already
been used in the literature \cite{aguilarnewcms,fabbrichesi,onofre}.
We have not used these results in our estimate for the following
reason: their value of $F_L$ is $0.35 \pm 0.01 \pm 0.024$ and it is
already more than $1\sigma$ above the SM prediction.  Therefore, the
allowed region for $g_L$ when $g_R \simeq 0$ is dramatically reduced
as compared to the region allowed by the previous year's data in
\cite{cms2013}.  See Table~I in \cite{fabbrichesi}.  It is likely that
future studies will shift the $F_L$ ratio back to the SM prediction,
and then the allowed region will look more like the one obtained with
the latest published report \cite{cms2013}.

%%%%%%%%%%%%%%%%%%%%%%%%%%%%%%%%%%%%%%%%%%%%%%%%

\subsection{Single top quark production}
\label{sec:single.top.prod}

Below we will discuss the constraints on the effective
couplings that come from $\sigma^{t+s}$ the inclusive
$t$-channel plus $s$-channel production of single top at the
Tevatron \cite{cdf2014}.  Also, we will use the inclusive
$\sigma^{t}$ t-channel production measured by CMS with
both $\sqrt{s}= \; 7$ and $8$ TeV \cite{cms2012,cms2014}.
From these measurements, the CMS collaboration obtains
a $68\%$ confidence level (CL)+ value for the SM $tbW$ coupling \cite{cms2014}:
$|V_{tb}| = 0.998 \pm 0.038 ({\rm exp.}) \pm 0.016 ({\rm theo.})$.
This is consistent with the value $V_L = 0.994 \pm 0.046$
obtained with our simplified method.  In addition, we will
make an estimate from a similar measurement at the future
HL-LHC with $\sqrt{s}=14$ TeV.
The SM values as well as the most recent experimental
measurements we use for our analysis are given in
Table~\ref{expresults}.

\begin{figure}[t]
  \centering
\includegraphics[scale=0.95]{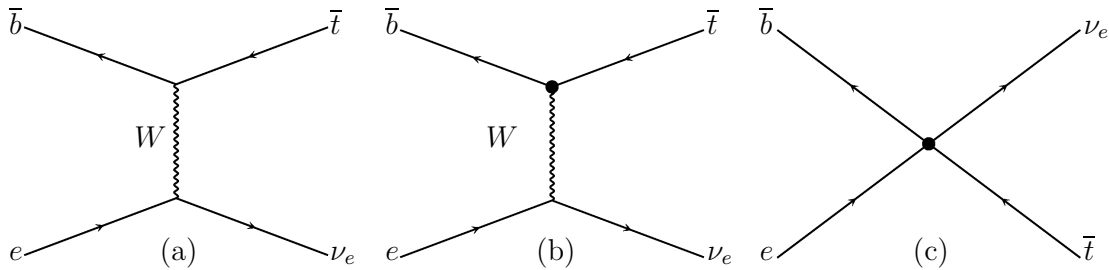}  
\caption{Feynman diagrams for single-top production in $e^-p$ collisions.
The top decay process is assumed to occur as in the SM.}
  \label{figsingle}
\end{figure}

%%%%%%%%%%%%%%%%%%%%%%%%%%%%%%%%%%%%%%%%%%%%%%%%%%%%%%%%%%
\begin{table}
\begin{center}
  \begin{tabular}{|c|c|c|c|c|c|}\hline
 & $\sigma^{t+s}(t+\bar t)$ & 
$\sigma^{t}(t+\bar t)$ & $\sigma^{t}(t+\bar t)$ & $F_L$ & $F_R$
\\  & Tevatron & LHC7 & LHC8 &  CMS & CMS  \\\hline\hline
Theory & $3.38$ & $65.9$ & $87.2$  & $0.311$ & $0.0017$ 
\\\hline
Experiment & $3.04 \pm 0.57$ & $67.2 \pm 6.1$ &
$83.6 \pm 7.8$ & $0.310 \pm 0.031$ & $0.008 \pm 0.018$
\\\hline
  \end{tabular}
\end{center}
\caption{Theoretical values and experimental measurements of cross
sections in units of pb and decay ratios used in this study.
References for the Tevatron $\sigma^{t+s}(t+\bar t)$, the LHC
$\sigma^{t}(t+\bar t)$ and the top decay $W$-boson helicity ratios
$F_L$ and $F_R$ are \cite{kidonakis1,kidonakis2,korner} (theory)
and \cite{cdf2014,cms2014,cms2013} (experiment).}
\label{expresults}
\end{table}
%%%%%%%%%%%%%%%%%%%%%%%%%%%%%%%%%%%%%%%%%%%%%%%%%%%%%%%%%%

We can write down the single top production cross section at hadron
colliders in terms of the effective couplings.  Since the $tbW$ vertex
occurs in the production and decay vertices, we expect a polynomial of
fourth degree.  However, in this simplified analysis we will take the
top decay exactly as in the SM with $V_L =1$.  Therefore, the
exppresion for the cross section contains only linear and square
terms.  We find it convenient to define the ratio $R_\sigma$ as
follows: \bea R_\sigma \equiv \frac{\sigma^{\rm SM}_{V_L}+\Delta
  \sigma} {\sigma^{\rm SM}_{V_L=1}} = V_L^2 + a_{12} g_R + a_2 g^2_R +
a_3 V^2_R + a_4 g^2_L + a_{15} g_{\times} + a_5 g^2_{\times} \; ,
\label{lhccross} 
\eea where by $\sigma^{\rm SM}$ we mean the SM prediction with $V_L$
not necessarily equal to one.  The coupling constant $g_{\times}$ in
(\ref{lhccross}) plays a role analogous to $C_1$ for four-quark
interactions, $g_{\times}/\Lambda^2$ being the coefficient of the
operator $O_{qq'}^{1331}=\bar b_L \gamma^\mu t_L \bar u_L \gamma_\mu
d_L$ in the effective Lagrangian. It was introduced in
Ref.~\cite{bach}, where the contribution of $O_{qq'}^{1331}$ was
considered along with the effective $tbW$ vertex.  For the sake of
simplicity we are considering interference terms with the SM amplitude
($V_L =1$) but we are disregarding interference terms between the
anomalous couplings (such as terms proportional to $V_R g_L$).  There
are no linear terms for $V_R$ and $g_L$ because their interference
with the SM is suppresed by $m_b$.  In this section we will only
consider terms that are the most significant.

%%%%%%%%%%%%%%%%%%%%%%%%%%%%%%%%%%%%%%%%%%%%%%%%%%
We will also define the ratio $R_\sigma$ for single
top production at the LHeC as shown in Fig.~\ref{figsingle}.
In this case we include the contribution from the four
lepton-quark operators of Eq.~(\ref{fourfermion}):
\bea
\label{lheccross0}
R_\sigma = V_L^2 + b_{12} g_R + b_2 {g_R}^2 +
b_3 {V_R}^2 + b_4 {g_L}^2 +
b_{15} C_1 + b_{5} C^2_1   
+ b_{6} {C_2}^2  + b_{7} {C_3}^2 + b_{8} {C_4}^2 
\; ,
\eea
where we have, as before, ignored the effects from top
decay.  Eventually, in Eq.~(\ref{lheccross})
below we will write again this cross section ratio in a general
context.  There, effects from the decay of top will be included.
In addition, the effective couplings which so far have been
considered real will be taken as complex.
%%%%%%%%%%%%%%%%%%%%%%%%%%%%%%%%%%%%%%%%%%%%%%%%%

The numerical values for the coefficients in (\ref{lhccross}) and
(\ref{lheccross0}), given in Appendix \ref{sec:param}, have been
obtained at LO with the program CalcHEP \cite{pukhov}.  For the
inclusive production at the Tevatron we obtain $\sigma^{t+s} (t+\bar
t)=2.5$ pb, and for the inclusive t-channel top production at the LHC
at $\sqrt{s}= \; 7,8,14$ TeV we obtain $\sigma^t (t+\bar t) = \; 56.6
\, , 74.8 \, , 221.0$ pb.  The values shown in Table~\ref{expresults}
include QCD corrections that typically increase the cross sections by
about $15\%$.  These are the values we actually use for $\sigma^{\rm
  SM}$, not the LO ones.  We obtain bounds on the effective couplings
by comparing $\sigma^{\rm SM} R_\sigma$ with the experimental values
and $1\sigma$ errors also given in Table~\ref{expresults}.  In the
case of the future LHC run at $14$ TeV and the proposed LHeC we assume
that the experimental central values will turn out to be exactly equal
to the SM prediction.  The bounds obtained will then be defined by
just the assumed experimental errors, which are listed in
Table~\ref{smvalues}.  For the case of the HL-LHC we assume that the
total error will go from the current $9.4 \%$ to a $7\%$.  For the
case of the LHeC the $4\%$ error is somewhat larger than the estimated
minimum of $3\%$ we will present in Section \ref{sec:sm} in a detailed
study of the semileptonic channel of single-top production and decay.

%%%%%%%%%%%%%%%%%%%%%%%%%%%%%%%%%%%%%%%%%%%%%%%%%%%%%%%%%%
\begin{table}
\begin{center}
  \begin{tabular}{|c|c|c|c|c|}\hline
 & HL-LHC $F_L$ & HL-LHC $F_R$ &
HL-LHC $\sigma^t (t+\bar t)$ & LHeC60
\\\hline\hline
Theory & $0.311$ & $0.0017$ & $248.0$ & $1.73$
\\\hline
Exp. Error & $\pm 0.016$ & $\pm 0.010$ & $\pm 7\%$ & $\pm 4\%$
\\\hline
  \end{tabular}
\end{center}
\caption{Assumed future experimental errors and asssociated
SM values for $W$-boson helicity ratios of top decay and
single top quark production cross sections in units of pb
at the HL-LHC and the LHeC.
The LHC (14 TeV) NNLO result is given in \cite{kidonakis2}.
The value for the LHeC ($E_e=60$ GeV) is at LO.}
\label{smvalues}
\end{table}
%%%%%%%%%%%%%%%%%%%%%%%%%%%%%%%%%%%%%%%%%%%%%%%%%%%%%%%%%%

\subsection{Current and future bounds}

Bounds obtained for each coupling and with our procedure
are shown in Table~\ref{prebounds}.  They are based on
$1\sigma$ deviations with $68\%$ CL.  For the sake of
comparing with other recent bounds in the literature we
point out that at $95\%$ CL our bounds from the current
LHC data would be $|\delta V_L|<0.09$, $|V_R|<0.42$,
$|g_L|<0.34$, and $g_R = 0.11 \pm 0.36$ which are consistent
with the bounds obtained in \cite{fabbrichesi} based on the
same CMS data of Table~\ref{expresults}  (see Figs. 2 and 4
in \cite{fabbrichesi}).  Let us discuss these
bounds for each coupling below.

Concerning the quartic operator coefficients $C_1$ and
$g_\times$.  Comparing with the bounds on the $tbW$ vertex these
coefficients may look to be weakly bounded.  However, this is
a matter of normalization.  In fact, if we compare with the
coefficients of the $tbW$ operators in
Eqs.~(\ref{tbwlagrangian})~and~(\ref{factors})
$C_{\phi q}=33.0 \delta V_L$, $C_{\phi \phi}=33.0 V_R$,
$C_{tW}=11.66 g_R$ and $C_{bW}=11.66 g_L$ we observe that
the four fermion operators are better constrained.
For instance, the LHC current bounds for $tbW$ are
$|C_{\phi q}|<1.5$, $|C_{\phi \phi}|<7.2$, $|C_{bW}|<2.1$ and
$C_{tW}=1.3\pm 2.8$, whereas the bound on
$\bar b_L \gamma^\mu t_L \bar u_L \gamma_\mu d_L$ is
$g_\times = -0.14 \pm 0.95$.  Concerning the bounds on the
coefficients $C_k$ with $k=2,3,4$ that are not shown in
Table~\ref{prebounds}.  For the LHeC at $E_e=60$ GeV they
are $|C_k|<4.5$ ($k=2,4$) and $|C_3|<2.6$ which are rather
weak compared with $|C_1|<0.34$.  The operators of
$C_k$ with $k=2,3,4$ in Eq.~(\ref{fourfermion})
are associated with a right-handed electron and there is
no interference with the SM.

Concerning $V_L$.  Since the decay ratios $F_L$ and $F_R$ change very
little with $V_L$ (or not at all if the other couplings are zero) we
do not obtain bounds from them. This also applies to the four fermion
operators.  The Tevatron result is about $11\%$ lower than the SM
prediction with $V_L=1$, so it is suggesting a lower value of $0.925$.
On the other hand, the LHC results for $7$ and $8$ TeV balance each
other.  One is $2\%$ above the theoretical values and the other is
$4\%$ below.  Their combined effect suggests $V_L=0.994\pm 0.046$
which is rather close to 1.  Notice that the allowed region we obtain
is somewhat weaker but still consistent with the more precise value of
$0.998 \pm 0.038$ obtained by CMS \cite{cms2014} ($68\%$ CL).  For the
LHC14 we assume that experimental and theoretical values coincide
within an error of $7\%$.  This directly translates to $V_L =1\pm
0.036$.  The lesson we learn here is that according to this estimate
the HL-LHC will not make a major improvement on $V_L$.  In this
scenario, if the LHeC reaches the $4\%$ error that we assume, it will
indeed yield a much better measurement of $V_L$.  On the other hand,
when we take into account the effects from the 4-fermion operators the
constraints on $V_L$ tend to relax.  This, of course also happens with
the LHC single top production \cite{bach}.  However, at the LHC there
are two independent channels of single top production ($s$-channel and
$t$-channel) that are sensitive to the 4-fermion interaction.  This
means that the combination of both measurements ends up constraining
again the allowed region of $V_L$ that is otherwise poorly constrained
by the $t$-channel measumerent alone \cite{bach,fabbrichesi}.
%%%%%%%%%%%%%%%%%%%%%%%%%%%%%%%  cambio 4 arbitro 1
Currently, the measurement of the $s$-channel cross section at the LHC
has proven to be very challenging \cite{cmsatlas2014,atlass}, and no
bounding regions can be obtained from this mode.  At the LHeC there is
no s-channel mode, and we have to rely on the t-channel only to
disentangle the effects of both interactions.
%%%%%%%%%%%%%%%%%%%%%%%%%%%%%%%%%%%%%%%%%%%%%%

Concerning $g_R$ that involves the left chirality of the bottom quark,
there is significant interference with the SM amplitude.  As shown in
the tables in Appendix \ref{sec:param} this is always a negative
interference, therefore there is more allowed space for positive
values of $g_R$.  The LHC yields stronger constraints (via single top
production) than the LHeC because the vertex proportional to $g_R$
depends on energy, which is larger at the LHC.  However, the bounds
obtained from the single top measurement are about one order of
magnitude lower than the bounds coming from the decay ratio $F_L$
whose measurement will be based on the much larger sample of $t\bar t$
events.  Concerning $g_L$, again the decay ratio $F_R$ is more likely
to be a better probe than the LHeC although not with a big difference.

Notice that the LHC14 bounds on $V_R$ and $g_L$ are actually
weaker than the current bounds based on the LHC run at $8$ TeV.
What happens is that for these (right handed bottom) couplings
there is no interference with the SM amplitude in the zero $m_b$
limit.  The contribution to single top production can only be
positive and proportional to the square of the couplings.  Since
the theoretical prediction is already above the experimental
value, there is little allowed region left for $V_R$ and $g_L$.
If, as we have assumed here, the $14$ TeV run yields a measurement
that is equal to the SM prediction, the allowed region will
actually be increased.  This is so even if the experiment achieves
a lower experimental error.
On the other hand, even though the ratio $F_R$ could yield a tight
bound on $V_R$, the LHeC we estimate could achieve a similar
constraint.
\begin{table}
\begin{center}
 \begin{tabular}{|c||c|c|c|c|c|}\hline
 & $V_L$ &  $V_R$ & $g_R$ & $g_L$ & $g_{\times}$;$C_1$
\\\hline
$F_L$ & -- & $\pm 0.34$ & $\pm 0.044$ & $\pm 0.28$ & --
\\\hline 
$F_R$ & -- & $\pm 0.30$ & -- & $\pm 0.14$ & --
\\\hline
Tevatron & $0.925 \pm 0.13$ & $\pm 0.32$ &
$0.25\pm 0.30$ & $\pm 0.17$ & $0.68 \pm 1.22$
\\\hline
LHC & $0.994 \pm 0.046$ & $\pm 0.22$ &
$0.11\pm 0.24$ & $\pm 0.18$ & $-0.14 \pm 0.95$
\\\hline\hline 
$F_L$ & -- & $\pm 0.23$ & $\pm 0.022$ & $\pm 0.19$ & --
\\\hline
$F_R$ & -- & $\pm 0.18$ & -- & $\pm 0.09$ & --
\\\hline
LHC14 & $1\pm 0.036$ & $\pm 0.27$ &
$0.09 \pm 0.22$ & $\pm 0.21$ & $\pm 0.53$
\\\hline
LHeC & $1\pm 0.02$ & $\pm 0.17$ &
$0.19 \pm 0.29$ & $\pm 0.13$ & $\pm 0.34$
\\\hline
  \end{tabular}
\end{center}
\caption{Bounds on couplings ($68\%$ CL).  The upper half is
from recent experimental results.  The lower half is an
estimate of the future results of the HL-LHC and LHeC.}
\label{prebounds}
\end{table}
%%%%%%%%%%%%%%%%%%%%%%%%%%%%%%%%%%%%%%%%%%%%%%%%%%%%%%%%%%

The results of the simplified analysis carried out in this section,
collected in Table~\ref{prebounds}, provide a semi-quantitative
picture of the most likely scenarios as to how much the LHeC could
improve the couplings analyses performed at the HL-LHC.  It appears
that the LHeC would give better constraints than the HL-LHC on $V_L$;
competitive and possibly better constraints on $V_R$; weaker, but
comparable bounds on $g_L$ and poorer constraints on $g_R$.  As for
contact interactions, the two machines will probe different sets:
quark-quark operators at the LHC and lepton-quark operators at the
LHeC.   We expect this semi-quantitative picture, obtained here by
means of a simplified calculational approach, to remain valid in the
context of a technically more detailed analysis as given in Section
\ref{sec:eff.oper} below.

%%%%%%%%%%%%%%%%%%%%%%%%%%%%%%%%%%%%%%%%%%%%%%%%%%%%%%%%%%%%%

\section{Single-top production at the LHeC in the Standard Model}
\label{sec:sm}

\begin{figure}[t]
  \centering
\includegraphics{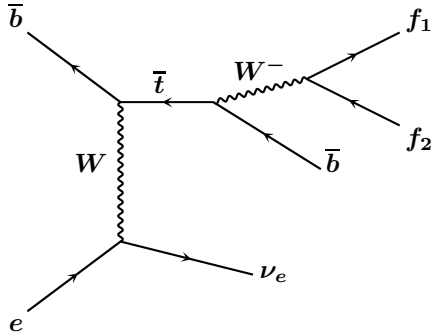}  
\caption{Feynman diagram for single-top production and decay in $e^-p$
  collisions.}
  \label{fig:sgn.dgrm}
\end{figure}

The scattering amplitude for single-antitop production in $e^-p$
collisions in the SM, followed by antitop decay, is given by the
Feynman diagram in Figure \ref{fig:sgn.dgrm}.  Throughout this paper
we restrict ourselves to light charged leptons in the final state,
which we take to be massless.  Thus, in the antitop-decay leptonic
channel the final-state fermions in the figure are $f_1,f_2=$
$e^-,\overline{\nu}_e$ or $\mu^-,\overline{\nu}_\mu$.  In the hadronic
channel $f_1,f_2=$ $d,\overline{u}$ or $s,\overline{c}$.  We work with
four massless flavors so, in particular, we neglect CKM mixing.

In the remainder of this section we discuss the cross sections for
single-antitop production in $e^-p$ collisions in the SM, in both the
leptonic and hadronic channels.  We analyze the signal process as well
as its irreducible and reducible backgrounds, and estimate the
statistical and systematical errors.  Our statistical error estimates
are based on the assumption that the LHeC will achieve a total
integrated luminosity $\mathcal{L}=100$ fb$^{-1}$.  In all cases we
set $m_t=172.5$ GeV, $m_b=4.7$ GeV, $m_Z=91.1735$ GeV, $m_W=80.401$
GeV, $m_h=125$ GeV, $\alpha=1/132.507$, $G_F=1.1664\times10^{-5}$
GeV$^{-2}$, $\alpha_S(m_Z)=0.118$, and the Higgs vacuum-expectation
value $v=246.2185$ GeV. The computation of the cross
sections for the various different processes considered in this
section were carried out with the matrix-element Monte Carlo simulation
program \textsc{MadGraph5\_aMC@NLO} version 2.1.0 \cite{mg5a,mg5b} at
tree level.  Where needed, the events generated by the simulation were
analyzed with \textsc{MadAnalysis5} version 1.1.9 \cite{ma5}.  We set
the renormalization and factorization scales fixed at
$\mu_R=m_t=\mu_F$ and used the parton-distribution functions CTEQ6--L1
as implemented in \textsc{MadGraph5}. 

An electron--proton collider such as the LHeC offers the oportunity of
performing measurements with a polarized electron beam.  Current
accelerator technology makes it possible to achieve longitudinal
polarization at the interaction points in electron storage rings, as
was done at HERA where polarizations of 65\% were reached
\cite{bar95}.  In linear accelerators, electron beam polarizations of
up to 80\% can be achieved reliably for long periods of time as was
done at SLC \cite{moo08}.  For our purposes in this paper the most
interesting case is that of right-handed electron polarization,
$0\leq\P\leq+1$, so that below we discuss the signal and backgounds
cross sections for electron polarizations $\P=0$, $+0.4$ and $+0.7$.

\subsection{Standard Model: Leptonic channel}
\label{sec:sm.lpt}

\begin{figure}[t]
  \centering
\includegraphics{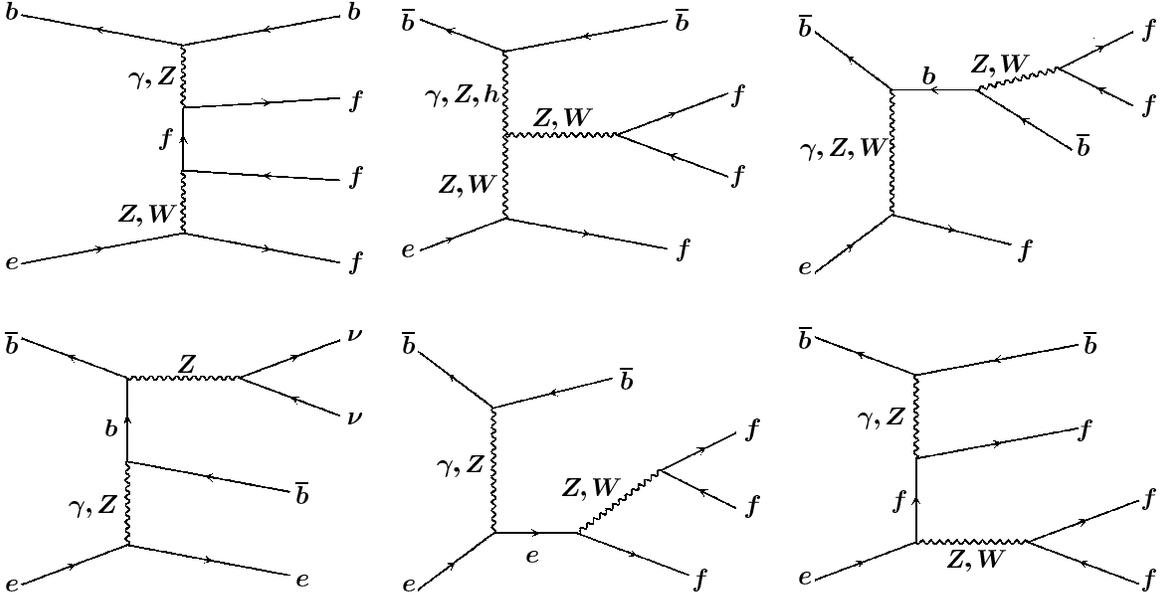}  
  \caption{Feynman diagram for the irreducible background to
    leptonic-channel single-top production. $\nu$ refers to
    $\nu_{e,\mu,\tau}$ and $f$ to $e^-$, $\mu^-$ or $\nu$.}
  \label{fig:lpt.ibk.dgrm}
\end{figure}

In the leptonic mode the signal ($S$) and signal plus irreducible
background ($S+B$) in the SM are defined as
\begin{equation}
  \label{eq:lpt.sig.bck}
  S: e^- p(\overline{b}) \rightarrow \overline{t} \nu_e 
  \rightarrow\overline{b} \ell^- \overline{\nu}_\ell \nu_e,
\qquad
  S+B: e^- p(\overline{b}) \rightarrow\overline{b} \ell^- \overline{\nu}\nu, 
\end{equation}
where the final-state charged leptons are restricted to the light
flavors $\ell=e$, $\mu$.  In the SM the signal process $S$ involves
only two Feynman diagrams, corresponding to Figure \ref{fig:sgn.dgrm}
with $f_1$, $f_2=$ $\ell^-$, $\overline{\nu}_\ell$.  The irreducible
background $B$ comprises 51 diagrams with four electroweak vertices,
see Figure \ref{fig:lpt.ibk.dgrm}, out of which 20 have $\ell^-
\overline{\nu}\nu = e^-\overline{\nu}_e\nu_e$ in
(\ref{eq:lpt.sig.bck}), 11 have $\ell^- \overline{\nu}\nu =
e^-\overline{\nu}_\mu\nu_\mu$, 11 have
$e^-\overline{\nu}_\tau\nu_\tau$, and 9 have
$\mu^-\overline{\nu}_\mu\nu_e$.  %
%%%%%%%%%%%%%%%%%%%%%%%%%%%%%%%  cambio 2 arbitro 1
Furthermore, there are 33 diagrams of the type
$e^-\overline{b}\rightarrow e^-\overline{b} ff$ in which the fermion
line beginning with the initial electron goes through NC vertices
only, and 18 of the form $e^-\overline{b}\rightarrow \nu_e\overline{b}
ff$ in which one CC vertex is attached to that fermion line.
For the computation of cross sections we apply phase-space cuts on the final-state
momenta.  We consider several sets of cuts defined as
\begin{equation}
  \label{eq:cuts.L}
  \begin{aligned}
    L_1 &:\quad |\vec{p}_T(\ell)| > 20\mbox{ GeV, }  
    |\vec{p}_T(\overline{b})| > 20,\\
    L_2 &:\quad  L_1\mbox{, } \met > 25\mbox{ GeV,}\\
    L_3 &:\quad  L_2\mbox{, } |\eta(\ell)| < 2.5\mbox{, } 
    |\eta(\overline{b})| < 2.5,\\
    L_4 &:\quad  L_3\mbox{, } \Delta R(\ell,\overline{b}) > 0.4,
  \end{aligned}
\end{equation}
where $\Delta R=\sqrt{(\Delta\eta)^2+(\Delta\varphi)^2}$ is the distance
in the $\eta$-$\varphi$ plane, in the laboratory frame.  The cuts
(\ref{eq:cuts.L}) are standard centrality and isolation cuts, needed
to ensure a hard partonic scattering and to render the background
processes free from infrared instabilities due to photon $t$-channel
exchange.  With cuts $L_4$, at the three electron energies
considered here more than 95\% of the irreducible background
originates in the 12 diagrams from Figure \ref{fig:lpt.ibk.dgrm} in
which $\ell^-$, $\overline{\nu}$ are the decay products of an on-shell
$W^-$, as happens also in the signal process.  Furthermore, we cannot
impose cuts on the would-be $\tbar$ decay products constraining them
to the $\tbar$ mass shell, as there are two neutrinos in the final
state.  Therefore, a small signal--background interference remains
that turns out to be destructive.  As a consequence, the ratios
$\sigma_B/\sigma_S$ and $(\sigma_{S+B}-\sigma_{S})/\sigma_{S}$ have
different sign, and the former is larger than the absolute value of
the latter. In what follows we make the more conservative choice of
using the larger ratio $\sigma_{B}/\sigma_{S}$ as a measure of the
systematic error originating in the irreducible background.  In Table
\ref{tab:cuts.L} we summarize the effects of the cuts
(\ref{eq:cuts.L}) on the SM signal and irreducible background.
\begin{table}[t]
  \centering
 \begin{tabular}{|c||c|c|c||c|c|c||c|c|c|}\hline
    & \multicolumn{3}{c||}{$E_e=60$ GeV} 
    & \multicolumn{3}{c||}{$E_e=140$ GeV} 
    & \multicolumn{3}{c|}{$E_e=300$ GeV} \\\hline
    & $\sigma_S$ & $\sigma_B$ & $\sigma_{S+B}$  &             
      $\sigma_S$ & $\sigma_B$ & $\sigma_{S+B}$ &
      $\sigma_S$ & $\sigma_B$ & $\sigma_{S+B}$     \\\hline\hline
$\es$&0.3701&      &      &1.1210&      &      &2.519&      &       \\\hline
$L_1$&0.2891&0.0028&0.2892&0.8754&0.0108&0.8688&1.967&0.0318&1.942\\\hline 
$L_2$&0.2482&0.0024&0.2477&0.7564&0.0095&0.7507&1.713&0.0289&1.693\\\hline 
$L_3$&0.2042&0.0018&0.2034&0.7022&0.0084&0.6971&1.647&0.0268&1.626\\\hline 
$L_4$&0.2030&0.0018&0.2028&0.6984&0.0084&0.6931&1.640&0.0268&1.618\\\hline 
 \end{tabular}
 \caption{Cross sections in pb for the signal $S$,
   irreducible background $B$ and signal-plus-background $S+B$
   processes defined in (\ref{eq:lpt.sig.bck}), with the cuts (\ref{eq:cuts.L}). $\es$ refers to the
   cross sections without cuts.}
  \label{tab:cuts.L}
\end{table}

The main source of reducible background to the signal process
(\ref{eq:lpt.sig.bck}) is $b$-jet mistagging from the the
flavor-diagonal processes
\begin{equation}
  \label{eq:lpt.red.bck}
e^- p(c)\rightarrow c \ell^- \nu \overline{\nu},
\qquad
e^- p(q)\rightarrow q \ell^- \nu \overline{\nu},
\end{equation}
where $c$ stands for $c$ or $\overline{c}$, $q$ for any of the quarks
or antiquarks lighter than $c$ and $\ell$, $\nu$ are as in
(\ref{eq:lpt.sig.bck}).  The processes in (\ref{eq:lpt.red.bck})
involve 96 diagrams with a charm final state, and 288 diagrams with
lighter flavors, for a total of 384 diagrams with four electroweak
vertices each.  We take the mistagging probability to be 0.1 for $c$
and 0.01 for $q$.  Thus, the associated systematical error is given
in terms of the cross sections for the processes
(\ref{eq:lpt.red.bck}) by $\delta\sigma_\mathrm{mis} = \sigma_c/10 +
\sigma_q/100$.  The results for reducible-background cross sections
are given in Table \ref{tab:lpt.red.bck}. 
\begin{table}[t]
  \centering
\begin{tabular}{|c|c|c|c|}\hline
  & $\sigma_c$ [pb] & $\sigma_q$ [pb] & $\delta_\mathrm{mis}\sigma$[pb] \\\hline\hline
 60 GeV & 0.006 & 0.080 & 0.0014 \\\hline
140 GeV & 0.021 & 0.220 & 0.0043 \\\hline
300 GeV & 0.055 & 0.451 & 0.0101 \\\hline
\end{tabular}
\caption{Cross sections for the reducible background to single-top
  production in leptonic channel and for $b$-mistagged
  events, $\delta\sigma_\mathrm{mis} = \sigma_c/10 +
  \sigma_q/100$, with cuts $L_4$ from (\ref{eq:cuts.L}).}
  \label{tab:lpt.red.bck}
\end{table}
Comparing with the signal cross sections given in Table
\ref{tab:cuts.L} we obtain mistagging errors of $\sim0.65$\% at the
three energies $E_e=60$, 140 and 300 GeV.

We have considered also the contributions of final states with one
additional neutrino pair $\nu\overline{\nu}$ ($\nu=\nu_{e,\mu,\tau}$),
which are found to be negligibly small, as expected.  For the signal
and irreducible background processes (\ref{eq:lpt.sig.bck}), their
contribution is less than 0.1\% of the cross sections $\sigma_S$ and
$\sigma_B$ with cuts $L_4$ in Table \ref{tab:cuts.L}.  Reducible
background processes of the form (\ref{eq:lpt.red.bck}) with one
additional neutrino pair are given by 16968 diagrams (4242 with a
charmed final state and 12726 with lighter flavors), with six
electroweak vertices.  Their contribution to the mistagging cross
section $\delta_\mathrm{mis}\sigma$ is less than 0.1\% than that in
Table \ref{tab:hdr.red.bck}.  Thus, we disregard final states with
additional neutrino pairs in what follows.

We can estimate the statistical error associated to the SM signal from
Table \ref{tab:cuts.L} as
$\delta\sigma_\mathrm{stat}=\sqrt{\sigma_S/\mathcal{L}}$, with
$\mathcal{L}=100$ fb$^{-1}$.  The systematical errors originating in
the irreducible background and the mistagging cross section have been
given in Tables \ref{tab:cuts.L} and \ref{tab:lpt.red.bck}.  Adding
those errors in quadrature we find total errors of 1.7\%, 1.9\% and
2.3\% at $E_e=$ 60, 140 and 300 GeV, respectively.  As seen from the
tables, whereas the three errors $\delta\sigma_\mathrm{stat}$,
$\sigma_B$ and $\delta\sigma_\mathrm{mis}$ at $E_e=60$ GeV have
similar sizes, at 140 and especially at 300 GeV, the dominant source
of error in this channel is the irreducible background.  From these
results, and taking into account other unspecified sources of
measurement error, we estimate a lower bound of 3\% on 
experimental errors in the leptonic channel.  In what follows we will
assume experimental uncertainties of 3\%, 6\% and 8\%.

%%%%%%%%%%%%  cambio 1 arbitro 1
In order to gain some perspective on the plausibility of these assumed
experimental-error levels for the detection of single top events at
the LHeC, we should bear in mind that the latest CMS analysis of
single top events at 8 TeV has already reached errors slightly below
9\% \cite{cms2014,pic13}.  In their case some of the most important
contributions to this error are jet energy scale, jet energy
resolution, missing $E_T$ and pileup \cite{cms2014}.  As mentioned
before, it is expected that the LHeC detector will achieve an
outstanding performance in these areas.  There will be an absence of
pileup and in contrast to the single top measurement at the LHC, there
is no $t\bar t$ background to take into consideration at the LHeC.
Thus, assuming an 8\% uncertainty at the LHeC corresponds to the
least-favorable scenario in which it barely manages to improve on the
precision already achieved by CMS in 2014.  As for the feasability of
the minimum of 3\%, we point out that the CMS and ATLAS collaborations
have reached errors as low as 4.1\% and 3.9\%, resp., in the
measurement of $\sigma(t\bar t)$ at $\sqrt{s}=7$ TeV in the dileptonic
$\mu e$ mode, and at $\sqrt{s}=8$ TeV both collaborations have also
reached similarly low errors \cite{pic13,cmsatlasttbar}. 

For electron polarizations $\P$ up to 90\%, the error estimates above
do not change significantly.  The SM cross section $\sigma_S$ for the
signal process (\ref{eq:lpt.sig.bck}) depends on electron polarization
as $\sigma_S(\P)=(1-\P)\sigma_S(0)$, $-1\leq\P\leq1$.  Thus, for the
absolute statistical error we have $\delta\sigma_\mathrm{stat}(\P)=
\sqrt{1-\P}\,\delta\sigma_\mathrm{stat}(0)$.  The
irreducible-background cross section $\sigma_B(\P)$ does not tend to
zero as $\P\rightarrow1$, even for a massless electron, due to the
diagrams in Figure \ref{fig:lpt.ibk.dgrm} in which the $e^-$ fermion
line is attached to the diagram by a $\gamma/Z$ vertex.  The value of
$\sigma_B(+1)$ is very small, however, being 0.04, 0.13, 0.23 fb at
$E_e=60$, 140 and 300 GeV respectively.  Therefore, for moderate
right-handed polarization values, the irreducible background cross
section scales with $\P$ as $\sigma_B(\P)=(1-\P)\sigma_B(0)$ to a good
approximation.  Analogous considerations hold for the $b$-mistagging
cross section $\delta\sigma_\mathrm{mis}$.  By adding the statistical
and systematical background errors in quadrature we find for $\P=0.4$
total errors of 1.8\%, 1.9\% and 2.3\% at $E_e=60$, 140 and 300 GeV,
respectively.  For $\P=0.7$ the errors are found to be 2.1\%, 2.1\%
and 2.4\% at those same electron energies.  Thus, for right-handed
polarization up to $\sim90\%$, we estimate experimental errors to be in the
same range 3--6\% as in the unpolarized case.

\subsection{Standard Model: Hadronic channel}
\label{sec:sm.hdr}

\begin{figure}[t]
  \centering
\includegraphics{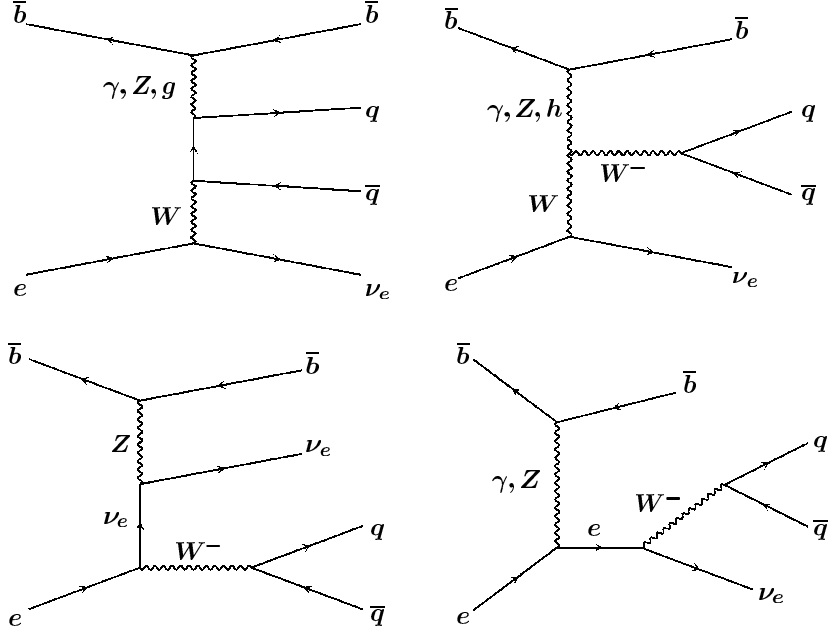}  
  \caption{Feynman diagram for the irreducible background in
    hadronic-channel single-top production.}
  \label{fig:hdr.ibk.dgrm}
\end{figure}

In the hadronic mode the signal ($S$) and signal plus irreducible
background ($S+B$) in the SM are defined as
\begin{equation}
  \label{eq:hdr.sig.bck.alt}
  S:\; e^- p(\overline{b}) \rightarrow \overline{t} \nu_e \rightarrow
  \overline{b}j_u j_d \nu_e,
%  \quad
%  \left(\left|\sqrt{p^2_{\overline{t}}} -  m_{\overline{t}}\right|
%  < 15\Gamma_{\overline{t}}\right),
\qquad
  S+B: e^- p(\overline{b}) \rightarrow\overline{b} j_u j_d \nu_e,
%  \quad
%  \left(\left|\sqrt{p^2_{\overline{t}}} -  m_{\overline{t}}\right|
%  > 15\Gamma_{\overline{t}}\;\; \mathrm{if\;\exists}\;\overline{t}\right),
\end{equation}
where $j_u=u$, $c$, $\overline{u}$, $\overline{c}$ and $j_d=d$, $s$,
$\overline{d}$, $\overline{s}$.  In the SM the signal process $S$
involves only two Feynman diagrams, corresponding to Figure
\ref{fig:sgn.dgrm} with $f_1$, $f_2=$ $\overline{u}$, $d$ and
$\overline{c}$, $s$.  The irreducible background $B$ comprises 24
diagrams: 20 of them with four electroweak vertices and no QCD vertex
(like the signal diagram), and 4 diagrams with two QCD vertices and
two electroweak vertices, see Figure \ref{fig:hdr.ibk.dgrm}.
For the computation of the cross section we impose on the final-state
momenta a set of appropriate phase-space cuts. As in the previous
section, we consider several progressively more restrictive cuts
defined as 
\begin{equation}
  \label{eq:cuts.H}
  \begin{aligned}
    H_1 &:\quad |\vec{p}_T(j)| > 20\mbox{ GeV, }  
    |\vec{p}_T(\overline{b})| > 20\mbox{ GeV,}\\
    H_2 &:\quad H_1\mbox{, } \met > 25\mbox{ GeV,}\\
    H_3 &:\quad H_2\mbox{, } |\eta(j)| < 2.5\mbox{, } 
    |\eta(\overline{b})| < 2.5,\\
    H_4 &:\quad H_3\mbox{, } \Delta R(j,\overline{b}) > 0.4\mbox{, }
    \Delta R(j,j) > 0.4,
  \end{aligned}
\end{equation}
with $\Delta R$ as in (\ref{eq:cuts.L}).  The cuts (\ref{eq:cuts.H})
are standard centrality and isolation cuts, needed to ensure a hard
partonic scattering and to render the background processes free from
infrared instabilities due to the emission of massless quarks.  In the
cut $H_3$ we could have set the limit on light-jet pseudorapidity to
$|\eta(j)|<5$, given the wide acceptance expected of hadronic
calorimeters; we use instead a tighter cut for reasons explained
in detail in Appendix \ref{sec:cuts.sensitiv} below.  Furthermore, in
all cases $H_{1,\ldots,4}$ we impose the additional cuts
\begin{equation}
  \label{eq:cuts.H.2}
  \left|\sqrt{(p_{q_1}^2+p_{q_2}^2+p_{b}^2)}-m_t\right| < 15 \Gamma_t
% \simeq 23 \mathrm{GeV}
  \quad
  \mathrm{and}
  \quad
  \left|\sqrt{(p_{q_1}^2+p_{q_2}^2)}-m_W\right| < 15 \Gamma_W,
%  \simeq 30 \mathrm{GeV},
\end{equation}
where $q_{1,2}$ refers to the light quarks in the final state. In
Table \ref{tab:cuts.H} we report the effects of the cuts
(\ref{eq:cuts.H}), together with (\ref{eq:cuts.H.2}), on the SM
signal, background and total cross sections.  Comparing
$\sigma_S+\sigma_B$ with $\sigma_{S+B}$ from the Table shows that a
small interference remains after the cuts (\ref{eq:cuts.H.2}) have
been applied, although at a lower level than in the leptonic channel.
We choose the larger ratio $\sigma_B/\sigma_S$, rather than
$(\sigma_{S+B}-\sigma_{S})/\sigma_S$, as a measure of the systematical
error caused by the irreducible background.  That choice turns out to
be immaterial in this case, however, since the systematical error is
dominated by the reducible background discussed next.
\begin{table}[t]
  \centering
    \begin{tabular}{|c||c|c|c||c|c|c||c|c|c|c|}\hline
    & \multicolumn{3}{c||}{$E_e=60$ GeV} 
    & \multicolumn{3}{c||}{$E_e=140$ GeV} 
    & \multicolumn{3}{c|}{$E_e=300$ GeV} \\\hline
Cuts &$\sigma_S$  &$\sigma_B$   &$\sigma_{S+B}$ &$\sigma_S$
&$\sigma_B$ &$\sigma_{S+B}$ &$\sigma_S$ &$\sigma_B$ &$\sigma_{S+B}$ \\\hline\hline
$\es$ &1.0880 &       &       &3.280 &       &      &7.345 &       &      \\\hline
$H_1$ &0.7594 &0.0059 &0.7606 &2.284 &0.0171 &2.287 &5.111 &0.0372 &5.114 \\\hline
$H_2$ &0.5590 &0.0042 &0.5599 &1.767 &0.0128 &1.769 &4.063 &0.0288 &4.069 \\\hline
$H_3$ &0.3932 &0.0022 &0.3941 &1.553 &0.0101 &1.555 &3.823 &0.0255 &3.827 \\\hline
$H_4$ &0.3912 &0.0022 &0.3921 &1.545 &0.0100 &1.546 &3.801 &0.0252 &3.804 \\\hline
    \end{tabular}
    \caption{Cross sections in pb for the signal $S$ and
      signal-plus-background $S+B$ processes defined in
      (\ref{eq:hdr.sig.bck.alt}), with the cuts (\ref{eq:cuts.H}) and
      (\ref{eq:cuts.H.2}). $\es$ refers to the
      cross sections without cuts.}
  \label{tab:cuts.H}
\end{table}

The main source of reducible background in this channel is
$b$-mistagging in the processes
\begin{equation}
  \label{eq:hdr.red.bck}
e^- p\rightarrow j j j \nu_e,
\end{equation}
where $j$ stands for a gluon or any of the quarks or antiquarks
lighter than $b$, and $\nu$ for any of the three neutrino flavors.
The processes in (\ref{eq:hdr.red.bck}) involve 528 diagrams in total,
of which 128 contain two QCD vertices and two electroweak ones, and
400 contain four electroweak vertices like the signal process $S$ in
(\ref{eq:hdr.sig.bck.alt}).  The diagrams containing two strong
vertices contribute about 80\% of the total cross section for
(\ref{eq:hdr.red.bck}), whereas the more numerous purely electroweak
diagrams supply the remaining 20\%.  As in the previous section, we
take the mistagging probability to be 1/10 for $c$ and 1/100 for
lighter jets.  Thus, we have to consider separately out of the
reactions (\ref{eq:hdr.red.bck}) those leading to 0, \ldots, 3
final-state charm quarks.  The cross sections for those processes are
summarized in Table \ref{tab:hdr.red.bck} with the cuts $H_4$ from
(\ref{eq:cuts.H}).  In this case we apply a modified version of
(\ref{eq:cuts.H.2}) in which the second equality in that equation must
be satisfied by at least one of the three possible pairs of
final-state quarks.
\begin{table}[t]
  \centering
  \begin{tabular}{|c||c|c|c|c|c|}\hline
   $E_e$[GeV] & $j_cj_cj_c\nu_e$ & $j_cj_cj  \nu_e$ & $j_cj  j  \nu_e$ & $j
   j  j  \nu_e$ & $b  b  j  \nu_e$\\\hline\hline
60  &0.0005&0.0085&0.0944&0.2413&0.0111\\\hline
140 &0.0016&0.0228&0.3386&0.6460&0.0263\\\hline
300 &0.0036&0.0451&0.7719&1.2700&0.0473\\\hline
  \end{tabular}
  \caption{Cross section in pb for reducible-background processes,
    with phase-space cuts $H_4$ defined in (\ref{eq:cuts.H}) and
    (\ref{eq:cuts.H.2}).} 
  \label{tab:hdr.red.bck}
\end{table}
Given the probability to mistagging as $b$ a single $c$ quark and a
single lighter parton, we have the mistagging probabilities for the
three-jet final state shown in Table \ref{tab:mistag}.  Also included
in Table \ref{tab:hdr.red.bck} is the cross section for the final
state $bbj\nu$ \cite{moretti} which, assuming a $b$-tagging efficiency
of 60\%, has a probability of 0.48 of being mistagged as $bjj\nu$.

\begin{table}[t]
  \centering
\begin{tabular}{|c||c|c|c|c|c|c|}\hline
  \parbox{55pt}{prob.\ of\\[-10pt]mistagging} & $j_c$ & $j_c$ & $j$ &
  $j_c$ & $j$ & $j$  \\\hline
out of & $j_cj_cj_c$ & $j_cj_cj$ & $j_cj_cj$ & $j_cjj$ & $j_cjj$ & $jjj$ \\\hline
$P$ &0.243 & 0.1782 & 0.0081 & 0.09801 & 0.01782 & 0.029403\\\hline
\end{tabular}
  \caption{Probability of mistagging as $b$ one parton out of the
    indicated final states, given that the probability of mistagging
    a $c$ quark is 1/10, and a lighter parton 1/100.}
  \label{tab:mistag}  
\end{table}
By combining the results of Table \ref{tab:hdr.red.bck} and
\ref{tab:mistag} we get mistagging cross sections
$\delta\sigma_\mathrm{mis}=25.1$, 75.5 and 158.7 fb at $E_e=$ 60, 140,
300 GeV, respectively.  We remark that with the cuts $H_4$ as defined
in (\ref{eq:cuts.H}), but without the cuts (\ref{eq:cuts.H.2}), the
mistagging cross sections would be 131.4, 424.9 and 998.6 fb at $E_e=$
60, 140, 300 GeV, respectively, corresponding to $\sim$30\% of the
signal cross section.  It is the top-mass cut given by the first
equality in (\ref{eq:cuts.H.2}) that plays a crucial role in taming
this large reducible background.

Processes with additional neutrino pairs in the final state yield
negligible cross sections, as expected.  Indeed, the cross section for
signal and irreducible background processes of the form
(\ref{eq:hdr.sig.bck.alt}) with one additional neutrino pair in the
final state are less than 0.1\% of $\sigma_S$ and less than 1\% of
$\sigma_B$ as given in Table \ref{tab:cuts.H}.  The reducible
background processes (\ref{eq:hdr.red.bck}) with an additional
final-state neutrino pair yield a scattering amplitude with 144
diagrams for three-$c$ final states, 432 for two-$c$, 1360 for one-$c$
and 2432 for no-$c$ final states.  Computation of the cross section in
this case, with cuts $H_4$, gives results that are less than 0.1\% of
those in Table \ref{tab:hdr.red.bck}.  Thus, we ignore processes with
multi-neutrino final states in what follows.

As stated at the beginning of this section, we assume a total
integrated luminosity $\mathcal{L}=100$ fb$^{-1}$.  Thus, for the
statistical error associated to the SM signal from Table
\ref{tab:cuts.H} we get $\delta\sigma_\mathrm{stat}=2$, 3.9, 6.2 fb at
$E_e=60$, 140, 300 GeV, respectively.  We consider the $b$-mistagging
cross section given above and the irreducible background $\sigma_B$
from Table \ref{tab:cuts.H} as systematical errors.  It is apparent
from these results that the dominant source of error in this channel
is the reducible background. By adding statistical and systematical
errors in quadrature we find total errors of 6.4\%, 4.9\% and 4.2\% at
$E_e=$ 60, 140 and 300 GeV, respectively, relative to the signal cross
sections with cuts $H_4$ from Table \ref{tab:cuts.H}.  From this
evaluation of statistical and background errors, and taking into
account other unspecified sources of measurement error, %we consider
%the range 7--12\% a credible estimate of experimental errors in the
%hadronic channel with an unpolarized electron beam.
we estimate a lower bound of 7\% on the experimental error.  
%%%%%%%%%%%%  cambio 1 arbitro 1
We cannot use existing experimental results on single-top production
as guidelines in our error estimates, since the hadronic channel has
never been observed so far.  We notice, however, that the cross section
for $t\tbar$ production in the semileptonic channel has been measured
by CMS at $\sqrt{s}=7$ TeV \mbox{\cite{cms.x}} with an error or 7\%,
and by ATLAS at $\sqrt{s}=8$ TeV with an error of 13\% \cite{atlas.x}.
Therefore, we believe that experimental errors in the range 7--12\%
for the hadronic channel could be achieved at the LHeC.

If the electron beam is polarized, the SM cross section $\sigma_S$ for
the signal process (\ref{eq:hdr.sig.bck.alt}) depends on electron
polarization as $\sigma_S(\P)=(1-\P)\sigma_S(0)$, $-1\leq\P\leq1$, as
is apparent from Figure \ref{fig:sgn.dgrm}.  Thus, for the statistical
error we have $\delta\sigma_\mathrm{stat}(\P)=
\sqrt{1-\P}\,\delta\sigma_\mathrm{stat}(0)$.  For a massless electron,
the dependence of the irreducible-background cross section
$\sigma_\mathrm{B}$ with $\P$ is the same as that of the signal cross
section, as can be seen from Figure \ref{fig:hdr.ibk.dgrm}, so
$\sigma_\mathrm{B}(\P)/\sigma_S(\P)=
\sigma_\mathrm{B}(0)/\sigma_S(0)$.  The same considerations hold for
the reducible background (\ref{eq:hdr.red.bck}).  Our results for the
systematical error originating in background processes in the
unpolarized case, therefore, remain unchanched in the polarized case
when expressed as a fraction of the signal cross section.  Since the
statistical-error contribution to the total error is much smaller than
that of the reducible background, the variation of
$\delta\sigma_\mathrm{stat}$ with $\P$ will not significantly change
our estimate of the total error as long as it remains subdominant
relative to $\delta\sigma_\mathrm{mis}$.  Thus, for $\P\lesssim +0.9$
we consider experimental errors in the same range 7-12\% as in the
unpolarized case.

%%%%%%%%%%%%%%%%% cambio 3 arbitro 2
\subsection{Theoretical uncertainties}
\label{sec:thr.uncr}

The computations of the SM cross sections for single-antitop
production and decay are affected by theoretical uncertainties that we
briefly discuss here.  
The use of LO PDFs leads to uncertainties arising from the choice of
renormalization and factorization scales that we estimate by varying
them as $\mu_R=\mu_F=m_t/2$, $m_t$, $2m_t$.  The scale uncertainty in
the unpolarized cross section is found to be 7.5\%, 9\%, 10.3\% at
$E_e=60$, 140, 300 GeV for both the leptonic and the hadronic channel.
The choice of PDF is also a source of uncertainty, which we have found
to be 2\%, 5\%, 6\% for both channels at the same three energies.
%There is, furthermore, a theoretical error stemming from the
%truncation of perturbation theory at tree level.
The NLO correction to the LO approximation depicted in Figure
\ref{fig:sgn.dgrm} has not yet been given in the literature, but it
could be of order $3\%$ as in the t-channel top production at the LHC
\cite{cao2011}.  We emphasize here, however, that these theoretical
uncertainties have only a minor impact on the results presented in
this paper.  This is so because the bounds and exclusion regions for
effective couplings presented in sections \ref{tdecandsingle} and
\ref{sec:eff.oper} depend on cross sections only through the ratio
$R=\sigma_\mathrm{eff}(\lambda)/\sigma_\mathrm{eff}(0)$, with
$\sigma_\mathrm{eff}(\lambda)$ the tree-level cross section in the
effective theory, depending on the effective couplings $\lambda$, and
$\sigma_\mathrm{eff}(\lambda=0)$ the tree-level SM cross section.  For
values of the effective couplings within the bounds established below,
we find the scale and PDF uncertainties in the ratio $R$ to be
$\lesssim 0.6\%$.  Notice that this uncertainty is significantly
smaller than the experimental errors assumed in the two previous
subsections.

%%%%%%%%%%%%%%%%%%%%%%%%%%%%%%%%%%%%%%%%%%%%%%%%

\section{Contribution from the effective operators}
\label{sec:eff.oper}

For the computation of the amplitudes in the effective theory we make
the same approximations---i.e., two massless generations and diagonal
CKM matrix---and the same choices of parameters, PDF and scales as in
the SM calculations of section \ref{sec:sm}.  We implemented the basis
of dimension-six $SU(2)_L\times U(1)$--invariant
effective operators involved in the anomalous $tbW$ couplings and
contact-interaction vertices in \textsc{Madgraph 5}  by means of
the program \textsc{FeynRules} 2.0 \cite{feynrul}. 
\begin{figure}[t]
  \centering
\includegraphics[scale=0.8]{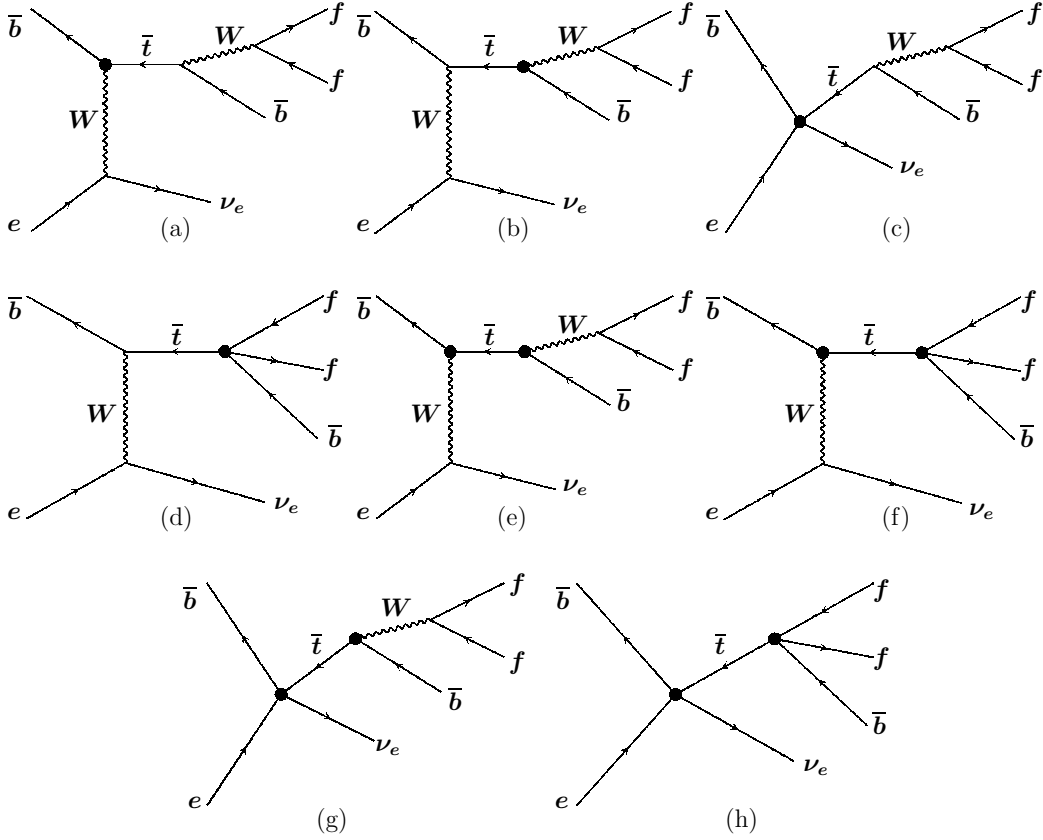}  
\caption{Feynman diagrams for single-top production and decay in $ep$
  collisions containing one or two effective vertices.  The
  $W^-$-decay products $f$ can be $e$, $\overline{\nu}_e$ or $\mu$,
  $\overline{\nu}_\mu$ in the leptonic channel, and $\overline{u}$,
  $d$, or $\overline{c}$, $s$ in the hadronic channel.}
  \label{fig:sgn.dgrm.eff}
\end{figure}

The full set of tree-level Feynman diagrams for single-top production
(including the top decay) in $ep$ collisions is given by the SM diagram
from Figure~\ref{fig:sgn.dgrm} together with the diagrams in
Figure~\ref{fig:sgn.dgrm.eff}.  There can be one or two effective
vertices of the $SU(2)_L\times U(1)$ gauge-invariant effective theory.
Diagrams with two effective vertices
must be taken into account, as they contribute with terms of second
order that come from the interference with the SM amplitude.
Thus, the cross section numerical expressions contain terms of up
to fourth order in the effective couplings.  However, it turns out
that within the bounds obtained below, the contributions from terms
of order higher than the second are neglibly small.  Below, we show
the single antitop production cross section ratio at the LHeC without
terms of third and fourth order:
\begin{equation}
  \label{lheccross}
  \begin{aligned}
\lefteqn{
\frac{\sigma^{\rm SM}+\Delta \sigma}
{\sigma^{\rm SM}_{V_L=1}}
= (1+\delta V_L)^4 + b_{12} g^r_R + b_2 {g^r_R}^2 + d_2 {g^i_R}^2
+ b_{13} V^r_R + b_3 {V^r_R}^2 + d_3 {V^i_R}^2 + b_{14} g^r_L 
}\\
&+ b_4 {g^r_L}^2 + d_4 {g^i_L}^2 + b_{15} C_1 + b_{5} C^2_1 +
b_{6} {C^r_2}^2 + d_{6} {C^i_2}^2  + b_{7} {C^r_3}^2 + d_{7} {C^i_3}^2
+ b_{8} {C^r_4}^2 + d_{8} {C^i_4}^2 \; .
  \end{aligned}
\end{equation}
Unlike the previous simple expression in Eq.~(\ref{lheccross0}), here
we have assumed that the effective coefficients are complex numbers,
except $V_L=1+\delta V_L$ and $C_1$ \cite{aguilar4}.  For a given
complex coupling $\lambda$ we denote its real and imaginary parts as
$\lambda = \lambda^r + i \lambda^i$.  The coefficients in
(\ref{lheccross}) for the leptonic- and hadronic-channel cross sections, with
unpolarized and polarized electron beams, are given in Appendix
\ref{sec:param}.  
Notice that the coefficients of the $tbW$ couplings in
(\ref{lheccross}) are larger than the ones in (\ref{lheccross0}), but
the coefficients of the contact-interaction couplings have roughly the
same values in both cases, due to the effect of the anomalous $tbW$
couplings on the decay vertex which is not taken into account in the
preliminary analysis of section \ref{sec:single.top.prod}.
Thus, as expected, the bounds on
$tbW$ couplings obtained in this section are stronger than the ones in
section~\ref{tdecandsingle}.  Notice also that $b_{13}$ and $b_{14}$
are small compared to $b_{12}$ because of the $m_b$ suppression.  For
simplicity, we do not include in (\ref{lheccross}) interference terms
involving two different effective couplings.  This is appropriate with
the goal of obtaining bounds by taking only one coupling non-zero at a
time.  On the other hand, we do take into account all interference
terms involving two anomalous couplings for the computation of the
correlated regions of allowed parameter space discussed below.

%%%%%%%%%%%%%%%%%%%%%%%%%%%%%%%%%%%%%%%%%%%%%%%%%%%%%%%%%%%%%%

In addition to the total cross section $\sigma$ we consider
various asymmetries $A(X,X_0)$, where $X$ is the kinematical
observable and $X_0$ the reference value:
$A(X,X_0) =
\left.
  \left(
   \sigma\left(X>X_0\right) - \sigma\left(X<X_0\right)\rule{0pt}{11pt}
  \right)
\right/
\left(
    \sigma\left(X>X_0\right) + 
\sigma\left(X<X_0\right)\rule{0pt}{11pt}\right).$ In
what follows, when the reference value is $X_0=0$ we simply write
$A(X)\equiv A(X,0)$.

Polarization of the electron beam does not change the sensitivity of
the cross-section ratios in (\ref{lheccross}), or in the various
asymmetries, to the couplings in the $tbW$ Lagrangian
(\ref{tbwlagrangian}), as the $\nu e W$ vertex in
Figure~\ref{fig:sgn.dgrm.eff} is always the same as in the SM. For the
same reason, it cannot change the sensitivity to the
contact-interaction coupling $C_1$ multiplying the operator $O_{\ell
  q'}^{1331}$ in the Lagrangian (\ref{fourfermion}), as the electron
there is left-handed.  The couplings $C_{2,3,4}$, on the other hand,
involve a right-handed electron.  The sensitivity of the cross section
and the asymmetries to those couplings is then strongly enhanced by
right-polarization of the electron beam, as discussed in detail in the
reminder of this section.

\subsection{Bounds from leptonic channel}
\label{sec:lpt.chn}

In the leptonic channel, due to the presence of two neutrinos in the
final state, the kinematics of the reaction cannot be fully determined
experimentally. In particular, the four-momenta of final-state charged
particles can only be determined in the lab frame, and the momenta of
the intermediate-state \tbar, $W^-$ and of the initial-state \bbar\
cannot be reconstructed.  The only experimentally accessible
quantities are the four-momenta of $\overline{b}$, $\ell$, the
missing energy
$\not\!\!E_T=|\vec{p}_{T\nu_e}+\vec{p}_{T\overline{\nu}}|$ and its
azimuthal angle $\varphi(\not\!\!E_T)$. The observables we consider
are the cross section $\sigma$ and the asymmetries associated with
$\Delta\eta(\bbar,\ell)$, $\Delta\varphi(\bbar,\ell)$,
$\Delta\varphi(\bbar, \met)$, $\Delta\varphi(\ell, \met)$,
$\cos(\bbar,\ell)$, and the leptonic fraction of the visible energy
$u(\bbar,\ell)=E_\ell/(E_\ell+E_{\bbar})$ \cite{she09}.  All of these
kinematic variables refer to the lab frame, although $\Delta\eta$,
$\varphi$ and $\Delta\varphi$ are obviously invariant under
longitudinal boosts. We compute all cross sections and asymmetries
with the cuts $L_4$ defined in (\ref{eq:cuts.L}).

Despite their not being shown in Eq.~(\ref{lheccross}), the
contributions from terms cubic and quartic in the effective $tbW$ and
contact-interaction couplings have been fully taken into account in
our computation of anomalous-coupling bounds.  These contributions are
at least one order of magnitude smaller than the quadratic terms for
the $tbW$ couplings within the bounds obtained below.  For the contact
terms they are even smaller, as contact-interaction effects on the
decay vertex are insignificant \cite{celine}.
% rescatado 1

In Table \ref{tab:lptsgm.bounds.pol00} we report the bounds on
effective couplings obtained from the unpolarized cross section at the
level of one standard deviation, assuming experimental errors of 3\%
and 6\% as discussed in section \ref{sec:sm.lpt}.  As seen in the
table, the sensitivity to the $tbW$ couplings and the contact
interactions is not very dependent on $E_e$.  Notice also that the
coefficients for Eq.~(\ref{lheccross}) shown in the appendix change
little with the three electron energies.  This is because the energy
of the collision goes as $\sqrt{s} = \sqrt{4E_e E_p} = 1.3$, $2.0$
and $2.9$ TeV which does not increase with the same proportion as $E_e$.
The allowed ranges for the imaginary parts of the anomalous $tbW$
couplings are symmetric about 0 due to the absence of interference of
those couplings with the SM.  In the case of the contact-interaction
couplings $C_{2,3,4}$, whose real parts do not interfere with the SM,
the allowed ranges for the imaginary parts $C^i_{2,3,4}$ are equal to
those for the real parts $C^r_{2,3,4}$.  We remark here, as done in
Section \ref{tdecandsingle}, that the
difference in order of magnitude between the bounds for the $tbW$ form
factors and those for the contact-interaction couplings is due to a
large extent to the normalization of the former.  Indeed, using for
example the results for $\delta V_L$ from Table
\ref{tab:lptsgm.bounds.pol00} together with (\ref{factors}), we get
$-0.25<C_{\phi q}<0.24$, to be compared with the bounds for $C_1$
given in the table which are essentially the same.

\begin{table}[t]
  \centering
\begin{tabular}{|c|cc|cc|cc||cc|cc|cc|}\cline{2-13}
\multicolumn{1}{c|}{}&\multicolumn{6}{c||}{$\epsilon_\mathrm{exp}=3\%$}
                     &\multicolumn{6}{c|}{$\epsilon_\mathrm{exp}=6\%$} \\\cline{2-13}
\multicolumn{1}{c|}{} & \multicolumn{2}{c|}{60 GeV} 
                      & \multicolumn{2}{c|}{140 GeV} 
                      & \multicolumn{2}{c||}{300 GeV}
                      & \multicolumn{2}{c|}{60 GeV} 
                      & \multicolumn{2}{c|}{140 GeV} 
                      & \multicolumn{2}{c|}{300 GeV} \\\hline
$\delta V_L\times 10^{2}$ 
           &-0.76&0.73&-0.76&0.73&-0.76&0.76&-1.55&1.48&-1.55&1.45&-1.55&1.48\\\hline
$V^r_R\times10$ 
           &-0.96&1.21&-0.98&1.25&-0.98&1.28&-1.40&1.66&-1.43&1.69&-1.45&1.73\\\hline
$V^i_R\times10$ 
           &-1.09&1.09&-1.12&1.12&-1.12&1.12&-1.53&1.53&-1.57&1.57&-1.58&1.58\\\hline
$g^r_R\times10$ 
           &-0.21&0.22&-0.22&0.23&-0.22&0.23&-0.42&0.45&-0.43&0.48&-0.43&0.49\\\hline
$g^i_R\times10$ 
           &-1.11&1.11&-1.05&1.05&-1.02&1.02&-1.56&1.56&-1.49&1.49&-1.44&1.44\\\hline
$g^r_L\times10$
           &-0.97&0.72&-0.96&0.72&-0.96&0.72&-1.30&1.05&-1.30&1.06&-1.30&1.05\\\hline
$g^i_L\times10$
           &-0.83&0.83&-0.83&0.83&-0.83&0.83&-1.17&1.17&-1.17&1.17&-1.17&1.17\\\hline
$C_1$   &-0.24&0.25&-0.22&0.23&-0.20&0.21 &-0.47&0.50&-0.44&0.46&-0.40&0.43\\\hline%
$C^{r,i}_2$ &-4.06&4.06&-3.60&3.60&-3.16&3.16 &-5.74&5.74&-5.09&5.09&-4.47&4.47\\\hline%
$C^{r,i}_3$ &-2.43&2.43&-2.08&2.08&-1.80&1.80 &-3.44&3.44&-2.95&2.95&-2.55&2.55\\\hline%
$C^{r,i}_4$ &-3.83&3.83&-3.45&3.45&-3.04&3.04 &-5.43&5.43&-4.88&4.88&-4.30&4.30\\\hline%
\end{tabular}
\caption{Bounds on effective couplings at the 68\% CL\ 
  obtained from the leptonic-channel unpolarized cross section
  by varying the couplings one at a time.  The assumed
  experimental errors are 3\% and 6\%. 
}
\label{tab:lptsgm.bounds.pol00}
\end{table}

For the contact-interaction couplings $C_{2,3,4}$, which involve
right-handed electrons, the bounds given in Table
\ref{tab:lptsgm.bounds.pol00} can be significantly improved if the
electron beam is right polarized.  In Table
\ref{tab:lptsgm.bounds.pol4070} we give the bounds obtained for those
couplings by assuming the initial-electron polarizations $\P=+0.4$
and $+0.7$.  As discussed in section \ref{sec:sm.lpt}, the
experimental errors for those polarizations are expected to be in the
same range as in the unpolarized case.
\begin{table}[t]
  \centering
\begin{tabular}{|c|c|c|c|c||c|c|c|}\cline{3-8}
\multicolumn{2}{c|}{}&\multicolumn{3}{c||}{$\epsilon_\mathrm{exp}=3\%$}
                     &\multicolumn{3}{c|}{$\epsilon_\mathrm{exp}=6\%$} \\\cline{3-8}
\multicolumn{2}{c|}{} &60 GeV&140 GeV&300 GeV&60 GeV&140 GeV&300 GeV \\\hline
&$C^{r,i}_2$&$\pm$2.66&$\pm$2.36&$\pm$2.07&$\pm$3.76&$\pm$3.34&$\pm$2.93\\\cline{2-8}
\raisebox{-15pt}[0pt][0pt]{\rotatebox{90}{$\P=0.4$}}
&$C^{r,i}_3$ &$\pm$1.59&$\pm$1.36&$\pm$1.18&$\pm$2.25&$\pm$1.93&$\pm$1.67\\\cline{2-8}
&$C^{r,i}_4$ &$\pm$2.54&$\pm$2.26&$\pm$1.99&$\pm$3.59&$\pm$3.20&$\pm$2.81\\\hline
&$C^{r,i}_2$ &$\pm$1.71&$\pm$1.52&$\pm$1.32&$\pm$2.41&$\pm$2.14&$\pm$1.87\\\cline{2-8}
\raisebox{-15pt}[0pt][0pt]{\rotatebox{90}{$\P=0.7$}}
&$C^{r,i}_3$ &$\pm$1.02&$\pm$0.87&$\pm$0.76&$\pm$1.45&$\pm$1.24&$\pm$1.07\\\cline{2-8}
&$C^{r,i}_4$ &$\pm$1.63&$\pm$1.45&$\pm$1.28&$\pm$2.30&$\pm$2.05&$\pm$1.81\\\hline
\end{tabular}
  \caption{Bounds at the 68\% CL\  on
    contact-interaction couplings involving initial right-handed
    electrons, obtained from the leptonic-channel polarized cross
    section by varying the couplings one at a time.  The assumed
  experimental errors are 3\% and 6\%. }
\label{tab:lptsgm.bounds.pol4070}
\end{table}

As mentioned above, besides the cross section we have considered as
well several asymmetries.  In the leptonic channel asymmetries turn
out not to possess much better sensitivity than the cross section to
anomalous couplings.  For $\delta V_L$, $g_L$ and $C_1$ the cross
section is the only observable having significant sensitivity. Of the
six asymmetries we considered, only three yielded better bounds on
some coupling than the cross section.  The asymmetry
$A(\Delta\eta(\bbar,\ell))$ yields significantly tighter bounds on
$C_2$, and slightly tighter bounds on $V_R$ and $g_R$. The bounds on
$C_3$ from this observable are the same as those from the cross
section.  $A(\Delta\varphi(\bbar,\ell),\pi/2)$ improves the bounds on
$C_3$ and $C_4$, and $A(\Delta\varphi(\ell,\met),\pi/2)$ improves the
bounds on $C_3$.  In Table \ref{tab:lptasym.bounds.pol00} we summarize
the bounds on the effective couplings obtained from asymmetries,
including only those results that are as restrictive or better than
the corresponding results in Table \ref{tab:lptsgm.bounds.pol00}.
\begin{table}[t]
  \centering
\begin{tabular}{|c|c|cc|cc|cc||cc|cc|cc|}\cline{3-14}
\multicolumn{2}{c|}{}&\multicolumn{6}{c||}{$\epsilon_\mathrm{exp}=3\%$}
                     &\multicolumn{6}{c|}{$\epsilon_\mathrm{exp}=6\%$} \\\cline{2-14}
\multicolumn{1}{c|}{}& Observable
                      & \multicolumn{2}{c|}{60 GeV} 
                      & \multicolumn{2}{c|}{140 GeV} 
                      & \multicolumn{2}{c||}{300 GeV}
                      & \multicolumn{2}{c|}{60 GeV} 
                      & \multicolumn{2}{c|}{140 GeV} 
                      & \multicolumn{2}{c|}{300 GeV} \\\hline
$V^r_R\times10$ &$\Delta\eta(\bbar,\ell)$ 
                &-0.93&0.89&-0.92&0.94&-0.96&0.97&-1.31&1.28&-1.32&1.34&-1.38&1.39\\\hline
$V^i_R\times10$ &$\Delta\eta(\bbar,\ell)$ 
                &-0.92&0.92&-0.94&0.94&-0.95&0.95&-1.32&1.32&-1.35&1.35&-1.36&1.36\\\hline
$g^r_R\times10$ &$\Delta\eta(\bbar,\ell)$
                &-0.20&0.19&-0.18&0.18&-0.18&0.17&-0.39&0.38&-0.38&0.35&-0.36&0.33\\\hline
$g^i_R\times10$ &$\Delta\eta(\bbar,\ell)$ 
                &-0.82&0.82&-0.79&0.79&-0.76&0.76&-1.17&1.17&-1.13&1.13&-1.09&1.09\\\hline
$C^{r,i}_2$         &$\Delta\eta(\bbar,\ell)$
                &-2.83&2.83&-2.63&2.63&-2.38&2.38&-4.04&4.04&-3.75&3.75&-3.40&3.40\\\hline%
                &$\Delta\varphi(\bbar,\ell)$ 
                &-2.18&2.18&-1.56&1.56&-1.21&1.21&-3.11&3.11&-2.22&2.22&-1.72&1.72\\\cline{2-14}
\raisebox{5pt}{$C^{r,i}_3$}&$\Delta\varphi(\ell,\met)$ 
                &-2.35&2.35&-2.05&2.05&-1.76&1.76&-3.37&3.37&-2.95&2.95&-2.53&2.53\\\cline{2-14}
                &$\Delta\eta(\bbar,\ell)$
                &-2.43&2.43&-2.11&2.11&-1.82&1.82&-3.49&3.49&-3.03&3.03&-2.61&2.61\\\hline%
$C^{r,i}_4$         &$\Delta\varphi(\bbar,\ell)$ 
                &-3.72&3.72&-2.75&2.75&-2.13&2.13&-5.34&5.34&-3.92&3.92&-3.03&3.03\\\hline%
\end{tabular}
\caption{Bounds on effective couplings at the 68\% CL\ 
  obtained from the indicated leptonic-channel unpolarized asymmetries
  by varying the couplings one at a time.  $\Delta\eta(\bbar,\ell)$
  stands for the asymmetry $A\left(\Delta\eta(\bbar,\ell),0\right)$, 
  $\Delta\varphi(\bbar,\ell)$ for
  $A\left(\Delta\varphi(\bbar,\ell),\pi/2\right)$ and 
  $\Delta\varphi(\ell,\met)$ for
  $A\left(\Delta\varphi(\ell,\met),\pi/2\right)$. 
  The assumed experimental errors are 3\% and 6\%.}
\label{tab:lptasym.bounds.pol00}
\end{table}
As in the case of the total cross section, the sensitivity of the
asymmetries to the couplings $C_{2,3,4}$, involving right-handed
initial electrons, can be significantly enhanced by assuming a
right-polarized electron beam.  In Table
\ref{tab:lptasym.bounds.pol4070} we summarize our results on these
couplings, using the same asymmetries as in Table
\ref{tab:lptasym.bounds.pol00}, with initial electron polarizations
$\P=+0.4$, $+0.7$, and with the same range of assumed experimental
errors.
\begin{table}[t]
  \centering
\begin{tabular}{|c|c||c|c|c|c||c|c|c|}\cline{4-9}
\multicolumn{2}{c}{}&\multicolumn{1}{c|}{} &\multicolumn{3}{c||}{$\epsilon_\mathrm{exp}=3\%$}
                     &\multicolumn{3}{c|}{$\epsilon_\mathrm{exp}=6\%$} \\\cline{3-9}
\multicolumn{2}{c|}{} & Observable
                     & \multicolumn{1}{c|}{60 GeV} 
                     & \multicolumn{1}{c|}{140 GeV} 
                     & \multicolumn{1}{c||}{300 GeV}
                     & \multicolumn{1}{c|}{60 GeV} 
                     & \multicolumn{1}{c|}{140 GeV} 
                     & \multicolumn{1}{c|}{300 GeV} \\\hline
&$C^{r,i}_2$ &$\Delta\eta(\bbar,\ell)$    &$\pm$1.87&$\pm$1.73&$\pm$1.56&$\pm$2.66&$\pm$2.46&$\pm$2.23\\\cline{2-9}
\raisebox{-40pt}[0pt][0pt]{\rotatebox{90}{$\P=0.4$}}
&        &$\Delta\varphi(\bbar,\ell)$ &$\pm$1.42&$\pm$1.02&$\pm$0.79&$\pm$2.04&$\pm$1.46&$\pm$1.13\\\cline{3-9}
&\raisebox{3pt}{$C^{r,i}_3$}
        &$\Delta\varphi(\ell,\met)$   &$\pm$1.54&$\pm$1.34&$\pm$1.15&$\pm$2.21&$\pm$1.93&$\pm$1.66\\\cline{3-9}
&        &$\Delta\eta(\bbar,\ell)$    &$\pm$1.62&$\pm$1.39&$\pm$1.20&$\pm$2.32&$\pm$1.99&$\pm$1.73\\\cline{2-9}
&$C^{r,i}_4$ &$\Delta\varphi(\bbar,\ell)$ &$\pm$2.45&$\pm$1.80&$\pm$1.40&$\pm$3.51&$\pm$2.57&$\pm$1.99\\\hline
&$C^{r,i}_2$ &$\Delta\eta(\bbar,\ell)$    &$\pm$1.20&$\pm$1.11&$\pm$1.00&$\pm$1.72&$\pm$1.58&$\pm$1.43\\\cline{2-9}
\raisebox{-40pt}[0pt][0pt]{\rotatebox{90}{$\P=0.7$}}
&        &$\Delta\varphi(\bbar,\ell)$ &$\pm$0.91&$\pm$0.66&$\pm$0.51&$\pm$1.30&$\pm$0.94&$\pm$0.72\\\cline{3-9}
&\raisebox{3pt}{$C^{r,i}_3$}
         &$\Delta\varphi(\ell,\met)$  &$\pm$0.99&$\pm$0.86&$\pm$0.74&$\pm$1.42&$\pm$1.23&$\pm$1.06\\\cline{3-9}
&        &$\Delta\eta(\bbar,\ell)$    &$\pm$1.04&$\pm$0.89&$\pm$0.77&$\pm$1.49&$\pm$1.28&$\pm$1.11\\\cline{2-9}
&$C^{r,i}_4$ &$\Delta\varphi(\bbar,\ell)$ &$\pm$1.56&$\pm$1.15&$\pm$0.90&$\pm$2.24&$\pm$1.64&$\pm$1.28\\\hline
\end{tabular}
\caption{Bounds at the 68\% CL\ on
  contact-interaction couplings involving initial right-handed
  electrons, obtained from the indicated leptonic-channel asymmetries
  by varying the couplings one at a time.  The definition of the
  asymmetries is as in table \ref{tab:lptasym.bounds.pol00}. The assumed
  experimental errors are 3\% and 6\%. }
\label{tab:lptasym.bounds.pol4070}
\end{table}

The asymmetry of $\Delta\varphi(\bbar, \met)$, with reference value
$\pi/2$, has only some marginal sensitivity to $C_3$ leading to bounds
much weaker than those in Table \ref{tab:lptasym.bounds.pol00}.
Similarly, the asymmetry related to $\cos(\bbar,\ell)$ leads to loose
bounds on $C_3$, $g_R$, $g_L$.  We considered also the asymmetry
$A(u(\bbar,\ell),u_0)$ with $u_0=m_W^2/m_t^2\simeq 0.215$, where the
distribution of $u$ has a shoulder \cite{she09}, and also with
$u_0=1/2$. The former reference value yields better result than the
latter, but even in that case we do not find this asymmetry to possess
any significant sensitivity to the effective couplings studied in this
paper, at the energies considered here.  For instance, for the
coupling $C_1$ we find the marginally interesting bounds
$-4.8<C_1<8.4$ at $\varepsilon=3\%$ and $-11.1<C_1<9.5$ at
$\varepsilon=6\%$, which are much weaker than the bounds from the
unpolarized cross section.

We have also studied the sensitivity of single top production in the
leptonic channel to the four-fermion operators ${\cal O}^{2332}_{\ell
  q'}$, ${\cal O}^{2233}_{qde}$, ${\cal O}^{3223}_{q\ell \epsilon}$,
${\cal O}^{2233}_{\ell q\epsilon}$, involving second-generation
leptons.  These operators are obtained from those in Table
\ref{leptonquark} by substituting $e$, $\nu_e$ by $\mu$, $\nu_\mu$.
At the LHeC, these operators can only enter single-top production and
decay through the top decay vertex (diagrams (d), (f), (h) in Figure
\ref{fig:sgn.dgrm.eff}).  Since the contribution of contact
interactions to the top decay vertex is known to be negligible
\cite{celine}, we do not expect to find any sensitivity to the coupling
constants associated to these operators.  That is, in fact, the case.
The bounds we find on four-fermion couplings involving muons are quite
weak, of order $C\sim 10^2$ (or, equivalently, new physics scales
$\Lambda\gtrsim 100$ GeV).

We should compare the bounds obtained here with those from
\cite{mellado}, which are based on a combined bin analysis of the
distributions of a similar set of kinematical observables as ours.  In
particular, from Figure 7 of \cite{mellado} with 10\% systematical
error we observe similar bounds for $\delta V_L$ ($\simeq 10^{-2}$),
$V_R$ and $g_L$ ($\simeq 10^{-1}$).  On the other hand, our bounds on
$g_R$ ($\simeq 0.02$) are somewhat tighter than theirs ($\simeq
0.03$).

\begin{figure}[t]
  \centering
%\fbox{
\includegraphics[scale=1]{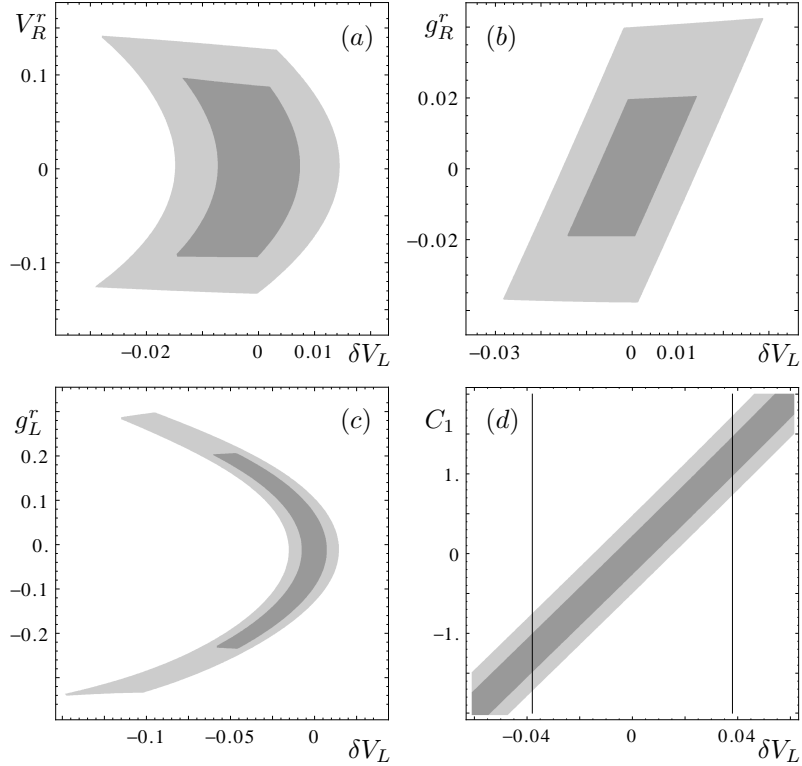}
%}
\caption{Allowed regions at 68\% CL\ at $E_e=60$ GeV for (a) $\delta
  V_L$, $V_R$ from the cross section and the asymmetry
  $A(\Delta\eta(\bbar,\ell))$, (b) $\delta V_L$, $g_R$ from the cross
  section and the asymmetry $A(\Delta\eta(\bbar,\ell))$, (c) $\delta
  V_L$, $g_L$ from the cross section and the asymmetry
  $A(\Delta\varphi(\bbar,\met))$, (d) $\delta V_L$, $C_1$ from the
  cross section. The vertical lines in (d) correspond to the current
  bounds $|\delta V_L| < 0.038$ \cite{cms2014}. The light-gray regions
  correspond to $\varepsilon_\mathrm{exp}=8\%$, the medium-gray
  regions to $\varepsilon_\mathrm{exp}=6\%$ and the dark-gray regions
  to $\varepsilon_\mathrm{exp}=3\%$.  }
  \label{fig:xcl.a}
\end{figure}

We turn next to the allowed regions of parameter space obtained by
letting two couplings to be non-zero simultaneously, and by allowing
for the necessary additional interference terms in
(\ref{lheccross}). In Figure \ref{fig:xcl.a} the allowed regions at
68\% confidence level (CL) and at $E_e=60$ GeV are shown, on the
planes $\delta V_L$--$V^r_R$, $g^r_R$, $g^r_L$ and $C_1$. In all cases
we assume an unpolarized electron beam, as is appropriate for these
couplings.  In the cases of $V^r_R$ and $g^r_R$ (figures
\ref{fig:xcl.a} (a) and (b)) the allowed regions are determined by the
level curves of the asymmetry $A(\Delta\eta(\bbar,\ell))$ and of the
cross section.  For $\delta V_L$ and $g^r_L$ the only available
observable with significant sensitivity to both couplings is the cross
section.  The resulting allowed region is not a neighborhood of the
origin, but an elliptical corona having the origin at its periapsis.
In order to obtain a neighborhood of the origin, in Figure
\ref{fig:xcl.a} (c) we used the level curves of the cross section and
the asymmetry $A(\Delta\varphi(\bbar,\met))$.  As a consequence, the
single-coupling bounds on $g^r_L$ determined by the intersection of
the allowed region in the figure with the axis $\delta V_L=0$ are less
restrictive than those obtained directly from the cross section (see
Table \ref{tab:lptsgm.bounds.pol00}).  Similarly, for the couplings
$\delta V_L$ and $C_1$ the only sensitive observable is the cross
section $\sigma$. For moderate values of these couplings, $\sigma$
does not depend on them independently but only through a linear
combination, as can be seen in Figure \ref{fig:xcl.a} (d).  Also shown
in that figure, for reference, is the current bound $V_L = 0.998 \pm
0.038$ (exp) from CMS \cite{cms2014}.

We determine the allowed region in the plane $V^r_R$--$V^i_R$ from the
asymmetry $A(\Delta(\eta(\bbar,\ell)))$, which gives the best bounds
on these couplings (see Table \ref{tab:lptasym.bounds.pol00}).  Since
the interference term with the SM proportional to $V^r_R$ is small
(see Appendix \ref{sec:param}), the resulting allowed region is
essentially a solid ellipse inscribed in the rectangle formed with the
single-coupling bounds from $A(\Delta(\eta(\bbar,\ell)))$.  Similarly,
the allowed region in the plane $g^r_L$--$g^i_L$ is a solid ellipse
inscribed in the rectangle determined by the single-coupling bounds
given in Table \ref{tab:lptsgm.bounds.pol00}.  A figure for these
allowed regions is therefore not needed. 

\begin{figure}[t]
  \centering
\includegraphics[scale=1]{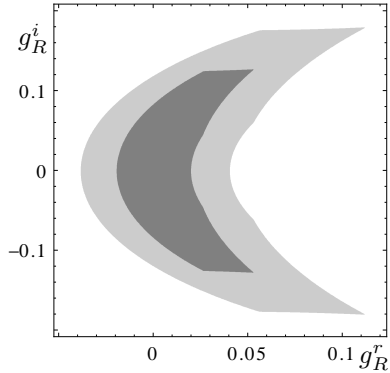}
\caption{Allowed regions at 68\% CL\ at $E_e=60$ GeV for $g_R$
  obtained from the leptonic cross section and the asymmetries
  $A(\Delta\eta(\bbar,\ell))$ and $A(\Delta\varphi(\bbar,\ell))$. The
  light-gray area corresponds to $\varepsilon_\mathrm{exp}=8\%$, the
  medium-gray one to $\varepsilon_\mathrm{exp}=6\%$ and the
  dark-gray one to $\varepsilon_\mathrm{exp}=3\%$.}
  \label{fig:xcl.25}
\end{figure}

There is a substantial interference term proportional to $g^r_R$ in
(\ref{lheccross}), however, as shown by its coefficients in Appendix
\ref{sec:param}. Thus the allowed region in the plane $g^r_R$--$g^i_R$
determined by the cross section alone or the asymmetry
$A(\Delta(\eta(\bbar,\ell)))$ alone, are elliptical coronas.  In
Figure \ref{fig:xcl.25} we show the allowed region in the plane
$g^r_R$--$g^i_R$ resulting from the intersection of the level curves
of the cross section and the asymmetries $A(\Delta\eta(\bbar,\ell))$
and $A(\Delta\varphi(\bbar,\ell))$.

\begin{figure}[t]
  \centering
\includegraphics[scale=1]{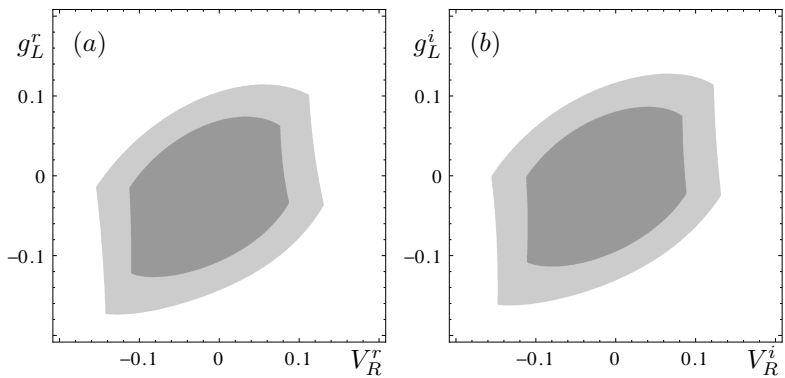}
\caption{Allowed regions at 68\% CL\ at $E_e=60$ GeV in the plane
  (a) $V^r_R$--$g^r_L$ and (b) $V^i_R$--$g^i_L$, obtained from the
  asymmetries $A(\Delta\eta(\bbar,\ell))$ and the cross section. The
  light-gray areas correspond to $\varepsilon_\mathrm{exp}=8\%$, the
  medium-gray ones to $\varepsilon_\mathrm{exp}=6\%$ and the
  dark-gray ones to $\varepsilon_\mathrm{exp}=3\%$.}
  \label{fig:xcl.26}
\end{figure}

The leptonic cross section and the asymmetry
$A(\Delta\eta(\bbar,\ell))$ yield the best bounds on the couplings
$g_L$ and $V_R$, respectively, as seen from Tables
\ref{tab:lptsgm.bounds.pol00} and \ref{tab:lptasym.bounds.pol00}.  The
allowed regions determined by those observables in the planes
$V^r_R$--$g^r_L$ and $V^i_R$--$g^i_L$ are shown in Figure
\ref{fig:xcl.26}.  We have taken into account in the figure 
interference terms of the form $V_R\times g_L$, which are not
suppressed by $m_b$.

\begin{figure}[t]
  \centering
\includegraphics[scale=1]{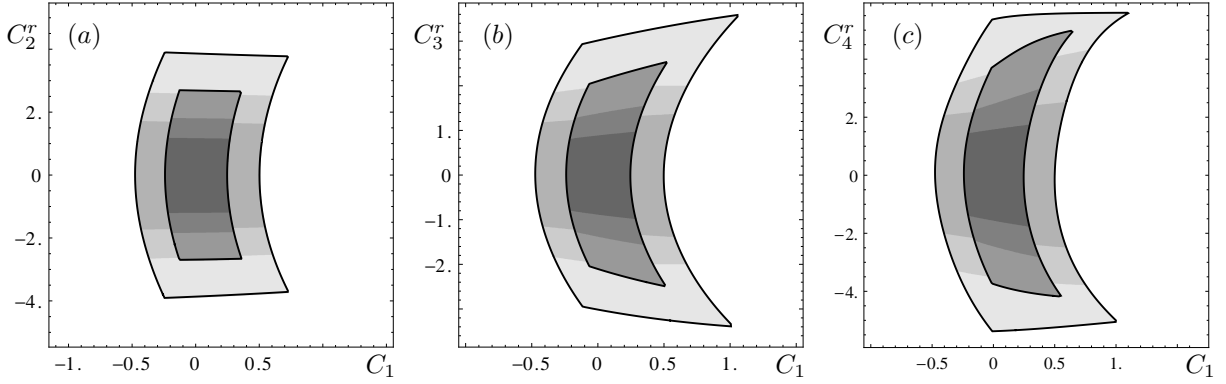}
\caption{Allowed regions at 68\% CL\ at $E_e=60$ GeV for (a)
  $C^r_1$, $C^r_2$ from the unpolarized cross section and the
  asymmetry $A(\Delta\eta(\bbar,\ell))$, (b) $C^r_1$, $C^r_3$ from the
  unpolarized cross section and the asymmetry
  $A(\Delta\varphi(\bbar,\ell))$, (c) $C^r_1$, $C^r_4$ from the
  unpolarized cross section and the asymmetry
  $A(\Delta\varphi(\bbar,\ell))$.  The larger quadrangles in each panel
  correspond to $\varepsilon_\mathrm{exp}=6\%$, the smaller ones to
  $\varepsilon_\mathrm{exp}=3\%$.  Within each quadrangle the
  lighter-gray region corresponds to $\P=0$, the medium-gray region
  to $\P=0.4$ and the darker-gray region to $\P=0.7$.  }
  \label{fig:xcl.b}
\end{figure}

The allowed regions for the contact-interaction couplings
$C^r_{2,3,4}$ versus $C_1$ are shown in Figure \ref{fig:xcl.b}, at
$E_e=60$ GeV for electron-beam polarizations $\P=0.0$, $+0.4$,
$+0.7$. The large increase in sensitivity to $C^r_{2,3,4}$, already
apparent from Tables \ref{tab:lptsgm.bounds.pol4070} and
\ref{tab:lptasym.bounds.pol4070}, is clearly seen in the figure.  
The figures for $C^i_2$, $C^i_3$, $C^i_4$ versus $C_1$ are essentially
identical to Figure \ref{fig:xcl.b}.

\begin{figure}[t]
  \centering
\includegraphics[scale=1]{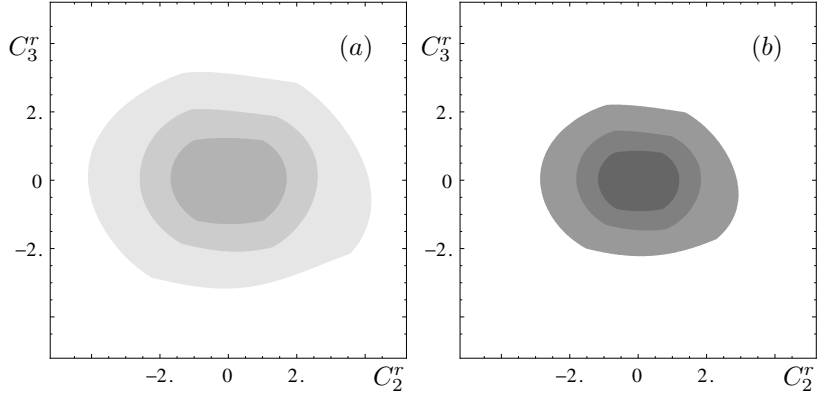}
\caption{Allowed regions at 68\% CL\ at $E_e=60$ GeV for $C^r_2$,
  $C^r_3$ from the leptonic-channel cross section and the asymmetries
  $A(\Delta\eta(\bbar,\ell))$ and $A(\Delta\varphi(\bbar,\ell))$ with
  (a) $\varepsilon_\mathrm{exp}=6\%$ and (b)
  $\varepsilon_\mathrm{exp}=3\%$. In each panel the lighter-gray
  region corresponds to $\P=0$, the medium-gray region to $\P=0.4$ and
  the darker-gray region to $\P=0.7$.}
  \label{fig:xcl.c}
\end{figure}

The best bounds we obtain on $C_2$ are those from the leptonic-channel
asymmetry $A(\Delta\eta(\bbar,\ell))$, and the bounds on $C_3$ from
the leptonic asymmetry $A(\Delta\varphi(\bbar,\ell))$ are equally
tight as those obtained in the hadronic channel (see section
\ref{sec:hdr.chn} below).  The allowed regions in the plane
$C^r_2$--$C^r_3$ determined by those asymmetries are displayed in
Figure \ref{fig:xcl.c} for the three electron polarizations $\P=0.0$,
$+0.4$, $+0.7$. 

\subsection{Bounds from hadronic channel}
\label{sec:hdr.chn}

In the hadronic channel it is possible to measure the four-momenta of
the three final-state jets and, therefore, to reconstruct the
four-momenta of the intermediate-state $W$ and $\tbar$.  From the
known four-momenta of the initial electron and of $\tbar$ the entire
kinematics of the process can be fully reconstructed.  In particular,
we can find the four-momenta of the final-state neutrino and the
initial-state $\bbar$ and thus also the total four momentum of the
process, which permits us to boost the event to other frames such as
the partonic center-of-mass frame, or the $\tbar$ rest frame.  We can
therefore obtain asymmetries for a large variety of kinematic
observables, and look for the ones most sensitive to the anomalous
couplings.  In all cases, we compute the required cross sections with
the cuts $H_4$ as defined in (\ref{eq:cuts.H}).

The sensitivity to contact interactions entering only through decay
vertices is negligible \cite{celine}, as is clearly illustrated by the
case of couplings involving muons in the leptonic channel discussed in
section \ref{sec:lpt.chn}.  For that reason, in the hadronic channel
we ignore diagrams (d), (f), (h) in Figure \ref{fig:sgn.dgrm.eff}.  On
the other hand, we do retain in our computations the contribution of
anomalous $tbW$ couplings to the top decay vertex, which yields a
non-negligible enhancement of the sensitivity to those
couplings.  Thus, the tree-level cross section we compute is quadratic
in $C_{1,\ldots,4}$ and quartic in $V_{L,R}$, $g_{L,R}$.  Within the
ranges for the anomalous couplings determined by the bounds we obtain
below, however, the contribution of the terms cubic and quartic in the
anomalous $tbW$ couplings is small.  Thus, within those ranges, the
cross section can be well approximated by a quadratic parameterization
of the form (\ref{lheccross}), with $(1+\delta V_L)^4\simeq 1+4\delta
V_L+6\delta V_L^2$.  The coefficients in (\ref{lheccross}) for the
hadronic-channel unpolarized cross section, and for the polarized one
with $\P=+0.4$ and $+0.7$, are given in Appendix \ref{sec:param}.  As
noted in section \ref{sec:lpt.chn}, since (\ref{lheccross}) does not
include interference terms involving two different effective
couplings, it is appropriate only for obtaining bounds on the
anomalous couplings taken one at a time.  Nevertheless, we do take
those missing interference terms into account in the computation of
exclusion regions for pairs of effective couplings.

In Table \ref{tab:hdrsgm.bounds.pol00} we report the bounds on the
effective couplings obtained from the unpolarized cross section at the
level of one standard deviation, assuming experimental errors of 7\%
and 12\% as discussed in section \ref{sec:sm.hdr}.  As discussed in
relation to Table \ref{tab:lptsgm.bounds.pol00} for the leptonic channel,
the bounds on effective $tbW$ couplings are almost independent of
$E_e$, and those on contact-interaction couplings show a mild
dependence on $E_e$ that makes them somewhat tighter at higher
energies.  Also as in the case of the leptonic channel, the bounds on
the imaginary parts $C^i_{2,3,4}$ are the same as those on the real
parts $C^r_{2,3,4}$.  The bounds from the hadronic cross section shown
in Table \ref{tab:hdrsgm.bounds.pol00} are weaker than those in Table
\ref{tab:lptsgm.bounds.pol00} from the leptonic channel, not
surprisingly, since the assumed experimental errors in the former
channel are about twice as large as those in the latter one.
\begin{table}[t]
  \centering
\begin{tabular}{|c|cc|cc|cc||cc|cc|cc|}\cline{2-13}
\multicolumn{1}{c|}{}&\multicolumn{6}{c||}{$\epsilon_\mathrm{exp}=7\%$}
                     &\multicolumn{6}{c|}{$\epsilon_\mathrm{exp}=12\%$} \\\cline{2-13}
\multicolumn{1}{c|}{} & \multicolumn{2}{c|}{60 GeV} 
                      & \multicolumn{2}{c|}{140 GeV} 
                      & \multicolumn{2}{c||}{300 GeV}
                      & \multicolumn{2}{c|}{60 GeV} 
                      & \multicolumn{2}{c|}{140 GeV} 
                      & \multicolumn{2}{c|}{300 GeV} \\\hline
$\delta V_L\times 10^{2}$&-1.79&1.70&-1.80&1.71&-1.80&1.71&-3.15&2.87&-3.15&2.88&-3.15&2.88\\\hline
$V^r_R\times10$          &-1.49&1.77&-1.53&1.82&-1.55&1.83&-1.98&2.26&-2.04&2.32&-2.05&2.35\\\hline
$V^i_R\times10$          &-1.63&1.63&-1.67&1.67&-1.69&1.69&-2.12&2.12&-2.17&2.17&-2.20&2.20\\\hline
$g^r_R\times10$          &-0.54&0.62&-0.50&0.59&-0.48&0.57&-0.88&1.12&-0.82&1.07&-0.78&1.05\\\hline
$g^i_R\times10$          &-1.61&1.61&-1.54&1.54&-1.48&1.48&-2.10&2.10&-2.01&2.01&-1.93&1.93\\\hline
$g^r_L\times10$          &-1.30&1.07&-1.36&1.12&-1.36&1.14&-1.66&1.43&-1.72&1.49&-1.74&1.52\\\hline
$g^i_L\times10$          &-1.18&1.18&-1.23&1.23&-1.24&1.24&-1.54&1.54&-1.60&1.60&-1.62&1.62\\\hline
$C_1$                    &-0.48&0.51&-0.45&0.48&-0.43&0.46&-0.81&0.89&-0.76&0.85&-0.72&0.82\\\hline%
$C^{r,i}_2$                  &-5.31&5.31&-5.06&5.06&-4.62&4.62&-6.95&6.95&-6.62&6.62&-6.05&6.05\\\hline%
$C^{r,i}_3$                  &-3.29&3.29&-2.97&2.97&-2.65&2.65&-4.31&4.31&-3.88&3.88&-3.47&3.47\\\hline%
$C^{r,i}_4$                  &-5.14&5.14&-4.88&4.88&-4.45&4.45&-6.73&6.73&-6.39&6.39&-5.83&5.83\\\hline%
\end{tabular}
\caption{Bounds on effective couplings at the 68\% CL\ 
  obtained from the hadronic-channel unpolarized cross section
  by varying the couplings one at a time.  The assumed
  experimental errors are 7\% and 12\%. }
\label{tab:hdrsgm.bounds.pol00}
\end{table}

In the hadronic channel some asymmetries yield significantly better
sensitivity than the cross section, unlike what happens in the
leptonic case.  In Table \ref{tab:hdrasym.bounds.pol00} we show bounds
on the effective couplings obtained from the asymmetries indicated
there, that are better than those from the cross section at the three
energies $E_e=60$, 140, 300 GeV.  The asymmetry
$A(\Delta\varphi(j_2,\met),\pi/2)$ gives more restrictive bounds on
$V^r_R$, $V^i_R$, $g^r_R$ than $\sigma$ (see Table
\ref{tab:hdrsgm.bounds.pol00}) at $E_e=60$ GeV, but not at 140 and 300
GeV, and for that reason they are not shown in Table
\ref{tab:hdrasym.bounds.pol00}.
On the other hand, the bounds given by this asymmetry for $g^i_R$
are slightly better than those from $\sigma$, as seen in the table.
Similarly, the asymmetry $A(\Delta\varphi(W^-,\met),\pi/2)$ gives
more restrictive bounds on $C^{r,i}_{3,4}$, than $\sigma$ at
$E_e=60$ GeV, but not at 140 and 300 GeV.  

\begin{table}[t]
  \centering
\begin{tabular}{|c|c|cc|cc|cc||cc|cc|cc|}\cline{3-14}
\multicolumn{2}{c|}{}&\multicolumn{6}{c||}{$\epsilon_\mathrm{exp}=7\%$}
                     &\multicolumn{6}{c|}{$\epsilon_\mathrm{exp}=12\%$} \\\cline{2-14}
\multicolumn{1}{c|}{}& Observable
                      & \multicolumn{2}{c|}{60 GeV} 
                      & \multicolumn{2}{c|}{140 GeV} 
                      & \multicolumn{2}{c||}{300 GeV}
                      & \multicolumn{2}{c|}{60 GeV} 
                      & \multicolumn{2}{c|}{140 GeV} 
                      & \multicolumn{2}{c|}{300 GeV} \\\hline
 &$\Delta\eta(\bbar,j_1)$
 &-1.21&1.31&-1.22&1.36&-1.20&1.37&-1.62&1.72&-1.64&1.78&-1.62&1.79 \\
 \cline{2-14}\raisebox{0pt}[0pt][0pt]{$V^r_R\times10$}
 &$\Delta y(\tbar,\bbar)$  &-1.29&1.49&-1.26&1.39&-1.27&1.41&-1.76&1.93&-1.70&1.83&-1.71&1.84\\ \cline{2-14}
 &$\Delta y(\tbar,j_2)$    &-1.32&1.44&-1.31&1.36&-1.27&1.37&-1.78&1.88&-1.75&1.79&-1.71&1.80\\ \cline{2-14}
 &$\Delta\eta(\bbar,j_2)$  &-1.35&1.44&-1.30&1.35&-1.27&1.38&-1.81&1.90&-1.73&1.79&-1.70&1.81 \\ \hline%\cline{2-14}
%&$\Delta\varphi(j_2,\met)$&-1.38&1.67&-1.59&1.92&-1.68&1.92&-1.90&2.24&-2.19&2.53&-2.31&2.62 \\ \hline

 &$\Delta\eta(\bbar,j_1)$  &-1.28&1.28&-1.29&1.29&-1.30&1.30&-1.70&1.70&-1.72&1.72&-1.72&1.72 \\\cline{2-14}\raisebox{-10pt}[0pt][0pt]{$V^i_R\times10$}
 &$\Delta y(\tbar,\bbar)$  &-1.37&1.37&-1.32&1.32&-1.32&1.32&-1.82&1.82&-1.76&1.76&-1.75&1.75\\ \cline{2-14}
 &$\Delta y(\tbar,j_2)$    &-1.38&1.38&-1.31&1.31&-1.31&1.31&-1.83&1.83&-1.74&1.74&-1.73&1.73\\\cline{2-14}
 &$\Delta\eta(\bbar,j_2)$  &-1.38&1.38&-1.32&1.32&-1.32&1.32&-1.83&1.83&-1.75&1.75&-1.75&1.75 \\\cline{2-14}
%&$\Delta\varphi(j_2,\met)$&-1.55&1.55&-1.76&1.76&-1.83&1.83&-2.09&2.09&-2.37&2.37&-2.48&2.48 \\\cline{2-14}
 &$\Delta\eta(j_1,j_2)$    &-1.59&1.59&-1.45&1.45&-1.36&1.36&-2.12&2.12&-1.93&1.93&-1.81&1.81 \\\hline

 &$\Delta\eta(\bbar,j_1)$  &-0.18&0.18&-0.18&0.18&-0.17&0.17&-0.32&0.31&-0.31&0.30&-0.30&0.29 \\\cline{2-14}\raisebox{0pt}[0pt][0pt]{$g^r_R\times10$}
 &$\Delta y(\tbar,\bbar)$  &-0.25&0.25&-0.21&0.21&-0.20&0.20&-0.43&0.43&-0.37&0.37&-0.34&0.34 \\\cline{2-14}
 &$\Delta\eta(\bbar,j_2)$  &-0.35&0.36&-0.30&0.30&-0.27&0.27&-0.60&0.63&-0.50&0.51&-0.46&0.47 \\\cline{2-14}
 &$\Delta y(\tbar,j_2)$    &-0.55&0.63&-0.45&0.50&-0.39&0.45&-0.90&1.22&-0.74&0.92&-0.66&0.81 \\\hline%\cline{2-14}
%&$\Delta\varphi(j_2,\met)$&-0.57&0.46&-0.69&0.49&-0.99&0.51&-1.10&0.76&-4.20&0.78&-3.50&0.81 \\\hline
                                                                                             
 &$\Delta\eta(\bbar,j_1)$  &-1.05&1.05&-1.02&1.02&-1.00&1.00&-1.39&1.39&-1.36&1.36&-1.32&1.32 \\\cline{2-14}\raisebox{-10pt}[0pt][0pt]{$g^i_R\times10$}
 &$\Delta y(\tbar,\bbar)$  &-1.19&1.19&-1.09&1.09&-1.05&1.05&-1.58&1.58&-1.45&1.45&-1.39&1.39 \\\cline{2-14}
 &$\Delta\eta(\bbar,j_2)$  &-1.32&1.32&-1.20&1.20&-1.14&1.14&-1.76&1.76&-1.60&1.60&-1.52&1.52 \\\cline{2-14}
 &$\Delta\varphi(j_2,\met)$&-1.35&1.35&-1.28&1.28&-1.23&1.23&-1.78&1.78&-1.70&1.70&-1.63&1.63 \\\cline{2-14}
 &$\Delta y(\tbar, j_2)$   &-1.55&1.55&-1.39&1.39&-1.32&1.32&-2.08&2.08&-1.85&1.85&-1.76&1.76 \\\hline
\end{tabular}
\caption{Bounds on effective couplings at the 68\% CL\ 
  obtained from the indicated hadronic-channel unpolarized asymmetries
  by varying the couplings one at a time.  The definition of the
  asymmetries is analogous to that in table
  \ref{tab:lptasym.bounds.pol00}; $j_{1}$ (resp.\ $j_2$) refers to the
  light non-$b$ jet with the larger (resp.\ smaller) $|\vec{p}_T|$.
  The assumed experimental errors are 7\% and 12\%. (Continued on next page.)}
\label{tab:hdrasym.bounds.pol00}
\end{table}
\addtocounter{table}{-1}
\begin{table}[t]
  \centering
\begin{tabular}{|c|c|cc|cc|cc||cc|cc|cc|}\cline{3-14}
\multicolumn{2}{c|}{}&\multicolumn{6}{c||}{$\epsilon_\mathrm{exp}=7\%$}
                     &\multicolumn{6}{c|}{$\epsilon_\mathrm{exp}=12\%$} \\\cline{2-14}
\multicolumn{1}{c|}{}& Observable
                      & \multicolumn{2}{c|}{60 GeV} 
                      & \multicolumn{2}{c|}{140 GeV} 
                      & \multicolumn{2}{c||}{300 GeV}
                      & \multicolumn{2}{c|}{60 GeV} 
                      & \multicolumn{2}{c|}{140 GeV} 
                      & \multicolumn{2}{c|}{300 GeV} \\\hline
 &$\Delta\eta(\bbar,j_1)$  &-3.75&3.75&-3.74&3.74&-3.51&3.51&-4.98&4.98&-4.96&4.96&-4.66&4.66\\\cline{2-14}\raisebox{-10pt}[0pt][0pt]{$C^{r,i}_2$}
 &$\Delta y(\tbar,\bbar)$  &-4.01&4.01&-3.84&3.84&-3.54&3.54&-5.33&5.33&-5.10&5.10&-4.71&4.71\\\cline{2-14}
 &$\Delta y(\tbar,j_2)$    &-4.03&4.03&-3.85&3.85&-3.59&3.59&-5.36&5.36&-5.12&5.12&-4.77&4.77\\\cline{2-14}
 &$\Delta\eta(\bbar,j_2)$  &-4.04&4.04&-3.85&3.85&-3.58&3.58&-5.36&5.36&-5.11&5.11&-4.77&4.77\\\cline{2-14}
 &$\Delta\eta(j_1,j_2)$    &-4.35&4.35&-3.97&3.97&-3.58&3.58&-5.79&5.79&-5.27&5.27&-4.77&4.77\\\hline%
 &$\Delta\varphi(j_2,\met)$&-2.16&2.16&-1.85&1.85&-1.61&1.61&-2.85&2.85&-2.44&2.44&-2.13&2.13\\\cline{2-14}\raisebox{10pt}[0pt][0pt]{$C^{r,i}_3$}
 &$\Delta\varphi(j_1,j_2)$ &-2.51&2.51&-1.88&1.88&-1.50&1.50&-3.33&3.33&-2.49&2.49&-1.97&1.97\\\hline%\cline{2-14}
%&$\Delta\varphi(W^-,\met)$&-3.01&3.01&-3.01&3.01&-2.80&2.80&-4.03&4.03&-4.05&4.05&-3.78&3.78\\\hline
%&$\Delta\met$             &-&&-&&-&&-&&-&&-&\\\hline%
 &$\Delta\varphi(j_2,\met)$&-3.70&3.70&-3.25&3.25&-2.82&2.82&-4.91&4.91&-4.31&4.31&-3.74&3.74\\\cline{2-14}\raisebox{10pt}[0pt][0pt]{$C^{r,i}_4$}
 &$\Delta\varphi(j_1,j_2)$ &-4.28&4.28&-3.30&3.30&-2.63&2.63&-5.70&5.70&-4.37&4.37&-3.47&3.47\\\hline%\cline{2-14}
%&$\Delta\varphi(W^-,\met)$&-4.99&4.99&-5.18&5.18&-4.85&4.85&-6.69&6.69&-6.98&6.98&-6.55&6.55\\\hline%
\end{tabular}
\caption{(Continued from previous page.)}
\end{table}

Notice that, as seen from Tables \ref{tab:hdrasym.bounds.pol00} and
\ref{tab:lptsgm.bounds.pol00}, the bounds on $g^r_R$, $g^i_R$ obtained
from $\Delta\eta(\bbar,j_1)$ with $\epsilon_\mathrm{exp}=7\%$ are
actually tighter than those obtained from the leptonic cross section
with $\epsilon_\mathrm{exp}=3\%$.  The bounds on $C^{r,i}_2$ from
those observables are essentially the same, despite the larger error
in the hadronic channel.  Similarly, the bounds on $C^{r,i}_3$,
$C^{r,i}_4$ obtained from $\Delta\varphi(j_2,\met)$ with
$\epsilon_\mathrm{exp}=7\%$ are tighter than those obtained from the
leptonic cross section with $\epsilon_\mathrm{exp}=3\%$.  Furthermore,
the bounds from hadronic asymmetries with $\epsilon_\mathrm{exp}=7\%$
in Table \ref{tab:hdrasym.bounds.pol00} are seen to be comparable to
those from leptonic asymmetries with $\epsilon_\mathrm{exp}=3\%$ in
Table \ref{tab:lptasym.bounds.pol00}.  In particular, the best
hadronic-channel bounds on $g^r_R$ with $\epsilon_\mathrm{exp}=7\%$
are those from $\Delta\eta(\bbar,j_1)$, which are better than the best
leptonic-channel bounds with $\epsilon_\mathrm{exp}=3\%$ coming from
$\Delta\eta(\bbar,\ell)$, at the three energies.  Likewise, the best
hadronic-channel bound on $C_3^{r,i}$ obtained from
$\Delta\varphi(j_2,\met)$ at $E_e=60$ GeV and
$\epsilon_\mathrm{exp}=7\%$ is better than the best leptonic-channel
bound from $\Delta\varphi(\bbar,\ell)$ at that energy with
$\epsilon_\mathrm{exp}=3\%$.

For the contact-interaction couplings $C^{r,i}_{2,3,4}$ involving
right-handed electrons, the sensitivity can be significantly improved
if the electron beam is right polarized.  In Table
\ref{tab:hdrsgm.bounds.pol4070} we give the bounds obtained for those
couplings from the cross section by assuming initial-electron
polarizations $\P=+0.4$ and $+0.7$.  Those bounds are tighter than the
ones from the unpolarized cross section in Table
\ref{tab:hdrsgm.bounds.pol00}.  At $\P=+0.4$ they are comparable to,
and at $+0.7$ better than, the bounds from the unpolarized hadronic
asymmetries in Table \ref{tab:hdrasym.bounds.pol00} and from the
unpolarized leptonic cross section, Table
\ref{tab:lptsgm.bounds.pol00}, and unpolarized leptonic asymmetries,
Table \ref{tab:lptasym.bounds.pol00}.  On the other hand, the bounds
from Table \ref{tab:hdrsgm.bounds.pol4070} are weaker than those from
the leptonic-channel polarized cross section in Table
\ref{tab:lptsgm.bounds.pol4070}.  
\begin{table}[t]
  \centering
\begin{tabular}{|c|c|c|c|c||c|c|c|}\cline{3-8}
\multicolumn{2}{c|}{}&\multicolumn{3}{c||}{$\epsilon_\mathrm{exp}=7\%$}
                     &\multicolumn{3}{c|}{$\epsilon_\mathrm{exp}=12\%$} \\\cline{3-8}
\multicolumn{2}{c|}{} &60 GeV&140 GeV&300 GeV&60 GeV&140 GeV&300 GeV \\\hline
&$C^{r,i}_2$ &$\pm$3.47&$\pm$3.31&$\pm$3.03&$\pm$4.55&$\pm$4.33&$\pm$3.97\\\cline{2-8}
\raisebox{-15pt}[0pt][0pt]{\rotatebox{90}{$\P=0.4$}}
&$C^{r,i}_3$ &$\pm$2.16&$\pm$1.95&$\pm$1.74&$\pm$2.83&$\pm$2.55&$\pm$2.27\\\cline{2-8}
&$C^{r,i}_4$ &$\pm$3.37&$\pm$3.20&$\pm$2.92&$\pm$4.41&$\pm$4.19&$\pm$3.82\\\hline
&$C^{r,i}_2$ &$\pm$2.23&$\pm$2.12&$\pm$1.94&$\pm$2.92&$\pm$2.78&$\pm$2.54\\\cline{2-8}
\raisebox{-15pt}[0pt][0pt]{\rotatebox{90}{$\P=0.7$}}
&$C^{r,i}_3$ &$\pm$1.39&$\pm$1.25&$\pm$1.11&$\pm$1.81&$\pm$1.63&$\pm$1.46\\\cline{2-8}
&$C^{r,i}_4$ &$\pm$2.16&$\pm$2.05&$\pm$1.87&$\pm$2.83&$\pm$2.69&$\pm$2.45\\\hline
\end{tabular}
  \caption{Bounds at the 68\% CL\ on
    contact-interaction couplings involving initial right-polarized
    electrons, obtained from the hadronic-channel polarized cross
    section by varying the couplings one at a time.  The assumed
  experimental errors are 7\% and 12\%. The asymmetries of
  $\Delta y(\tbar,j_2)$ and $\Delta y(\tbar, \bbar)$ yield the same
  bounds on $C^{r,i}_2$ as $\Delta\eta(\bbar,j_2)$.}
\label{tab:hdrsgm.bounds.pol4070}
\end{table}

As happens in the unpolarized case, with a right-polarized electron
beam the sensitivity of some asymmetries is significantly better than
that of the polarized cross section.  In Table
\ref{tab:hdrasym.bounds.pol4070} we summarize the best bounds on
$C^{r,i}_{2,3,4}$ for the relevant asymmetries indicated there, for
initial-electron polarizations $\P=+0.4$ and $+0.7$.  We omit for
brevity the bounds on $C^{r,i}_2$ obtained from the asymmetries of
$\Delta y(\tbar,j_2)$, $\Delta y(\tbar, \bbar)$, which are essentially
the same as those from $\Delta\eta(\bbar,j_2)$, as is the case also
for the unpolarized asymmetries in Table
\ref{tab:hdrasym.bounds.pol00}.  Due to the polarization, the bounds
in that table are significantly better than those from the unpolarized
asymmetries, Table \ref{tab:hdrasym.bounds.pol00}, and due to the
enhanced sensitivity of the asymmetries, also significantly better
than the bounds from the polarized hadronic cross section, Table
\ref{tab:hdrasym.bounds.pol4070}.  Furthermore, the bounds on
$C^{r,i}_{2,3,4}$ from polarized hadronic asymmetries in Table
\ref{tab:hdrasym.bounds.pol4070} are tighter than those from the
unpolarized leptonic asymmetries in Table
\ref{tab:lptasym.bounds.pol00}, most of them are better than those
coming from the polarized leptonic cross section in Table
\ref{tab:lptsgm.bounds.pol4070}, and they are only slightly weaker
than the bounds from polarized leptonic asymmetries, Table
\ref{tab:lptasym.bounds.pol4070}, despite the fact that the
experimental errors assumed in the hadronic channel are twice as large
as those in the leptonic channel.
\begin{table}[t]
  \centering
\begin{tabular}{|c|c||c|c|c|c||c|c|c|}\cline{4-9}
\multicolumn{2}{c}{}&\multicolumn{1}{c|}{} &\multicolumn{3}{c||}{$\epsilon_\mathrm{exp}=7\%$}
                     &\multicolumn{3}{c|}{$\epsilon_\mathrm{exp}=12\%$} \\\cline{3-9}
\multicolumn{2}{c|}{} & Observable
                     & \multicolumn{1}{c|}{60 GeV} 
                     & \multicolumn{1}{c|}{140 GeV} 
                     & \multicolumn{1}{c||}{300 GeV}
                     & \multicolumn{1}{c|}{60 GeV} 
                     & \multicolumn{1}{c|}{140 GeV} 
                     & \multicolumn{1}{c|}{300 GeV} \\\hline
&        &$\Delta\eta(\bbar,j_1)$
         &$\pm$2.47&$\pm$2.43&$\pm$2.29&$\pm$3.27&$\pm$3.23&$\pm$3.05\\\cline{3-9}
&$C^{r,i}_2$ &$\Delta\eta(\bbar,j_2)$
         &$\pm$2.64&$\pm$2.52&$\pm$2.35&$\pm$3.50&$\pm$3.35&$\pm$3.13\\\cline{3-9}
&        &$\Delta\eta(j_1,j_2)$
         &$\pm$2.83&$\pm$2.59&$\pm$2.33&$\pm$3.76&$\pm$3.45&$\pm$3.09\\\cline{2-9}
\raisebox{-20pt}[0pt][0pt]{\rotatebox{90}{$\P=0.4$}}&  &$\Delta\varphi(j_2,\met)$
         &$\pm$1.42&$\pm$1.21&$\pm$1.05&$\pm$1.88&$\pm$1.60&$\pm$1.39\\\cline{3-9}
&\raisebox{10pt}[0pt][0pt]{$C^{r,i}_3$} &$\Delta\varphi(j_1,j_2)$
         &$\pm$1.64&$\pm$1.23&$\pm$0.98&$\pm$2.18&$\pm$1.63&$\pm$1.29\\\cline{2-9}
&\raisebox{-10pt}[0pt][0pt]{$C^{r,i}_4$} &$\Delta\varphi(j_2,\met)$ 
         &$\pm$2.46&$\pm$2.13&$\pm$1.86&$\pm$3.26&$\pm$2.82&$\pm$2.45\\\cline{3-9}
&        &$\Delta\varphi(j_1,j_2)$ 
         &$\pm$2.81&$\pm$2.16&$\pm$1.72&$\pm$3.74&$\pm$2.86&$\pm$2.27\\\hline      
&        &$\Delta\eta(\bbar,j_1)$
         &$\pm$1.58&$\pm$1.57&$\pm$1.47&$\pm$2.09&$\pm$2.08&$\pm$1.95\\\cline{3-9}
&$C^{r,i}_2$ &$\Delta\eta(\bbar,j_2)$
         &$\pm$1.69&$\pm$1.62&$\pm$1.51&$\pm$2.25&$\pm$2.16&$\pm$2.01\\\cline{3-9}
&        &$\Delta\eta(j_1,j_2)$
         &$\pm$1.81&$\pm$1.66&$\pm$1.49&$\pm$2.41&$\pm$2.21&$\pm$1.99\\\cline{2-9}
\raisebox{-20pt}[0pt][0pt]{\rotatebox{90}{$\P=0.7$}}&&$\Delta\varphi(j_2,\met)$
         &$\pm$0.91&$\pm$0.78&$\pm$0.68&$\pm$1.20&$\pm$1.03&$\pm$0.89\\\cline{3-9}
&\raisebox{10pt}[0pt][0pt]{$C^{r,i}_3$} &$\Delta\varphi(j_1,j_2)$
         &$\pm$1.05&$\pm$0.79&$\pm$0.63&$\pm$1.40&$\pm$1.04&$\pm$0.83\\\cline{2-9}
&\raisebox{-10pt}[0pt][0pt]{$C^{r,i}_4$}&$\Delta\varphi(j_2,\met)$ 
         &$\pm$1.57&$\pm$1.37&$\pm$1.19&$\pm$2.08&$\pm$1.81&$\pm$1.57\\\cline{3-9}
&       &$\Delta\varphi(j_1,j_2)$ 
         &$\pm$1.80&$\pm$1.39&$\pm$1.10&$\pm$2.40&$\pm$1.84&$\pm$1.46\\\hline
\end{tabular}
\caption{Bounds at the 68\% CL\ on
  contact-interaction couplings involving initial right-handed
  electrons, obtained from the indicated hadronic-channel asymmetries
  by varying the couplings one at a time.  The definition of the
  asymmetries is as in table \ref{tab:lptasym.bounds.pol00}. The assumed
  experimental errors are 7\% and 12\%.}
\label{tab:hdrasym.bounds.pol4070}
\end{table}

The asymmetries discussed so far are all based on longitudinal-boost
invariant kinematic observables measured in the lab frame not
involving longitudinal neutrino momenta.  We have considered several
other asymmetries of the same type, that we briefly mention here.  We
have not included in tables
\ref{tab:hdrsgm.bounds.pol00}--\ref{tab:hdrasym.bounds.pol4070} the
bounds obtained from the asymmetry $A(\Delta y(\tbar,j_1))$.  At
$E_e=60$ GeV this asymmetry gives the best bounds we have found for
$C^{r,i}_2$, $V^{r,i}_R$, $g^{r,i}_R$, $g^{r,i}_L$.  At $E_e=140$, 300
GeV, however, the bounds from this asymmetry are significantly less
tight.  This phenomenon suggests that those bounds may not be fully
reliable, as is discussed in more detail in Appendix
\ref{sec:cuts.sensitiv}.
The asymmetries of $\Delta y(\tbar,W)$ and $\Delta y(\bbar,W)$ yield
bounds on the effective couplings that are the same as, or weaker
than, those from $\Delta y(\tbar,\bbar)$ given in the tables.  The
asymmetries $\Delta y(j_{1,2},W)$ give the same bounds as
$\Delta\eta(j_1,j_2)$, and the asymmetry
$A(\Delta\varphi(j_1,W),\pi/2)\equiv 1$.  The asymmetries of
$\Delta\varphi(\bbar,j_{1,2})$, $\Delta\varphi(\bbar,W)$,
$\Delta\varphi(j_2,W)$, $\Delta\varphi(\bbar,\met)$,
$\Delta\varphi(j_1,\met)$, all give bounds on the effective couplings
that are weaker than those from the cross section and therefore not
worth examining in detail.

We have also considered the asymmetries associated with the lab frame
observables $\cos(\tbar,j_{1,2})$, $\cos(\tbar,W)$,
$\cos(\tbar,\bbar)$, $\cos(\bbar,j_{1,2})$, $\cos(\bbar,W)$,
$\cos(j_{1,2},W)$, $\cos(j_1,j_2)$.  We have not found any significant
sensitivity to the anomalous couplings for any of them.

Another class of lab-frame kinematic observables giving rise to
asymmetries involves the reconstructed longitudinal momentum of the
final-state neutrino.  Notice that to measure experimentally the
asymmetries of these observables a full reconstruction of the
hard-event kinematics is needed, which can give rise to systematic
errors additional to those involved in pure lab frame measurements.
For the purpose of obtaining bounds at the one-sigma level, we
nevertheless assume the same experimental errors as in the case of lab
frame observables in order to compare the sensitivities of the two
classes of observables.  Within this class we have considered the
longitudinal-boost invariants $\Delta\eta(x,\nu_e)$ with $x=j_{1,2},$
$\bbar$, $\Delta y(\tbar,\nu_e)$ and $\Delta y(W,\nu_e)$.  Notice that
the asymmetry $A(\Delta y(\tbar,\nu_e))$ is equal to the asymmetry of
the longitudinal momentum of $\tbar$ in the center of mass frame,
$A(p_z(\tbar)_\mathrm{c.m.})$.  These five asymmetries possess poor
sensitivity to anomalous $tbW$ couplings, but they are sensitive to
the contact-interaction couplings $C^{r,i}_{2,3,4}$ for which all of
them give similar results, better than those from $\sigma$ in Table
\ref{tab:hdrsgm.bounds.pol00}.  On the other hand, the best bounds for
$C^{r,i}_{2,3,4}$ given in Table \ref{tab:hdrasym.bounds.pol00} are
better than those obtained from these asymmetries, and for this reason
we did not include them in that table.

Another class of asymmetries is based on the observables
$\cos(\tbar_\mathrm{c.m.},x_\mathrm{c.m.})$ with $x=j_{1,2},$ $\bbar$,
$W$. Since both momenta are measured in the center-of-mass frame,
these observables also require full reconstruction of the event
kinematics.  We do not find significant sensitivity to any of the
anomalous couplings in these asymmetries.

We have also studied the class of observables of the form
$\cos(\tbar_\mathrm{c.m.},x_*)$, $x=j_1$, $j_2$, $\bbar$, $W$, where
$\tbar_\mathrm{c.m.}$ refers to the momentum of $\tbar$ in the c.m.
frame and $x_*$ to the momentum of $x$ in the rest frame of $\tbar$.
This class of observables has been considered in the literature in
connection with top quark polarization \cite{tpol}.  They
obviously require a full reconstruction of the partonic kinematics so,
as mentioned above, they may be affected by systematical errors beyond
those involved in pure lab frame measurements.  The asymmetries
$A(\cos(\tbar_\mathrm{c.m.},x_*))$ are found to have good sensitivity
to $C^{r,i}_{3,4}$ and $V^{r,i}_R$, and to a lesser extent to
$C^{r,i}_2$.  The most sensitive to these anomalous couplings is
$\cos(\tbar_\mathrm{c.m.},j_{1*})$, which yields bounds that are
tighter than those from lab-frame asymmetries in Table
\ref{tab:hdrasym.bounds.pol00}.  A summary of the results obtained
from this asymmetry is given in Table \ref{tab:hdrasym.tcmj1*.pol00}.
The asymmetries $\cos(\tbar_\mathrm{c.m.},W_*)$ and
$\cos(\tbar_\mathrm{c.m.},\bbar_*)$ also have good sensitivity,
leading to bounds that are weaker than the tightest ones in Table
\ref{tab:hdrasym.bounds.pol00}, but stricter than those from the cross
section in Table \ref{tab:hdrsgm.bounds.pol00}. The asymmetry
$\cos(\tbar_\mathrm{c.m.},j_{2*})$ does not yield better bounds than
$\sigma$.
\begin{table}[t]
  \centering
\begin{tabular}{|c|c|c|c|c||c|c|c|}\cline{3-8}
\multicolumn{2}{c|}{}&\multicolumn{3}{c||}{$\epsilon_\mathrm{exp}=7\%$}
                     &\multicolumn{3}{c|}{$\epsilon_\mathrm{exp}=12\%$} \\\cline{3-8}
\multicolumn{2}{c|}{} & \multicolumn{1}{c|}{60 GeV} 
                      & \multicolumn{1}{c|}{140 GeV} 
                      & \multicolumn{1}{c||}{300 GeV}
                      & \multicolumn{1}{c|}{60 GeV} 
                      & \multicolumn{1}{c|}{140 GeV} 
                      & \multicolumn{1}{c|}{300 GeV} \\\hline
\raisebox{-60pt}[0pt][0pt]{\rotatebox{90}{$\P=0.0$}}
 &$V^r_R\times10$  &-0.87 0.96&-0.63 0.74&-0.47 0.51&-1.16 1.25&-0.85 0.95&-0.62 0.66 \\\cline{2-8}
 &$V^i_R\times10$  &$\pm$0.90 &$\pm$0.67 &$\pm$0.49 &$\pm$1.18 &$\pm$0.88 &$\pm$0.65  \\\cline{2-8}
 &$C^{r,i}_2$      &$\pm$4.16 &$\pm$2.39 &$\pm$1.48 &$\pm$5.53 &$\pm$3.14 &$\pm$1.94  \\\cline{2-8}
 &$C^{r,i}_3$      &$\pm$1.36 &$\pm$0.83 &$\pm$0.54 &$\pm$1.79 &$\pm$1.09 &$\pm$0.71  \\\cline{2-8}
 &$C^{r,i}_4$      &$\pm$2.32 &$\pm$1.46 &$\pm$0.95 &$\pm$3.05 &$\pm$1.91 &$\pm$1.24  \\\hline
\raisebox{-37pt}[0pt][0pt]{\rotatebox{90}{$\P=0.4$}}
 &$C^{r,i}_2$      &$\pm$2.71 &$\pm$1.53 &$\pm$0.96 &$\pm$3.60 &$\pm$2.02 &$\pm$1.26  \\\cline{2-8}
 &$C^{r,i}_3$      &$\pm$0.89 &$\pm$0.54 &$\pm$0.35 &$\pm$1.17 &$\pm$0.71 &$\pm$0.46  \\\cline{2-8}
 &$C^{r,i}_4$      &$\pm$1.51 &$\pm$0.94 &$\pm$0.62 &$\pm$1.99 &$\pm$1.24 &$\pm$0.81  \\\hline
\raisebox{-37pt}[0pt][0pt]{\rotatebox{90}{$\P=0.7$}}
 &$C^{r,i}_2$      &$\pm$1.73 &$\pm$0.99 &$\pm$0.62 &$\pm$2.30 &$\pm$1.30 &$\pm$0.81  \\\cline{2-8}
 &$C^{r,i}_3$      &$\pm$0.57 &$\pm$0.35 &$\pm$0.23 &$\pm$0.75 &$\pm$0.46 &$\pm$0.30 \\\cline{2-8}
 &$C^{r,i}_4$      &$\pm$0.97 &$\pm$0.61 &$\pm$0.40 &$\pm$1.28 &$\pm$0.80 &$\pm$0.52 \\\hline
\end{tabular}
  \caption{Bounds on effective couplings at the 68\% CL\ 
  obtained by varying the couplings one at a time in the asymmetry
  $A(\cos(\tbar_\mathrm{c.m.},j_{1*}))$, with $j_{1}$ the
  light non-$b$ jet with the larger  $|\vec{p}_T|$.  
  The assumed experimental errors are 7\% and 12\%.}
  \label{tab:hdrasym.tcmj1*.pol00}
\end{table}

Finally, the single-coupling bounds obtained from a further class of
observables not considered here is briefly discussed at the end of
Appendix \ref{sec:cuts.sensitiv}.

\begin{figure}[t]
  \centering
%\fbox{
\includegraphics[scale=1]{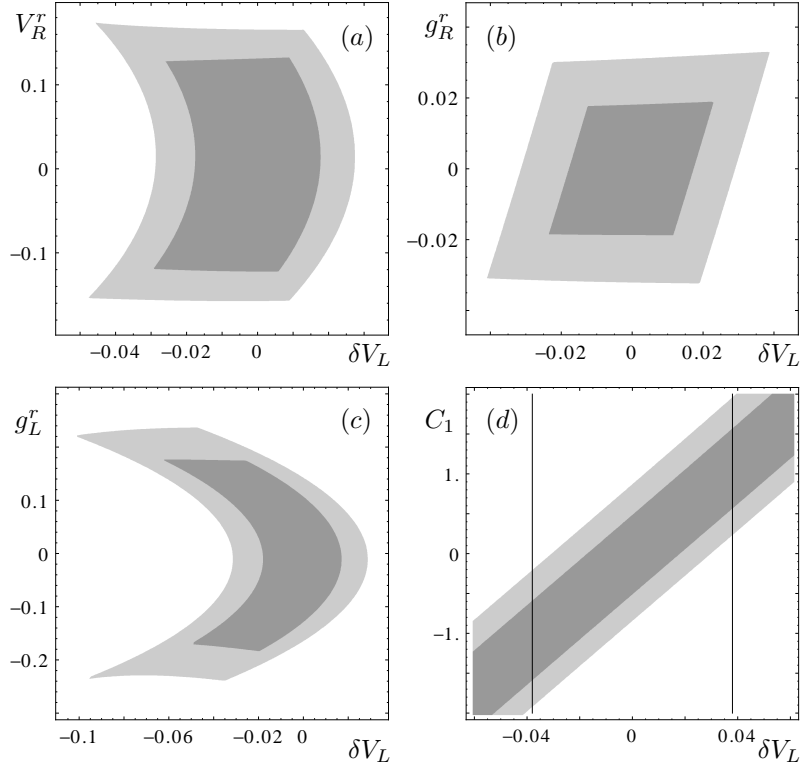}
%}
\caption{Allowed regions at 68\% CL\ at $E_e=60$ GeV for (a) $\delta
  V_L$, $V_R$ from the cross section and the asymmetry
  $A(\Delta\eta(\bbar,j_1))$, (b) $\delta V_L$, $g_R$ from the cross
  section and the asymmetry $A(\Delta\eta(\bbar,j_1))$, (c) $\delta
  V_L$, $g_L$ from the cross section and the asymmetry
  $A(\Delta\varphi(j_2,\met))$, (d) $\delta V_L$, $C_1$ from the cross
  section. The vertical lines in (d) correspond to the current bounds
  $|\delta V_L| < 0.038$ \cite{cms2014}. The light-gray
  regions correspond to $\varepsilon_\mathrm{exp}=12\%$, the darker
  regions to $\varepsilon_\mathrm{exp}=7\%$.  }
  \label{fig:xcl.1}
\end{figure}

We now turn to correlated regions of allowed parameter space, obtained
by considering two effective couplings to be simultaneously
non-vanishing and by supplying the necessary additional interference
terms in (\ref{lheccross}).  Figure \ref{fig:xcl.1} shows the allowed
regions at the 68\% confidence level (CL) and at $E_e=60$ GeV,
assuming an unpolarized electron beam in all cases, for $\delta V_L$
versus $V^r_R$, $g^r_R$, $g^r_L$ and $C_1$. Figures \ref{fig:xcl.1}
(a) and (b) show the allowed regions for $V^r_R$ and $g^r_R$,
respectively, as determined by the level curves of the cross section
and $A(\Delta\eta(\bbar,j_1))$.  As seen in the figure, those
observables yield closed regions consistent with the bounds given in
tables \ref{tab:hdrsgm.bounds.pol00} and
\ref{tab:hdrasym.bounds.pol00}.  In the case of $\delta V_L$ .vs.\
$g^r_L$ the only observable with significant sensitivity is the cross
section, which gives an elliptical corona containing the origin at its
periapsis as allowed region.  In order to restrict that region to a
neighborhood of the origin we use the asymmetry
$A(\Delta\varphi(j_2,\met))$ which has a somewhat poor sensitivity to
$g^r_L$, as shown in Figure \ref{fig:xcl.1} (c).  As a result, the
bounds on $g^r_L$ displayed in the figure are less restrictive than
those in Table \ref{tab:hdrsgm.bounds.pol00}.  For $C_1$ and $\delta
V_L$ the cross section is the only available observable with any
sensitivity which, as seen in Figure \ref{fig:xcl.1} (d), for small
values of those couplings depends on them only through the linear
combination $\simeq \delta V_L+(5/4)\,v^2/(2\Lambda^2)C_1$.  As in
Figure \ref{fig:xcl.a} for the leptonic channel, we include in Figure
\ref{fig:xcl.1} (d), for reference, the current bounds $V_L = 0.998
\pm 0.038$ (exp) from CMS \cite{cms2014}.

\begin{figure}[t]
  \centering
\includegraphics[scale=1]{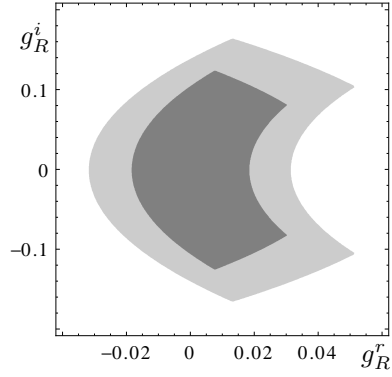}
\caption{Allowed regions at 68\% CL\ at $E_e=60$ GeV for $g_R$
  obtained from the asymmetries $A(\Delta\eta(\bbar,j_1))$ and
  $A(\Delta\varphi(j_2,\met))$. The light-gray area corresponds to
  $\varepsilon_\mathrm{exp}=12\%$, the darker one to
  $\varepsilon_\mathrm{exp}=7\%$.}
  \label{fig:xcl.5}
\end{figure}

We have also considered the allowed regions on the complex plane of
the $tbW$ couplings $V_R$, $g_R$ and $g_L$.  For $V_R$, the allowed
region is obtained from the asymmetry $A(\Delta(\eta(\bbar,j_1)))$
which gives the best bounds on that coupling. It consists of a solid
ellipse inscribed in the rectangle formed with the single-coupling
bounds in Table \ref{tab:hdrasym.bounds.pol00}.  Similarly, the best
bounds for $g_L$ are obtained from the unpolarized cross section,
which leads to an allowed region in the $g^r_L$--$g^i_L$ plane
consisting of a solid ellipse inscribed in the rectangle formed by the
bounds in Table \ref{tab:hdrsgm.bounds.pol00}.  A figure is clearly
not needed for these ellipses.  The best bounds on $g_R$ are obtained
from $A(\Delta\eta(\bbar,j_1))$, which determines an allowed region
shaped as a two-dimensional toroidal region containing the origin.
The asymmetry $A(\Delta(\varphi(j_2,\met)))$ has some sensitivity to
$g^i_R$ and the allowed region it determines cleanly intersects the
previous toroid.  The allowed region in the complex $g_R$ plane at the
68\% CL\ at $E_e=60$ GeV is shown in Figure \ref{fig:xcl.5} for both
$\varepsilon_\mathrm{exp}=12\%$ and $\varepsilon_\mathrm{exp}=7\%$.

The allowed regions for $C^r_2$--$C^i_2$ (from
$A(\Delta(\eta(\bbar,j_1)))$) and for $C^r_3$--$C^i_3$ and
$C^r_4$--$C^i_4$ (from $A(\Delta(\varphi(j_2,\met)))$) are ellipses
inscribed in the rectangle formed with the single-coupling bounds from
tables \ref{tab:hdrasym.bounds.pol00} and
\ref{tab:hdrasym.bounds.pol4070}.  There is, then, no need to display
them explicitly.

\begin{figure}[t]
  \centering
\includegraphics[scale=1]{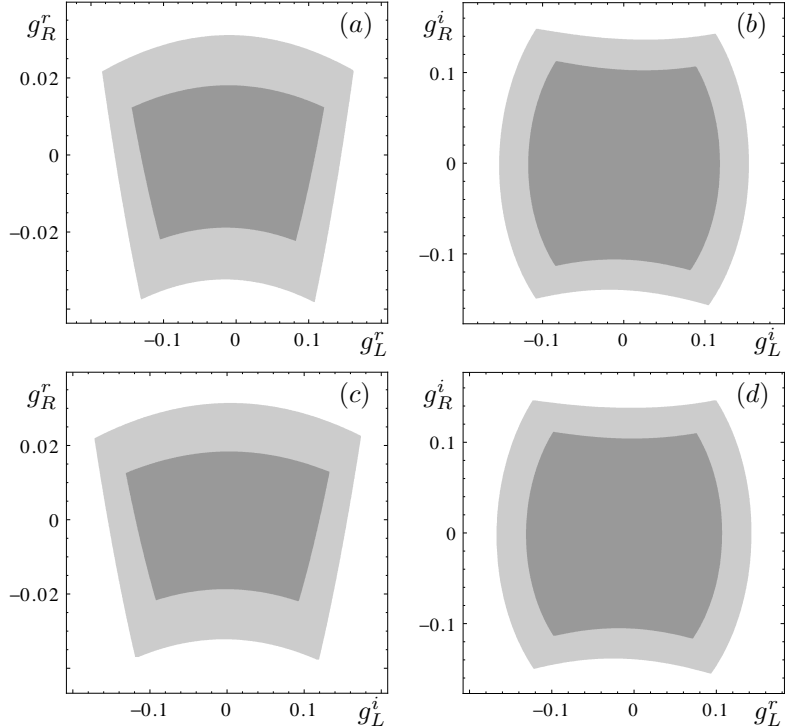}
\caption{Allowed regions at 68\% CL\ at $E_e=60$ GeV, from the
  hadronic cross section and the asymmetry $A(\Delta\eta(\bbar,j_1))$,
  with $\varepsilon_\mathrm{exp}=12\%$ (light gray) and
  $\varepsilon_\mathrm{exp}=7\%$ (dark gray). }
  \label{fig:xcl.4}
\end{figure}

Figure \ref{fig:xcl.4} shows the allowed regions in the four planes
$g_L$--$g_R$, determined by the cross section and the asymmetry
$A(\Delta\eta(\bbar,j_1))$ at 68\% CL\ and $E_e=60$ GeV. 

\begin{figure}[t]
  \centering
\includegraphics[scale=1]{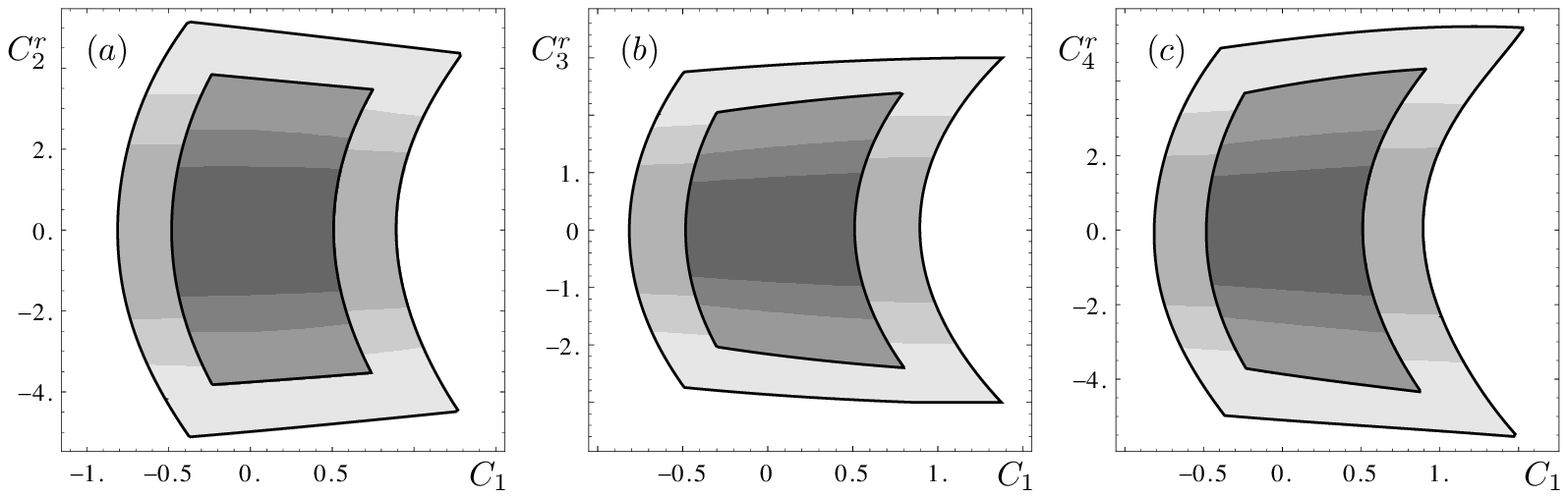}
\caption{Allowed regions at 68\% CL\ at $E_e=60$ GeV for (a)
  $C^r_1$, $C^r_2$ from the unpolarized cross section and the
  asymmetry $A(\Delta\eta(\bbar,j_1))$, (b) $C^r_1$, $C^r_3$ from the
  unpolarized cross section and the asymmetry
  $A(\Delta\varphi(j_2,\met))$, (c) $C^r_1$, $C^r_4$ from the
  unpolarized cross section and the asymmetry
  $A(\Delta\varphi(j_2,\met))$.  The larger quadrangles in each panel
  correspond to $\varepsilon_\mathrm{exp}=12\%$, the smaller ones to
  $\varepsilon_\mathrm{exp}=7\%$.  Within each quadrangle the
  lighter-gray region corresponds to $\P=0$, the medium-gray region
  to $\P=0.4$ and the darker-gray region to $\P=0.7$.  }
  \label{fig:xcl.2}
\end{figure}

In Figure \ref{fig:xcl.2} we display the allowed regions for the
contact--interaction couplings $C^r_{2,3,4}$ versus $C_1$, determined
by the unpolarized cross section, which bounds $C_1$, and the
polarized asymmetries $A(\Delta\eta(\bbar,j_1))$ (which bounds
$C^r_2$) and $A(\Delta\varphi(j_2,\met))$ (which bounds $C^r_{3,4}$).
The effect of polarization on the sensitivity of the asymmetries on
$C_{2,3,4}$ is clearly shown in the figure.  Notice that the areas
defined by each gray tone are (simply) connected, though they look
disconnected in the figure because they are stacked on one another. 

\begin{figure}[t]
  \centering
\includegraphics[scale=1]{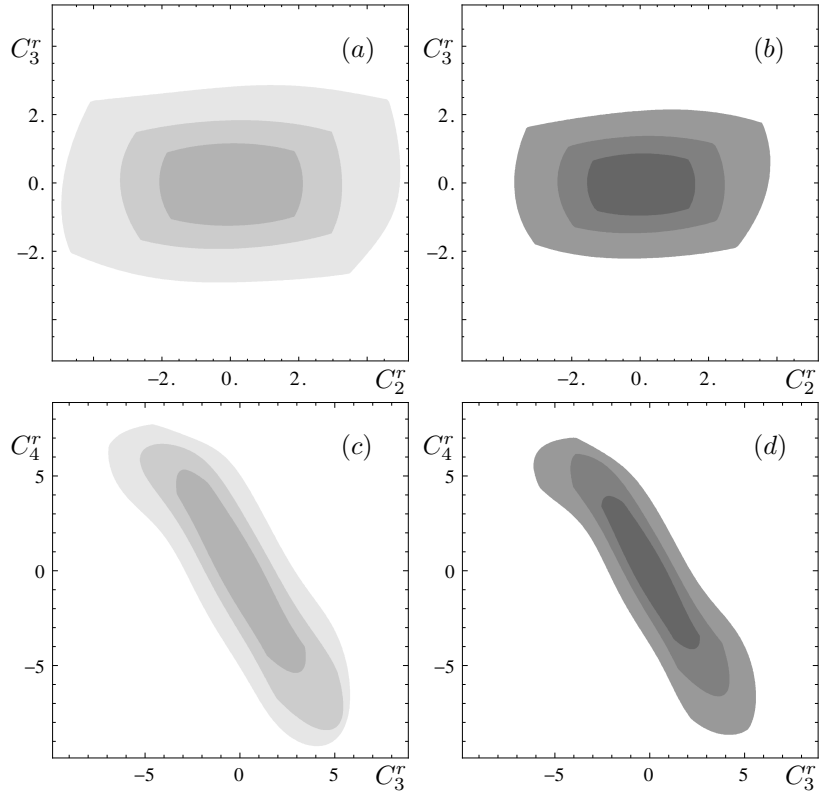}
\caption{Upper row: allowed regions at 68\% CL\ at $E_e=60$ GeV for
  $C^r_2$, $C^r_3$ from the cross section and the asymmetries
  $A(\Delta\eta(\bbar,j_1))$ and $A(\Delta\varphi(j_2,\met))$ with (a)
  $\varepsilon_\mathrm{exp}=12\%$ and (b)
  $\varepsilon_\mathrm{exp}=7\%$.  Lower row: allowed regions at 68\% CL\ at $E_e=60$ GeV for
  $C^r_3$, $C^r_4$ from the cross section and the asymmetry
  $A(\Delta\varphi(j_2,\met))$ with (c)
  $\varepsilon_\mathrm{exp}=12\%$ and (d)
  $\varepsilon_\mathrm{exp}=7\%$.  In each panel the lighter-gray
  region corresponds to $\P=0$, the medium-gray region to $\P=0.4$ and
  the darker-gray region to $\P=0.7$.}
  \label{fig:xcl.3}
\end{figure}

Figure \ref{fig:xcl.3} shows the allowed regions in the planes
$C^r_2$--$C^r_3$ (determined by the cross section and the asymmetries
$A(\Delta\eta(\bbar,j_1))$ and $A(\Delta\varphi(j_2,\met))$) and
$C^r_3$--$C^r_4$ (determined by the cross section and the asymmetry
$A(\Delta\varphi(j_2,\met))$), at 68\% CL\ and $E_e=60$ GeV, for
$\varepsilon_\mathrm{exp}=7\%$ and 12\%, and for the three
polarizations $\P=0$, $+0.4$, $+0.7$.  Here the effect of the
polarization on the sensitivity is apparent, and also the fact that
the interference terms proportional to $C^r_3C^r_4$ are significantly
larger than those proportional to $C^r_2C^r_3$.  The allowed regions
in the planes of the imaginary parts $C^i_2$--$C^i_3$ and
$C^i_3$--$C^i_4$ are identical to those in the figure for the real
parts.  Although the interference terms proportional to $C^r_3C^i_4$
and $C^i_3C^r_4$ do not vanish, they turn out to be small.  As a
result, the allowed regions in the planes $C^r_3$-$C^i_4$ and
$C^i_3$-$C^r_4$ are solid ellipses inscribed in the rectangles
determined by the bounds from $A(\Delta\varphi(j_2,\met))$ in tables
\ref{tab:hdrasym.bounds.pol00} and \ref{tab:hdrasym.bounds.pol4070}.

%%%%%%%%%%%%%%%%%%%%%%%%%%%%%%%%%%%%%%%%%%%%%%%%%%%%%%%%%%
\section{Conclusions}
\label{conclusions}

In this paper we have investigated the sensitivity of the LHeC to
probe new physics in single-antitop production within the framework of
the most general $SU(2)_L\times U(1)$-gauge invariant effective
Lagrangian of dimension six. In this theoretical context, a systematic
study of single-antitop production must take into account the fact
that the usual operator basis associated with the top-quark
charged-current interaction Lagrangian (\ref{bachfactors}) is complete
only up to four-fermion operators
\cite{aguilaroperators,aguilar4,bach}.  The appropriate complete
Lagrangian is discussed in Section \ref{secoperators}.

Since the LHeC will necessarily run simultaneously with the HL-LHC, it
is of interest to try to estimate the potential sensitivity to the
effective couplings of both programs. We do that in Section
\ref{tdecandsingle}, in a simplified approach involving $W$ helicity
fractions in top decay (in the approximation $t\rightarrow bW$) at the
LHC and single-top production (in the two-body approximation) at the
LHC and LHeC, and taking input from recent experimental data from CMS.
For simplicity, we consider only cross-section measurements and
$CP$-even couplings.  Our estimates indicate that the LHeC should
significantly improve the bounds of the HL-LHC on $V_L$, and lead to
somewhat tighter bounds on $V_R$. The tensor couplings $g_{L,R}$ would
be moderately better constrained by the HL-LHC than the LHeC.  The
bounds on the contact-interaction coupling $C_1$ at the LHeC are
expected from our estimates to be smaller than those on $g_\times$ at
the HL-LHC by a factor of 2.

In Section \ref{sec:sm} we considered single-top production in the
Standard Model in both leptonic and hadronic channel.  We found that
backgrounds in leptonic channel are quite mild, leading to a lower
bound on the experimental error as low as 3\%.  In the hadronic
channel, on the other hand, a strong reducible background results in
expected experimental uncertainties about twice as large as in the
leptonic channel.  For polarizations less than 90\%, our error
estimates are roughly independent of electron beam polarization.

In Section \ref{sec:eff.oper} we obtained bounds from the cross
section and several asymmetries on the eight effective couplings, for
three values of the electron beam energy ($E_e=60$, 140, 300 GeV),
three values of polarization ($\P=0$, 40\%, 70\%), and for the various
values of experimental uncertainty assumed in the previous section.
The cross section is the only observable we found to be significantly
sensitive to the couplings $V_L$, $g_L$ and $C_1$. For them, the
tightest bounds come from the leptonic channel cross section due to
its small expected experimental error.  In the leptonic channel,
discussed in Section \ref{sec:lpt.chn}, there are only three
asymmetries that give better sensitivity to some of the remaining
couplings $V_R$, $g_R$, $C_{2,3,4}$ than that of the cross section.
These are the asymmetries of $\Delta\eta(\bbar,\ell)$ (sensitive to
$C_2$, $V_r$ and $g_R$), $\Delta\varphi(\bbar,\ell)$ (sensitive to
$C_{3,4}$) and $\Delta\varphi(\ell,\met)$ (sensitive to $C_3$).  The
bounds obtained from these asymmetries, however, are only modestly
tighter than the ones from the cross section. (With the only exception
of $C_2$, for which the bounds resulting from the cross section are
about twice as large as those from the asymmetry of
$\Delta\eta(\bbar,\ell)$.)

%%%%%%%%%%% cambio
The hadronic channel of single-top production is especially
interesting because it has not been experimentally observed until now.
The bounds obtained from the cross section in that channel are looser
than those from the leptonic one because the errors assumed for the
former are twice as large as those of the latter.  In this channel,
however, it is possible to completely determine the kinematics of the
underlying partonic event, which gives rise to a large spectrum of
measurable asymmetries.  In section \ref{sec:hdr.chn}, we made an
extensive survey of kinematic observables and assessed the sensitivity
of their asymmetries to the anomalous couplings.  Unlike the leptonic
case, in the hadronic one several asymmetries were found to possess
much higher sensitivity than the cross section.  Those asymmetries
yield bounds on some effective couplings that are almost as tight as
the corresponding bounds in the leptonic channel with half the
experimental error.  In a few cases, the hadronic asymmetries yield
even slightly better bounds than the leptonic ones. That is the case
of the bounds on $g^r_R$ from the asymmetry of
$\Delta\eta(\bbar,j_1)$, and those on $C^r_{3,4}$ from
$\Delta\varphi(j_2,\met)$.  For the four-fermion couplings
$C_{2,3,4}$, involving right-handed electrons, a right-polarized beam
results in a strong sensitivity enhancement.

In summary, the LHeC will provide a clean experimental environment in
which it will be possible to obtain bounds on the $tbW$ vertex that
will be competitive with those from the HL-LHC, and in the case of the
vector form-factors arguably even better.  Furthermore, the LHeC will
provide unique information on four-fermion contact interactions
involving third-generation quarks and first-generation leptons.  Thus,
the input from the LHeC will be very useful in studying the effective
dimension six operators relevant to top-quark physics.

%%%%%%%%%%%%%%%%%%% reconocimientos %%%%%%%%%%%%%%%%%%%

\noindent {\bf Acknowledgments}~~~ We acknowlegde support from Conacyt
and Sistema Nacional de Investigadores de M\'exico.

%%%%%%%%%%%%%%%%%%%%%%%%%%%%%%%%%%%%%%%%%%%%%%%%

\appendix

\section{Cross section parameterization}
\label{sec:param}

In this appendix we gather, for reference, the numerical values of the
coefficients in Eqs.\ (\ref{lhccross}), (\ref{lheccross0}) and
(\ref{lheccross}).  See Tables
\ref{coefsinglelhc}--\ref{tab:hdrsgm.coeff.pol}.   

%%%%%%%%%%%%%%%%%%%%%%%%%%%%%%%%%%%%%%%%%%%%%%%%%%%%%%%%%%
\begin{table}
\begin{center}
  \begin{tabular}{|c|c|c|c|c|c|c|}\hline
 & $a_{12}$ & $a_{2}$ & $a_{3}$ & $a_{4}$ & $a_{15}$ & $a_{5}$
\\\hline\hline
Tevatron & $-1.90$ & $3.85$ & $0.89$ & $3.39$ & $-0.12$ & $0.09$
\\\hline
LHC7 & $-0.38$ & $1.68$ & $0.95$ & $1.45$ & $0.141$ & $0.017$
\\\hline
LHC8 & $-0.36$ & $1.73$ & $0.95$ & $1.51$ & $0.143$ & $0.019$
\\\hline
LHC14 & $-0.31$ & $1.84$ & $0.97$ & $1.67$ & $0.148$ & $0.027$
\\\hline
  \end{tabular}
\end{center}
\caption{Coefficients of Eq.~(\ref{lhccross}).}
\label{coefsinglelhc}
\end{table}
%%%%%%%%%%%%%%%%%%%%%%%%%%%%%%%%%%%%%%%%%%%%%%%%%%%%%%%%%%

%%%%%%%%%%%%%%%%%%%%%%%%%%%%%%%%%%%%%%%%%%%%%%%%%%%%%%%%%%
\begin{table}
\begin{center}
  \begin{tabular}{|c|c|c|c|c|c|c|c|c|c|c|}\hline
 & $b_{12}$ & $b_{2}$ & $b_{3}$ & $b_{4}$ & $b_{34}$
& $b_{15}$ & $b_{5}$ & $b_{6}$ & $b_{7}$ & $b_{8}$
\\\hline\hline 
60 GeV & $-0.33$ & $0.86$ & $1.34$ & $2.43$ & $-1.12$ &
$-0.12$ & $0.006$ & $0.002$ & $0.006$ & $0.002$ 
\\\hline
140 GeV & $-0.31$ & $1.09$ & $1.29$ & $2.45$ &
$-0.96$ & $-0.13$ & $0.008$ & $0.002$ & $0.007$ & $0.003$
\\\hline
300 GeV & $-0.29$ & $1.29$ & $1.26$ & $2.50$ &
$-0.84$ & $-0.14$ & $0.01$ & $0.003$ & $0.009$ & $0.003$
\\\hline
  \end{tabular}
\end{center}
\caption{Coefficients of Eq.~(\ref{lheccross0}).}
\label{coefsinglelhec}
\end{table}
%%%%%%%%%%%%%%%%%%%%%%%%%%%%%%%%%%%%%%%%%%%%%%%%%%%%%%%%%%

\begin{table}[t]
  \centering
  \begin{tabular}{|c|r|r|r|}\hline
\multicolumn{4}{|c|}{$\P=0.0$}\\\hline
$E_e$[GeV] & 60 & 140 & 300\\\hline
$b_2$&1.23&1.56&1.82 \\\hline
$b_3$&2.52&2.38&2.34 \\\hline
$b_4$&4.31&4.28&4.30 \\\hline
$b_5$&0.0062&0.0086&0.0119\\\hline
$b_6$&0.0018&0.0023&0.0030\\\hline
$b_7$&0.0051&0.0069&0.0092\\\hline
$b_8$&0.0020&0.0025&0.0033\\\hline
  \end{tabular}
  \begin{tabular}{|c|c|c|c|}\hline
\multicolumn{4}{|c|}{$\P=0.0$}\\\hline
$E_e$[GeV] & 60 & 140 & 300 \\\hline
$d_2$&2.43&2.65&2.86\\\hline
$d_3$&2.52&2.38&2.37\\\hline
$d_4$&4.33&4.30&4.30\\\hline
$b_{12}$&-1.37&-1.32&-1.32\\\hline
$b_{13}$&-0.061&-0.065&-0.072\\\hline
$b_{14}$&0.107&0.103&0.104\\\hline
$b_{15}$&-0.123&-0.134&-0.145\\\hline
  \end{tabular}
  \begin{tabular}{|c|c|c|c|}\hline
\multicolumn{4}{|c|}{$\P=0.4$}\\\hline
$E_e$[GeV] & 60 & 140 & 300\\\hline
$b_6\times10^2$& $0.425$& $0.537$ & $0.701$\\\hline
$b_7\times10^2$& $1.18 $& $1.62 $ & $2.15 $\\\hline
$b_8\times10^2$& $0.471$& $0.586$ & $0.753$\\\hline
\multicolumn{4}{|c|}{$\P=0.7$}\\\hline
$b_6\times10^2$&$1.03$ &$1.31$  &$1.71$  \\\hline
$b_7\times10^2$&$2.86$ &$3.93$  &$5.22$  \\\hline
$b_8\times10^2$&$1.14$ &$1.43$  &$1.83$  \\\hline
  \end{tabular}
  \caption{Coefficients in Eq.~(\ref{lheccross})
for unpolarized and polarized cross section in leptonic 
channel. The values of $d_{6,7,8}$ are the same as those
of $b_{6,7,8}$ shown in this Table.}
  \label{tab:lptsgm.coeff.pol}
\end{table}

\begin{table}[t]
  \centering
  \begin{tabular}{|c|r|r|r|}\hline
\multicolumn{4}{|c|}{$\P=0.0$}\\\hline
$E_e$[GeV] & 60 & 140 & 300\\\hline
$b_2$&1.45&1.87&2.19\\\hline
$b_3$&2.61&2.46&2.44\\\hline
$b_4$&4.93&4.56&4.43\\\hline
$b_5$&0.0076&0.010&0.013\\\hline
$b_6$&0.0025&0.0027&0.0033\\\hline
$b_7$&0.0065&0.0079&0.010\\\hline
$b_8$&0.0027&0.0029&0.0035\\\hline
  \end{tabular}
  \begin{tabular}{|c|c|c|c|}\hline
\multicolumn{4}{|c|}{$\P=0.0$}\\\hline
$E_e$[GeV] & 60 & 140 & 300 \\\hline
$d_2$   &2.64&2.89&3.12\\\hline
$d_3$   &2.60&2.47&2.41\\\hline
$d_4$   &4.95&4.56&4.46\\\hline
$b_{12}$&-1.22&-1.30&-1.35\\\hline
$b_{13}$&-0.074&-0.074&-0.069\\\hline
$b_{14}$&0.111&0.105&0.096\\\hline
$b_{15}$&-0.142&-0.150&-0.158\\\hline
  \end{tabular}
  \begin{tabular}{|c|c|c|c|}\hline
\multicolumn{4}{|c|}{$\P=0.4$}\\\hline
$E_e$[GeV] & 60 & 140 & 300\\\hline
$b_6\times10^2$&0.581&0.640&0.765\\\hline
$b_7\times10^2$&1.501&1.847&2.323\\\hline
$b_8\times10^2$&0.617&0.683&0.822\\\hline
\multicolumn{4}{|c|}{$\P=0.7$}\\\hline
%$E_e$[GeV] & 60 & 140 & 300\\\hline
$b_6\times10^2$&1.408&1.551&1.850 \\\hline
$b_7\times10^2$&3.649&4.490&5.635 \\\hline
$b_8\times10^2$&1.500&1.664&1.998 \\\hline
  \end{tabular}
%  \begin{tabular}{|c|c|c|c|}\hline
%\multicolumn{4}{|c|}{$\P=0.7$}\\\hline
%%$E_e$[GeV] & 60 & 140 & 300\\\hline
%$b_6\times10^2$&1.408&1.551&1.850 \\\hline
%$b_7\times10^2$&3.649&4.490&5.635 \\\hline
%$b_8\times10^2$&1.500&1.664&1.998 \\\hline
%  \end{tabular}
  \caption{Coefficients in Eq.~(\ref{lheccross}) for unpolarized and
    polarized cross section in hadronic 
    channel. The values of $d_{6,7,8}$ are the same as those of
    $b_{6,7,8}$ shown in this Table.}
  \label{tab:hdrsgm.coeff.pol}
\end{table}

\section{Stability of parton-level bounds}
\label{sec:cuts.sensitiv}

\begin{figure}[t]
  \centering
\includegraphics[scale=1]{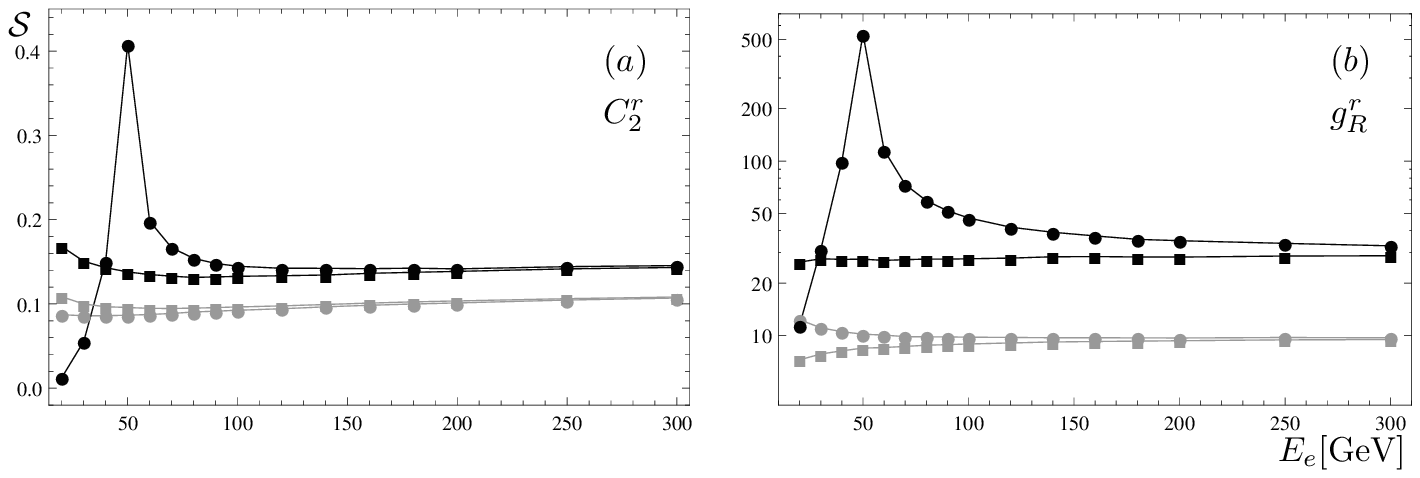}    
\caption{Sensitivity $\mathcal{S}$ of the cross section (gray markers)
  and the asymmetry $A(\Delta\eta(\bbar,j_1))$ (black markers) to the
  effective couplings (a) $C^r_2$ and (b) $g^r_R$, computed with the
  cuts $H_4$ defined in (\ref{eq:cuts.H}) with $|\eta(j_1)|<2.5$
  (solid boxes) and $|\eta(j_1)|<5$ (solid circles). The experimental
  error is assumed to be $\varepsilon_\mathrm{exp}=7\%$. The lines are
  only to guide the eye.}
  \label{fig:sens1}
\end{figure}

We have focused in the main body of this paper on obtaining bounds on
the anomalous couplings appearing in the effective Lagrangians
(\ref{fourfermion}) and (\ref{tbwlagrangian}).  We considered a
certain observable $X$ such as the cross section $\sigma$ or an
asymmetry associated with a kinematical variable and, by assuming that
an experimental measurement of $X$ was consistent with the SM result
within the experimental error $\Delta\sigma_\mathrm{exp}$, we obtained
bounds $\lambda_\mathrm{min}<0<\lambda_\mathrm{max}$ on an anomalous
coupling $\lambda$ at the level of one standard deviation.  As
remarked in the previous sections, the bounds obtained from the cross
section have a very mild dependence on $E_e$.  For other observables,
such as the various asymmetries considered above, that dependence may
be more complex and it may be desirable to characterize it.  For that
purpose, we introduce a simple measure of the sensitivity of an
observable $X$ to an anomalous coupling $\lambda$ by
\begin{equation}
  \label{eq:sensitivity}
  \S(X,\lambda) = \frac{1}{\lambda_\mathrm{max}-\lambda_\mathrm{min}}.
\end{equation}
In Figure \ref{fig:sens1} we show the sensitivity of the
hadronic-channel cross section (gray lines in the figure) to $C^r_2$
and $g^r_R$ as a function of $E_e$.  We computed the bounds assuming
an experimental error $\varepsilon_\mathrm{exp}=7\%$, with the cuts
$H_4$ defined in (\ref{eq:cuts.H}) (solid gray squares in the figure),
and also with a modified set of cuts obtained from $H_4$ by
substituting the requirement $|\eta(j)|<2.5$ in (\ref{eq:cuts.H}) by
the less restrictive one $|\eta(j)|<5$ (solid gray circles in the
figure). As seen in the figure both sets of cuts lead to essentially
the same sensitivity of the cross section to both couplings,
especially at $E_e>50$ GeV.  In all cases the sensitivity of the cross
section shows little variation with the energy $E_e$.
Also shown in Figure \ref{fig:sens1} is the sensitivity of
$A(\Delta\eta(\bbar,j_1))$ to $C^r_2$ and $g^r_R$ (black lines in the
figure), which is seen to be larger than that of $\sigma$ over the
entire range of $E_e$, as is also apparent from tables
\ref{tab:hdrsgm.bounds.pol00} and \ref{tab:hdrsgm.bounds.pol00}.
Whereas the dependence on $E_e$ of the sensitivity of this asymmetry
is seen to be essentially flat when computed with the cuts $H_4$ with
$|\eta(j)|<2.5$ (solid black squares in the figure), it has a sharp
peak about $E_e=50$ GeV when that cut is relaxed to $|\eta(j)|<5$
(solid black circles).  The peak is especially pronounced in the
sensitivity to $g^r_R$. 

It is not obvious that a sensitivity profile with such rapid
variations with $E_e$ would not be drastically altered by the
incorporation of radiative corrections, showering and hadronization in
the computation of the asymmetry.  Since those more detailed
computations are beyond the scope of the present preliminary study, we
have not considered in this paper observables whose sensitivity has a
strong dependence with $E_e$.  In the particular case of
$A(\Delta\eta(\bbar,j_1))$, the very tight bounds obtained at $E_e=60$
GeV on $g^r_R$ and other effective couplings when the cut
$|\eta(j)|<5$ is used are probably an artifact of the tree-level
partonic approximation used here, and therefore potentially
misleading.  It is for this reason that we chose the more restrictive
cut on $|\eta(j)|<2.5$ in (\ref{eq:cuts.H}), with which we obtain
bounds that are somewhat less tight at low energies but also more
reliable.

\begin{figure}[t]
  \centering
\includegraphics[scale=1]{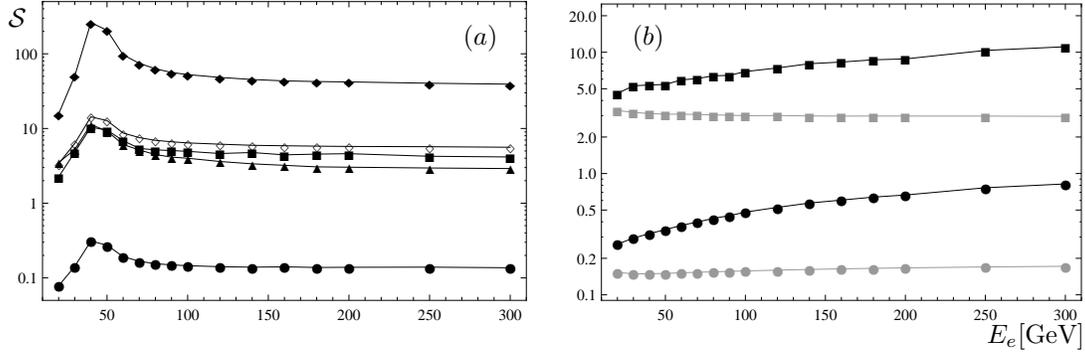}    
\caption{(a) Sensitivity of the asymmetry $A(\Delta y(\tbar,j_1))$ to
  $C^r_{2}$ (solid circles), $V^r_R$ (solid squares), $g^r_L$ (solid
  triangles), $g^r_R$ (solid diamonds) and $g^i_R$ (open
  diamonds). (b) Sensitivity of the asymmetry
  $A(\cos(\tbar_\mathrm{c.m.},j_{1*})$ (black markers) and of the
  cross section (gray markers) to $C^r_{3}$ (solid circles) and
  $V^r_R$ (solid squares).  All observables computed with the cuts
  $H_4$ defined in (\ref{eq:cuts.H}).  The experimental error is
  assumed to be $\varepsilon_\mathrm{exp}=7\%$. The lines are only to
  guide the eye.}
  \label{fig:sens2}
\end{figure}

In Figure \ref{fig:sens2} (a) we show the sensitivity of the asymmetry
$A(\Delta y(\tbar,j_1))$ to the anomalous couplings $V^r_R$, $g^r_R$,
$g^i_R$, $g^r_L$, $C^r_2$ as a function of $E_e$.  As already remarked
in section \ref{sec:hdr.chn}, the sensitivity of this asymmetry at
$E_e=60$ GeV is much larger than at 140 and 300 GeV.  The figure shows
that the sensitivity has a peak at about $E_e=40$ GeV where it is
larger than the average sensitivity at higher energies by a factor
3--10 depending on the coupling.  Thus, the bounds obtained from this
asymmetry at $E_e=60$ GeV are enhanced by the proximity of the peak
and therefore possibly unreliable.  For energies $E_e\geq 120$ GeV the
dependence of the sensitivity on $E_e$ is much weaker, as seen in the
figure, which suggests to us that the bounds in that region are more
realistic than at lower energies.

Figure \ref{fig:sens2} (b) shows the sensitivity of the asymmetry
$A(\cos(\tbar_\mathrm{c.m.},j_{1*}))$ (black markers) to the anomalous
couplings $V^r_R$ (solid squares), and $C^r_3$ (solid circles) as a
function of $E_e$. Also shown for reference is the sensitivity of the
cross section (gray markers).  As is already apparent from Table
\ref{tab:hdrasym.tcmj1*.pol00}, and can be seen in more detail in the
figure, the sensitivity of this asymmetry increases monotonically with
$E_e$ (by a factor $\gtrsim2$ as $E_e$ varies from 30 to 300 GeV)
without peaks or rapid oscillations.

A class of asymmetries not discussed in section \ref{sec:hdr.chn} is
based on longitudinal-boost non-invariant kinematic observables in lab
frame involving longitudinal neutrino momenta.  Among those, we have
considered $\cos(\tbar,\nu_e)$, $\cos(j_{1,2},\nu_e)$,
$\cos(\bbar,\nu_e)$, $\cos(W,\nu_e)$.  These observables yield tight
bounds on $C^{r,i}_{2,3,4}$ at $E_e=60$ GeV.  Their sensitivity,
however, seems to vary rapidly with $E_e$, becoming very small at
$E_e=140$ GeV but large at the other two energies.  For instance, for
$C^{r}_2$ from the asymmetry of $\cos(j_{2},\nu_e)$ we find the bounds
$\pm2.87$, $\pm5.54$, $\pm0.91$ at $E_e=60$, 140 and 300 GeV,
respectively, with an assumed experimental error of
$\varepsilon_\mathrm{exp}=7\%$.  For $C^{r}_3$ from the asymmetry of
$\cos(\tbar,\nu_e)$ we find the bounds $\pm2.15$, $\pm7.67$,
$\pm1.87$, and for $C^{r}_4$ from the asymmetry of $\cos(\bbar,\nu_e)$
we get $\pm3.11$, $\pm6.0$, $\pm5.16$, at the same energies and with
the same experimental error.  It is because of that suppressed
sensitivity at the intermediate energy that bounds from these
observables are not discussed in section \ref{sec:hdr.chn}.

\end{document}